\documentclass[11pt]{article}
\usepackage{a4wide,amsmath,amssymb,epsfig,slashed}
\allowdisplaybreaks[2]

\begin{document}

\begin{titlepage}

\indent\indent
\begin{flushright}
CPT-2004/P.081
\end{flushright}
\vspace*{0.2cm}

\begin{center}
{\Large {\bf The Dalitz decay $\pi^0 \rightarrow e^+e^-\gamma$
revisited}}\\[2 cm]

{\bf Karol Kampf}~$^{a,}$\footnote{karol.kampf@mff.cuni.cz},
 {\bf Marc Knecht}~$^{b,}$\footnote{knecht@cpt.univ-mrs.fr} and
{\bf Ji\v{r}\'{\i} Novotn\'y}~$^{a,}$\footnote{jiri.novotny@mff.cuni.cz}\\[1 cm]
$^a${\it Institute of Particle and Nuclear Physics, Charles
University,}\\{\it V Holesovickach 2, 180 00 Prague 8, Czech
Republic}
\\[0.5 cm]
$^b${\it Centre de Physique Th\'eorique}\footnote{Unit\'e mixte de
recherhce (UMR 6207) du CNRS et des universit\'es Aix-Marseille I,
Aix-Marseille II, et du Sud Toulon-Var; laboratoire affili\'e \`a
la FRUMAM (FR 2291).}, {\it  CNRS-Luminy, Case 907}\\ \it{F-13288
Marseille Cedex 9, France}
\end{center}

\vspace*{1.0cm}

\begin{abstract} The amplitude of the Dalitz decay $\pi
^0\rightarrow e^{+}e^{-}\gamma $ is studied and its
model-independent properties are discussed in detail.
A calculation of radiative corrections is performed
within the framework of two-flavour chiral perturbation
theory, enlarged by virtual photons and leptons. The lowest
meson dominance approximation, motivated by large $N_C$
considerations, is used for the description
of the $\pi^0$-$ \gamma^{*} $-$ \gamma^{*}$ transition form
factor and for the estimate of the NLO low
energy constants involved in the analysis.
The two photon reducible contributions is included
and discussed. Previous calculations are extended to
the whole kinematical range of the soft-photon approximation,
thus allowing for the possibility to consider various experimental
situations and observables.
\end{abstract}

\end{titlepage}
\setcounter{footnote}{0}

\renewcommand{\theequation}{\arabic{section}.\arabic{equation}}
\setcounter{equation}{0}

\section{Introduction}

With a branching ratio of $(1.198\pm 0.032)\%$ \cite{pdg}, the
three body decay $\pi ^0\rightarrow e^{+}e^{-} \gamma$ is the
second most important decay channel\footnote{ The process $\pi
^0\rightarrow e^{+}e^{-} \gamma$ is currently referred to as the
\textit{Dalitz decay\/}, after R.~H.~Dalitz who first studied it
more than fifty years ago~\cite{dalitz51}, and who was the first
to realize its connection with two-photon production in the
emulsion events of cosmic rays. For a nice and instructive
historical retrospective, see~\cite{dalitz87}.} of the neutral
pion. The dominant decay mode, $\pi ^0\rightarrow \gamma \gamma $,
with its overwhelming branching ratio of ($98.798\pm 0.032)\,\%$,
is deeply connected to this three body decay. The other decay
channels related to the anomalous $\pi ^0$-$\gamma$-$ \gamma $
vertex, like $\pi ^0\rightarrow e^{+}e^{-}$ and $\pi ^0\rightarrow
e^+ e^- e^+ e^-$, are suppressed approximately by factors of
$10^{-7}$ and $10^{-5}$, respectively. Another interest of the
Dalitz decay lies in the fact that it provides information on the
semi off-shell $\pi^0$-$ \gamma $-$ \gamma^{*}$ transition form
factor ${\cal F}_{\pi^0 \gamma \gamma^{*}}(q^2)$ in the time-like
region, and more specifically on its slope parameter $a_\pi$. The
most recent determinations of $a_\pi$ obtained from measurements
\cite{exp1,exp2,exp3} of the differential decay rate of the Dalitz
decay,
\begin{eqnarray*}
a_\pi &=& -0.11 \pm 0.03 \pm 0.08
\quad\protect{\cite{exp1}}
\\
a_\pi &=& +0.026 \pm 0.024 \pm 0.0048
\quad\protect{\cite{exp2}}
\\
a_\pi &=& +0.025 \pm 0.014 \pm 0.026
\quad\protect{\cite{exp3}}
,
\end{eqnarray*}
 are endowed
with large error bars, as compared to the values extracted from
the extrapolation of data at higher energies in the
space-like region, $Q^2=-q^2> 0.5$ GeV$^2$,
obtained by CELLO \cite{cello} and CLEO \cite{cleo},
\begin{eqnarray*}
a_\pi &=& +0.0326 \pm 0.0026 \pm 0.0026
\quad\protect{\cite{cello}}
\\
a_\pi &=& +0.0303 \pm 0.0008 \pm 0.0009 \pm 0.0012
\quad\protect{\cite{cleo}}
\end{eqnarray*}
These extrapolations are however model dependent, and a direct and
accurate determination of $a_\pi$ from the decay $\pi
^0\rightarrow e^{+}e^{-} \gamma$ would offer a complementary
source of information. Let us mention, in this context, the
proposal \cite{PrimEx} of the PrimEx experiment at TJNAF to study
the reaction $e^- \gamma  \to e^- \pi^0$, where the neutral pion
is produced in the field of a nucleus through virtual photons from
electron scattering \cite{hadji89}. Although this process concerns
again virtualities in the space-like region, very low values of
$Q^2$, in the range well below the lowest values attained by the
CELLO experiment, can be achieved upon selecting the events
according to the emission angles of the produced pion and of the
scattered electron \cite{hadji89,PrimEx}.

On the theoretical side, several studies have addressed
the issue of the radiative corrections
to the decay $\pi ^0\rightarrow e^{+}e^{-} \gamma$
in the past. At lowest order, the decay amplitude
is of order ${\cal O}(e^3)$. The next-to-leading radiative
corrections to the {\it total decay rate} were first evaluated
numerically by D.~Joseph~\cite{joseph60}, with the result
\[
\frac{\Gamma ^{\text{rad}}(\pi ^0\rightarrow e^{+}e^{-}
\gamma)}{\Gamma (\pi ^0\rightarrow \gamma \gamma )}\approx
1.0\times 10^{-4}.
\]
This shows that the radiative contribution is tiny and can be
neglected in the total decay rate. However, the {\it differential
decay rate}, which provides
the relevant observable for the determination of $a_\pi$,
is sensitive to these radiative corrections. This problem was
extensively studied in~\cite{lautrup71} and \cite{mikaelian72}.
In all cases, the two-photon exchange terms were neglected, and
some further approximations were made (e.g. restrictions in the
kinematical region and on the energy of the bremsstrahlung
photon). Subsequently, the Dalitz decay was further discussed in
 connection with the omission of the two photon exchange contributions.
Particularly, during the 1980s, the controversial question of the
actual size of these contributions was under debate, as well as
the relevance of Low's theorem in this context, cf. the articles
quoted in \cite{tpe}. Eventually, the non interchangeability of
the limits of vanishing electron mass and photon momentum was identified
\cite{tpe2} as the origin of the apparent puzzle raised by
the contradictory results obtained previously by various authors.

Our purpose is to provide a complete treatment of the
next-to-leading radiative corrections to the Dalitz decay, taking
into account the theoretical progresses accomplished in various
aspects related to this issue. For instance, in the studies quoted
so far, the pion was taken as point-like. On the other hand, the
leading order amplitude involves the form factor ${\cal F}_{\pi^0
\gamma \gamma^{*}}(q^2)$ with virtualities up to $q^2\sim
M_{\pi^0}^2$, which is within the realm where chiral perturbation
theory (ChPT) \cite{weinberg79,gasser,leutwyler94} is applicable.
The details of the one-loop calculation of the $\pi^0$-$\gamma$-$
\gamma ^{*}$ vertex in ChPT can be found in~\cite{donoghue}.
However, we also need to consider (among other contributions) the
electromagnetic corrections to ${\cal F}_{\pi^0 \gamma
\gamma^{*}}(q^2)$. We are thus led to reformulate and extend the
results described above within the unified and self-contained
framework of ChPT with virtual photons, as it was formulated in
\cite{urech} and in \cite{ruperts}. Actually, it is also quite
straightforward to include light leptons in the effective theory,
as described in \cite{knecht02a}, or, in the context of
semileptonic decays of the light mesons, in \cite{neufeld}.
Throughout, we shall work within the framework of two light quark
flavours, $u$ and $d$. The corresponding extension to virtual
photons is to be found in Refs.~\cite{knecht98} and
\cite{meissner97}. However, contributions of next-to-leading order
${\cal O}(e^5)$ to the amplitude now involve the doubly off-shell
$\pi^0 $-$ \gamma^{*} $-$ \gamma^{*}$ transition form factor
${\cal A}_{\pi^0\gamma^{*}\gamma^{*}}(q_1^2,q_2^2)$, but for
arbitrarily large virtualities, a situation which cannot be dealt
with within ChPT. We shall introduce and use a representation
\cite{knechtcpt,knecht01} of the form factor ${\cal
A}_{\pi^0\gamma^{*}\gamma^{*}}(q_1^2,q_2^2)$ that relies on
properties of both the large-$N_C$ limit and the short-distance
regime of QCD. The same framework also allows us to supplement our
analysis with estimates of the relevant low energy constants,
along the lines of, for instance, Refs.~\cite{moussallam97} and
\cite{knecht01}.

The material of this article is organized as follows. The general
properties (kinematics, diagram topologies,...) of the amplitude are
discussed in Section 2. Section 3 is devoted to the computation of
the differential decay rate at next-to-leading order (NLO).
Numerical results are presented in Section 4. A brief summary and
conclusions are gathered in Section 5. For reasons of convenience,
various technical details have been included in the form of
appendices. Preliminary reports of the present work have appeared in
Refs. \cite{kkn,kn}.

\section{General properties of the Dalitz decay amplitude}
\setcounter{equation}{0}

In this section we describe the general structure of the
amplitude for the Dalitz decay $\pi ^0\rightarrow e^{+}e^{-} \gamma$,
relevant for the discussion of the contributions both at leading order, ${\cal O}(e^3)$,
and at next-to-leading order, ${\cal O}(e^5)$.

\subsection{Notation and kinematics}

The Dalitz decay amplitude
$\mathcal{M}_{\pi ^0\rightarrow e^{+}e^{-} \gamma}$
is defined as
\begin{equation}
\langle e^{+}(p_+,s_+)e^{-}(p_-,s_-)\gamma (k,\lambda );\,{out}
\vert\pi^0(P);\,{in}\rangle = i(2\pi )^4\delta ^{(4)}(P-p_+ -p_-
-k) \mathcal{M}_{\pi ^0\rightarrow e^{+}e^{-} \gamma},
\end{equation}
where the transition matrix element has to be evaluated in the
presence of the strong {\it and} the electromagnetic interactions.
Lorentz covariance allows to express the amplitude
$\mathcal{M}_{\pi ^0\rightarrow e^{+}e^{-} \gamma}$ in the form
\begin{equation}
\mathcal{M}_{\pi ^0\rightarrow e^{+}e^{-} \gamma}
=\bar{u}(p_-,s_-)\Gamma _\mu (p_+,p_-,k)v(p_+,s_+) \varepsilon ^{*\mu }(k),
\label{Gamma}
\end{equation}
with
\begin{equation}
\bar{u}(p_-,s_-)\Gamma _\mu (p_+,p_-,k)v(p_+,s_+) = \lim_{k^2\to 0}\, i e
\langle e^{+}(p_+,s_+)e^{-}(p_-,s_-);\,{out}\vert
j_{\mu}(0)\vert\pi^0(P);\,{in}\rangle  \label{Gamma2}
\end{equation}
given in terms of the electromagnetic current
\begin{equation}
j_\mu = \frac{2}{3} \bar{u} \gamma_\mu u - \frac{1}{3} \bar d
\gamma_\mu d - {\bar{\psi_e}} \gamma_\mu \psi_e + \ldots
\end{equation}
Invariance under parity, charge conjugation, and gauge symmetry,
\begin{equation}
k^\mu \bar{u}(p_-,s_-)\Gamma _\mu (p_+,p_-,k)v(p_+,s_+)=0,  \label{G}
\end{equation}
implies a transverse structure and a decomposition in terms of
four independent form factors\footnote{We have omitted additional
structures, proportional to $k_\mu$, which vanish upon contraction
with the polarization vector $\varepsilon ^{*\mu }(k)$.
Implicitly, we only consider electromagnetic and strong
interactions, and we assume that there is no P and CP violating
$\theta$ term.}
\begin{align}
\Gamma ^\mu (p_+,p_-,k) &=\phantom{+}P(x,y)[(k\cdot p_+)p_-^\mu
-(k\cdot
p_-)p_+^\mu]\gamma_5  \notag \\
&\phantom{=}\;+A_{+}(x,y)[\slashed{k}\, p_+^\mu -(k\cdot
p_+)\gamma ^\mu ]\gamma_5 -A_{-}(x,y)[\slashed{k}\, p_-^\mu
-(k\cdot p_-)\gamma ^\mu ]\gamma_5
\notag \\
&\phantom{=}\;-\mathrm{i}T(x,y)\sigma ^{\mu \nu }k_\nu \gamma_5.
\label{Formfactors}
\end{align}
The invariant form factors $P(x,y)$, $A_{\pm }(x,y)$ and
$T(x,y)$ are functions
of two independent kinematical variables, which we have chosen as
($m$ denotes the electron mass, $p_-^2 = p_+^2 = m^2$)
\begin{eqnarray*}
x &=&\frac{(p_+ +p_-)^2}{M_{\pi ^0}^2},\qquad \nu ^2\leq x\leq 1,
\qquad \nu^2\,=\frac{4m^2}{M_{\pi ^0}^2}, \\
y &=&\frac{2P\cdot (p_+ -p_-)}{M_{\pi ^0}^2(1-x)},\,\,\,\,-\sigma
_e(M_{\pi ^0}^2x)\leq y\leq \sigma _e(M_{\pi ^0}^2x),\quad \sigma
_e(s)=\sqrt{1-\frac{4m^2}s}.
\end{eqnarray*}
In the pion rest frame, these invariants can be expressed in terms
of the energies of the photon ($\omega$), of the positron ($E_+$)
and of the electron ($E_{-}$) as
\begin{eqnarray*}
1-x &=&2\,\frac \omega {M_{\pi ^0}}, \\
y &=&\frac{E_+ -E_-}\omega .
\end{eqnarray*}
In terms of the variables $x$ and $y$,
charge conjugation invariance implies that the
form factors satisfy the symmetry relations
$$
P(x,y) = P(x,-y),
\ A_\mp(x,y) = A_{\pm}(x,-y),
\ T(x,y) = T(x,-y).
$$
Let us note that the form factors $P(x,y)$, $A_{\pm }(x,y)$
and $T(x,y)$ can be
projected out from $\Gamma ^\mu $ by means of the formula
\begin{equation}
F=\mathrm{Tr}\bigl( \Lambda _F^\mu (\slashed{p}_- +m)\Gamma _\mu
(\slashed{p}_+ -m)\bigr) ,  \label{projection}
\end{equation}
where $\Lambda _F^\mu $, with $F=P$, $A_{\pm }$ , $T$, are
projectors satisfying $k\cdot \Lambda _F=0$. Explicit expressions
of these projectors are given in Appendix A.

In terms of the variables $x$, $y$, the differential decay rate
is given by the formula
\begin{equation}
\mathrm{d}\Gamma =\frac 1{(2\pi )^3}
\frac{M_{\pi^0}}{64}(1-x)\overline{|
\mathcal{M}_{\pi ^0\rightarrow e^{+}e^{-} \gamma}|^2}
\,\mathrm{d}x\mathrm{d}y .
\end{equation}
Expressed in terms of the form factors $P$, $A_{\pm }$ and
$T$, the square of the invariant amplitude (summed over
polarizations) reads
\begin{align}
\overline{|\mathcal{M}_{\pi ^0\rightarrow e^{+}e^{-} \gamma}|^2}
&=\sum_{\mathrm{polarizations}}
|\mathcal{M}_{\pi ^0\rightarrow e^{+}e^{-} \gamma}|^2
=\frac 18\,M_{\pi^0}^4(1-x)^2  \notag \\
&\phantom{=}\;\times \bigl\{[M_{\pi^0}^2 x(1-y^2)-4m^2]
[|P|^2 x M_{\pi^0}^2-2mP(A_{+}+A_{-})^{*}
\notag \\
&\phantom{=\times}\;\;-2mP^{*}(A_{+}+A_{-})+2PT^{*}+2P^{*}T]  \notag \\
&\phantom{=\times}\;\;+2(xM_{\pi^0}^2 - 4m^2)[|A_{+}|^2(1+y)^2+|A_{-}|^2(1-y)^2]  \notag \\
&\phantom{=\times}\;\;-8m^2y^2(A_{+}^{*}A_{-}+A_{+}A_{-}^{*})+8my(1+y)(A_{+}^{*}T+A_{+}T^{*})
\notag \\
&\phantom{=\times}\;\;-8my(1-y)(A_{-}^{*}T+A_{-}T^{*})+8(1-y^2)|T|^2\bigr\}.
\label{MMbar}
\end{align}
In the case $m=0$, this reduces to
\begin{multline*}
\overline{|\mathcal{M}_{\pi ^0\rightarrow e^{+}e^{-} \gamma}|^2}
=\frac 18\,M_{\pi^0}^4(1-x)^2
\bigl\{M_{\pi^0}^2 x(1-y^2)[|P|^2 x M_{\pi^0} ^2+2PT^{*}+2P^{*}T] \\
+2M_{\pi^0}^2x[|A_{+}|^2(1+y^2)+|A_{-}|^2(1-y^2)]+8(1-y^2)|T|^2\bigr\}.
\end{multline*}
As usual, higher order corrections induced by virtual photon
contributions generate infrared singularities, even for a
nonvanishing electron mass $m$. In order to obtain an infrared
finite and physically observable (differential) decay rate, the
emission processes of real soft photons have also to be
considered.

\subsection{Anatomy of the Dalitz decay amplitude}

The contributions to the amplitude $\mathcal{M}_{\pi ^0\rightarrow
e^{+}e^{-} \gamma}$ rather naturally separate into two main classes.
The first one corresponds to the Feynman graphs where the
electron-positron pair is produced by a single photon (Dalitz pair).
The leading contribution, of order ${\cal O}(e^3)$, to the decay
amplitude belongs to these one-photon reducible graphs. They involve
the semi-off-shell $\pi ^0$-$\gamma$-$\gamma^{*} $ vertex ${\cal
F}_{\pi^0 \gamma \gamma^{*}}(q^2)$, see Fig.~\ref{figure1}. The
second class of contributions corresponds to the one-photon
irreducible topologies. They can be further separated into the
one-fermion reducible contributions, which represent the radiative
corrections to the $\pi ^0\rightarrow e^{+}e^{-}$ process (see
Fig.~\ref{figure2}), and the remaining one-particle irreducible
graphs (Fig.~\ref{figure3}), starting with the two photon exchange
box diagram, see the second graph on Fig. \ref{figure4}. Both types
of these one-photon irreducible contributions to the amplitude
involve the doubly off-shell $\pi ^0$-$\gamma^{*}$-$\gamma ^{*}$
vertex ${\cal A}_{\pi^0 \gamma^{*} \gamma^{*}}(q_1^2,q_2^2)$. They
are suppressed with respect to the lowest order one-photon reducible
contribution, starting at the order ${\cal O}(e^5)$ with the
contributions depicted on Fig. \ref{figure4}. Let us now discuss
consecutively these different topologies in greater detail.

\subsubsection{The one-photon reducible contributions}

\begin{figure}[htb]
\begin{center}
\epsfig{file=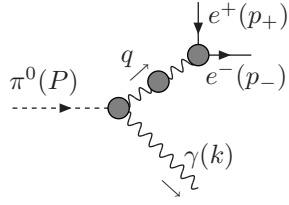}
\end{center}
\caption{One-photon reducible diagrams} \label{figure1}
\end{figure}
The one-photon reducible topologies are shown on Fig.~\ref{figure1}.
They contain the leading order contribution to $\mathcal{M}_{\pi
^0\rightarrow e^{+}e^{-} \gamma}$, and involve only low virtualities
of the semi off-shell form factor ${\cal F}_{\pi^0 \gamma
\gamma^{*}}(q^2)$. The contribution at leading, ${\cal O}(e^3)$, but
also at next-to-leading, ${\cal O}(e^5)$, orders can thus be fully
treated within the framework of ChPT, extended to virtual photons.
The general expression for this one-photon reducible part of the
Dalitz decay amplitude has the form\footnote{Henceforth, we simply
write ${\bar u}$ instead of ${\bar u}(p_-,s_-)$, and $v$ instead of
$v(p_+,s_+)$, whenever no confusion arises.}
\[
\mathcal{M}_{\pi ^0\rightarrow e^{+}e^{-} \gamma}^{1\gamma R}=
\bar{u}\Gamma _\mu ^{1\gamma
R}(p_+,p_-,k)v \varepsilon ^{*\mu }(k),
\]
where
\[
\Gamma ^{1\gamma R}_\mu (p_+,p_-,k)=+\mathrm{i}e^2
\varepsilon _\mu^{\ \nu\alpha \beta }
q_\alpha k_\beta
\,\mathcal{F}_{\pi^0 \gamma \gamma^{*}}(q^2)
\,\mathrm{i}D_{\nu \rho
}^T(q)(-\mathrm{i}e)\Lambda ^\rho (p_-,-p_+).
\]
In this and the following expressions, $q=p_+ + p_-$. The
form factor $\mathcal{F}_{\pi^0 \gamma \gamma^{*}}(q^2)$
is related to the doubly off-shell form factor
$\mathcal{A}_{\pi^0 \gamma^{*} \gamma^{*}}(q_1^2,q_2^2)$, defined
as\footnote{$\mathcal{A}_{\pi^0 \gamma^{*} \gamma^{*}}(q_1^2,q_2^2)=
\mathcal{A}_{\pi^0 \gamma^{*} \gamma^{*}}(q_2^2,q_1^2)$.}
\begin{equation}
\int d^4x\,e^{il\cdot x}\langle 0|T(j^\mu (x)j^\nu (0)|\pi ^0(P)\rangle
=-\mathrm{i}\varepsilon ^{\mu \nu \alpha \beta }l_\alpha P_\beta
\,\mathcal{A}_{\pi^0 \gamma^{*} \gamma^{*}}(l^2,(P-l)^2),
\label{Aff}
\end{equation}
by
\[
\mathcal{F}_{\pi^0 \gamma \gamma^{*}}(q^2) =
\mathcal{A}_{\pi^0 \gamma^{*} \gamma^{*}}(0,q^2).
\]
Here the matrix element on the left hand side can be obtained by
means of the LSZ formula from the three point
Green's function $\langle VVA\rangle$,
calculated within QCD+QED (\emph{i.e}. with the
QED corrections included). Furthermore, $D_{\mu \nu }^T(q)$ is
the transverse part of the photon propagator (the longitudinal,
gauge dependent, part of the photon propagator does not contribute),
\[
\mathrm{i}D_{\mu \nu }^T(q)=-\mathrm{i}\,\frac{g_{\mu \nu }-q_\mu
q_\nu /q^2}{q^2[1+\overline{\Pi }(q^2)]}\, ,
\]
where $\overline{\Pi }(q^2)$ is the renormalized vacuum
polarization function (in the on-shell renormalization scheme with
$\overline{\Pi}(0)=0$), and $\Lambda^\rho (q_1,q_2)$ stands for the
off-shell one-particle irreducible $e^{+}$-$e^{-}$-$\gamma $ vertex function.
For on-shell momenta, $q_1=p_-$, $q_2=-p_+$, it can be decomposed
in terms of the Dirac and Pauli form factors $\overline{F}_{1,2}(q^2)$,
\[
\bar{u}\Lambda ^\mu (p_-,-p_+)v
=\bar{u}
[\overline{F}_1(q^2)\gamma ^\mu +\frac
1{2m}\overline{F}_2(q^2)\mathrm{i}\sigma ^{\mu \nu }q_\nu
]v ,
\]
with $\overline{F}_1(0)=1$ and $\overline{F}_2(0)=a_e$, where
$a_e$ is the anomalous magnetic moment of the electron.

Note that the one-photon reducible part
$\Gamma _\mu ^{1\gamma R}(p_+,p_-,k)$ is gauge invariant by itself,
\[
k^\mu \Gamma _\mu ^{1\gamma R}(p_+,p_-,k)=0,
\]
and therefore it can be expressed in terms of form factors $P$,
$A_{\pm }$, $T$. Using e.g. the formulae~(\ref{projection}),
(\ref{projexpP}), (\ref{projexpA}), and (\ref{projexpT}), one
obtains
\begin{align}
P^{1\gamma R}(x,y) &=-e^3\mathcal{F}_{\pi^0 \gamma \gamma^{*}}(xM_{\pi ^0}^2)\frac
1{xM_{\pi ^0}^2[1+\overline{\Pi }(xM_{\pi^0}^2)]}
\frac{\mathrm{i}}m\,\overline{F}_2(xM_{\pi ^0}^2), \notag\\
A^{1\gamma R}_\pm(x,y) &=e^3\mathcal{F}_{\pi^0 \gamma \gamma^{*}}(xM_{\pi ^0}^2)
\frac
1{xM_{\pi ^0}^2[1+\overline{\Pi }(xM_{\pi ^0}^2)]}
\mathrm{i}\overline{F}_1(xM_{\pi ^0}^2), \label{1GR}\\
T^{1\gamma R}(x,y) &=e^3\mathcal{F}_{\pi^0 \gamma \gamma^{*}}(xM_{\pi ^0}^2)\frac
1{xM_{\pi ^0}^2[1+ \overline{\Pi
}(xM_{\pi^0}^2)]}
\mathrm{i}\bigl[2m\overline{F}_1(xM_{\pi^0}^2)+
\frac{xM_{\pi^0}^2}{2m}\,\overline{F}_2(xM_{\pi^0}^2)\bigr].
\notag
\end{align}

\subsubsection{One-fermion reducible and one-particle irreducible contributions}

\begin{figure}[htb]
\begin{center}
\epsfig{file=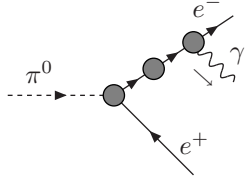}
\end{center}
\caption{One-fermion reducible diagrams} \label{figure2}
\end{figure}

\begin{figure}[b]
\begin{center}
\epsfig{file=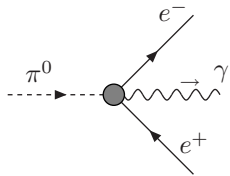}
\end{center}
\caption{One-particle irreducible diagrams} \label{figure3}
\end{figure}

The one-fermion reducible and the one-particle irreducible
topologies, shown on Fig.~\ref{figure2} and on Fig.~\ref{figure3},
respectively, both start at order ${\cal O}(e^5)$.

\noindent
Since the one-photon reducible part
$\mathcal{M}_{\pi ^0\rightarrow e^{+}e^{-} \gamma}^{1\gamma R}$
of the invariant amplitude is transverse by itself, the one-fermion
reducible and one particle irreducible contributions
$\mathcal{M}_{\pi ^0\rightarrow e^{+}e^{-} \gamma}^{1\psi R}+
\mathcal{M}_{\pi ^0\rightarrow e^{+}e^{-} \gamma}^{1PI}$
together also represent a transverse subset. However,
these two types of contributions are not transverse when
taken separately.

Let us first concentrate on the one-fermion
reducible topology. These contributions
 can be expressed in the form
\[
\mathcal{M}_{\pi ^0\rightarrow e^{+}e^{-} \gamma}^{1\psi R}=
\overline{u}\Gamma _\mu ^{1\psi R}(p_+,p_-,k)v \varepsilon ^\mu (k)^{*},
\]
where
\begin{multline}
\mathrm{i}\Gamma _\mu ^{1\psi R}(p_+,p_-,k) = (-\mathrm{i}e)\Lambda
_\mu (p_-,p_- + k)\mathrm{i}S(p_-+k)\mathrm{i}\Gamma _{\pi
^0e^{-}e^{+}}(p_- +k,p_+)\\
\phantom{=}+\mathrm{i}\Gamma _{\pi
^0e^{-}e^{+}}(p_-,p_+ +k)\mathrm{i}S(-p_+ -k)(-\mathrm{i}e)\Lambda
_\mu (-p_+ -k,-p_+).  \label{1FI}
\end{multline}
In this formula,
\[
\mathrm{i}S(q)=\frac{\mathrm{i}}{\slashed{q}-m-\Sigma (q)}
\]
is the full fermion propagator, with
\begin{equation}
\Sigma (q)=\slashed{q}\,\Sigma _V(q^2)+\Sigma _S(q^2)
\label{Selfenergy}
\end{equation}
the fermion self-energy, while $\Lambda _\mu (q_1,q_2) = \gamma_\mu +
{\cal O}(\alpha)$
and $\Gamma_{\pi ^0e^{-}e^{+}}(q_1,q_2)$ are the (off-shell)
one-particle irreducible
$e^{+}$-$e^{-}$-$\gamma $ and $\pi^0$-$e^{-}$-$e^{+}$ vertices, respectively.
As we have already mentioned, $\mathrm{i}\Gamma _\mu ^{1\psi R}(p_+,p_-,k)$
is not transverse, since $\Lambda _\mu (q_1,q_2)$ satisfies the
Ward-Takahashi identity
\begin{equation}
(q_1 - q_2)\cdot \Lambda (q_1,q_2)=S^{-1}(q_1)-S^{-1}(q_2).
 \label{WTI}
\end{equation}
\begin{figure}[b]
\center\epsfig{figure=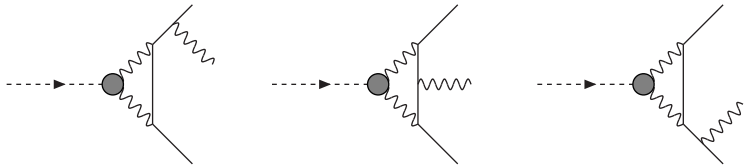} \caption{One-photon irreducible
contributions at one-loop level.} \label{figure4}
\end{figure}
The solution of this identity reads
\begin{equation}
\Lambda _\mu (q_1,q_2)=\Lambda _\mu ^L(q_1,q_2)+\Lambda _\mu ^T(q_1,q_2),
\label{solWTI}
\end{equation}
with
\begin{equation}
(q_1 - q_2)\cdot \Lambda^T (q_1,q_2)= 0,
\end{equation}
and the longitudinal part, which consists of any particular solution
of Eq.~(\ref{WTI}), may conveniently be chosen \cite{Ball&Chiu1980}
as
\begin{eqnarray}
\Lambda _\mu ^L(q_1,q_2) &=& \frac
12(\slashed{q}_1+\slashed{q}_2)(q_1+q_2)_\mu \,\frac{\Sigma _V(q_2^2)-\Sigma
_V(q_1^2)}{q_1^2-q_2^2}\nonumber\\
&&\quad
+\gamma _\mu \Bigl[ 1-\frac 12 \Sigma _V(q_2^2)- \frac 12 \Sigma
_V(q_1^2)\Bigr] +(q_1 + q_2)_\mu \,\frac{\Sigma _S(q_2^2)-\Sigma _S(q_1^2)}{
q_1^2-q_2^2}.
\label{longitudinal}
\end{eqnarray}
The transverse part $\Lambda _\mu ^T(q_1,q_2)$ is then parameterized
in terms of eight form factors $A_i$, $i=1,\ldots ,8$,
corresponding to the eight available independent transverse tensor
structures $T_\mu ^i$, $i=1,\ldots ,8$ (we will not reproduce them
here, for a detailed discussion and explicit
expressions, see Refs. \cite{Ball&Chiu1980} and
\cite{Kizilersu1995})
\begin{equation}
\Lambda _\mu ^T(q_1,q_2)=\sum_{i=1}^8
A_i(q_1^2,q_2^2,(q_1 - q_2)^2)T_\mu^i(q_1,q_2). \label{LambdaT}
\end{equation}
The decomposition (\ref{solWTI}) of the vertex function
$\Lambda (q_1,q_2)$ induces the corresponding decomposition
of $\Gamma _\mu ^{1\psi R}(p_+,p_-,k)$,
\begin{equation}
\Gamma _\mu ^{1\psi R}(p_+,p_-,k) = \Gamma _\mu ^{1\psi
R;T}(p_+,p_-,k) + \Gamma _\mu ^{1\psi R;L}(p_+,p_-,k),
\end{equation}
with
\begin{equation}
k^{\mu} \Gamma _\mu ^{1\psi R;T}(p_+,p_-,k) = 0
\end{equation}
and
\begin{eqnarray}
k^{\mu} \Gamma _\mu ^{1\psi R;L}(p_+,p_-,k) &=&
e \Gamma _{\pi^0e^{-}e^{+}}(p_-+k,p_+) -
e \Gamma _{\pi ^0e^{-}e^{+}}(p_-,p_+ +k)
\nonumber\\
&&
- S^{-1}(p_-) S(p_-+k) e\Gamma _{\pi^0e^{-}e^{+}}(p_- +k,p_+)
\nonumber\\
&&
+ e\Gamma _{\pi^0 e^{-}e^{+}}(p_-,p_+ +k) S(-p_+ -k) S^{-1}(-p_+).
\end{eqnarray}
Therefore
\begin{equation}
\overline{u} \,k\cdot \Gamma ^{1\psi R;L}(p_+,p_-,k)v =
e\overline{u} [\Gamma _{\pi^0 e^{-}e^{+}}(p_- +k,p_+)-
\Gamma _{\pi ^0 e^{-}e^{+}}(p_-,p_++k)]v
.
\label{WI1psiR}
\end{equation}
This non transverse piece should be cancelled by the contribution
$\overline{u}(k\cdot \Gamma ^{1PI})v$ of the
one-particle irreducible graphs.
In addition, the
transverse part $\Gamma_{\mu} ^{1\psi R;T}(p_+,p_-,k)$ admits a
representation of the type (\ref{Formfactors}), with appropriate
form factors $F^{1\psi R;T}(x,y)$, $A_{\pm}^{1\psi R;T}(x,y)$, and
$T^{1\psi R;T}(x,y)$, up to possible terms proportional to
$k_\mu$, which cancel when contracted with $\varepsilon ^\mu
(k)^{*}$.

As for the vertex $\Gamma _{\pi ^0e^{-}e^{+}}(q_2,q_1)$, it can
be decomposed (using Lorentz invariance, the Dirac structure of
the inverse fermion propagator $S^{-1}(q)$, and charge
conjugation invariance) as
\begin{multline}
\Gamma _{\pi ^0e^{-}e^{+}}(q_2,q_1) = P_{\pi
^0e^{-}e^{+}}(q_2^2,q_1^2)\gamma_5+\gamma_5 S^{-1}(-q_1)
A_{\pi^0e^{-}e^{+}}(q_2^2,q_1^2)\\
+S^{-1}(q_2)\gamma_5
A_{\pi^0e^{-}e^{+}}(q_1^2,q_2^2)+S^{-1}(q_2)
\gamma_5 S^{-1}(-q_1)T_{\pi ^0e^{-}e^{+}}(q_2^2,q_1^2),
\label{Peeformfactors}
\end{multline}
where $P_{\pi ^0e^{-}e^{+}}$, $A_{\pi ^0e^{-}e^{+}}$, and
$T_{\pi ^0e^{-}e^{+}}$ are scalar form factors, which, as a
consequence of charge conjugation invariance,
 satisfy the additional relations
\begin{equation}
P_{\pi ^0e^{-}e^{+}}(q_2^2,q_1^2) = P_{\pi
^0e^{-}e^{+}}(q_1^2,q_2^2),
\ T_{\pi ^0e^{-}e^{+}}(q_2^2,q_1^2) = T_{\pi ^0e^{-}e^{+}}(q_1^2,q_2^2).
\label{pieesymmetry}
\end{equation}
The form factor $P_{\pi ^0e^{-}e^{+}}$ is then related to the
on-shell $\pi ^0\rightarrow e^{+}e^{-}$ amplitude,
\[
\mathcal{M}_{\pi ^0\rightarrow e^{+}e^{-}}=
\overline{u}(p_-,s_-)
\gamma_5 v(p_+,s_+) P_{\pi ^0e^{-}e^{+}}(m^2,m^2).
\]
In terms of the form factors (\ref{Peeformfactors}) we can write
\begin{eqnarray}
\mathcal{M}_{\pi ^0\rightarrow e^{+}e^{-} \gamma}^{1\psi R} &=&
e\,\varepsilon ^\mu(k)^{*}\overline{u} \bigl\{\Lambda _\mu
(p_-,p_-+k)S(p_- +k)P_{\pi ^0e^{-}e^{+}}(m^2+2(k\cdot
p_-),m^2)\gamma_5  \nonumber \\
&&\;+P_{\pi ^0e^{-}e^{+}}(m^2,m^2+2(k\cdot p_+))
\gamma_5S(-p_+ -k))\Lambda
_\mu (-p_+ -k,-p_+)  \nonumber \\
&&\;+\Lambda _\mu (p_-,p_- +k)\gamma_5
A_{\pi^0e^{-}e^{+}}(m^2,m^2+2(k\cdot p_-))  \nonumber \\
&&\;+\gamma_5\Lambda _\mu (-p_+ -k,-p_+)A_{\pi
^0e^{-}e^{+}}(m^2,m^2+2(k\cdot p_+))\bigr\}v.
\label{PATamplitude}
\end{eqnarray}
At leading order in the fine structure constant $\alpha $, $\Gamma
_{\pi ^0e^{+}e^{-}}$ is given by
\begin{eqnarray}
\Gamma _{\pi ^0e^{-}e^{+}}(q_2,q_1) \label{1FRL}
&=&-e^4\varepsilon ^{\mu \nu \alpha \beta }\int
\frac{\mathrm{d}^4l}{(2\pi
)^4}\,\mathcal{A}_{\pi^0\gamma^{*}\gamma^{*}}(l^2,(q_1+q_2-l)^2)
\nonumber\\
&&\times
\frac{l_\alpha (q_1 + q_2)_\beta
}{(l^2+i0)[(q_1+q_2-l)^2+i0]}\,\gamma _\mu\,
\frac{\mathrm{i}}{\slashed{q}_2-\slashed{l}-m+i0}\,\gamma _\nu\,
,
\end{eqnarray}
where $\mathcal{A}_{\pi^0\gamma^{*}\gamma^{*}}(q_1^2,q_2^2)$
is now restricted to its pure QCD part.
The corresponding expression of $\Gamma _\mu ^{1\psi R}(p_+,p_-,k)$
then reads
\begin{eqnarray}
\mathrm{i}\Gamma _\mu ^{1\psi R}(p_+,p_-,k) &=&
e^5\varepsilon ^{\rho \sigma \alpha \beta }\int \frac{\mathrm{d}^4l}{(2\pi
)^4}\,\mathcal{A}_{\pi^0\gamma^{*}\gamma^{*}}(l^2,(P-l)^2)\,
\frac{l_\alpha P_\beta }{(l^2+\mathrm{i}0)[(P-l)^2+\mathrm{i}0]} \nonumber\\
&&\times \bigg[
\gamma _\mu \, \frac{\mathrm{i}}{\slashed{p}_- +\slashed{k}
-m+\mathrm{i}0}\, \gamma _\rho \,
\frac{\mathrm{i}}{\slashed{p}_- +\slashed{k}-\slashed{l}
-m+\mathrm{i}0}\,\gamma _\sigma \nonumber\\
&& \ -
\gamma _\rho \, \frac{\mathrm{i}}{\slashed{p}_- -\slashed{l}
-m+\mathrm{i}0}\, \gamma _\sigma \,
\frac{\mathrm{i}}{\slashed{p}_+ +\slashed{k}
+m+\mathrm{i}0}\,\gamma _\mu
\bigg].
\label{M1psiR2}
\end{eqnarray}
The general properties of the form factor
$\mathcal{A}_{\pi^0\gamma^{*}\gamma^{*}}(q_1^2,q_2^2)$
are summarized in Appendix B.
Here we only note that the short distance behaviour of
$\mathcal{A}_{\pi^0\gamma^{*}\gamma^{*}}(l^2,(P-l)^2)$
in QCD makes it act as an ultraviolet cut-off,
so that the loop integral on the right-hand side
of (\ref{1FRL}) actually converges.

Finally, the one-particle irreducible part of the amplitude
\[
\mathcal{M}_{\pi ^0\rightarrow e^{+}e^{-} \gamma}^{1PI}
=\overline{u}\Gamma _\mu
^{1PI}(p_+,p_-,k)v \varepsilon ^\mu (k)^{*}
\]
starts at the order $e^5$ with the box diagram of Fig.
\ref{figure4},
\begin{eqnarray}
\mathrm{i}\Gamma _\mu ^{1PI}(p_+,p_-,k) &=&e^5\varepsilon ^{\rho
\sigma \alpha \beta }\int \frac{\mathrm{d}^4l}{(2\pi
)^4}\,\mathcal{A}_{\pi^0\gamma^{*}\gamma^{*}}(l^2,(P-l)^2)\,
\frac{l_\alpha P_\beta }{(l^2+\mathrm{i}0)[(P-l)^2+\mathrm{i}0]} \nonumber\\
&&\times \gamma _\rho \, \frac{\mathrm{i}}{\slashed{p}_- -\slashed{l}
-m+\mathrm{i}0}\, \gamma _\mu \,
\frac{\mathrm{i}}{\slashed{p}_- +\slashed{k}-\slashed{l}
-m+\mathrm{i}0}\,\gamma _\sigma ,
\label{M1GIR2}
\end{eqnarray}
which is also ultraviolet finite. At this order, one
 verifies that the sum $\overline{u}\Gamma _\mu ^{1\psi R}(p_+,p_-,k)v
+ \overline{u}\Gamma _\mu ^{1PI}(p_+,p_-,k)v $ is indeed transverse.

\section{The NLO differential decay rate  }
\setcounter{equation}{0}

The leading order amplitude corresponds to the
${\cal O}(e^3)$ one-photon reducible contribution,
evaluated at lowest order in the extended chiral expansion,
i.e. with $\overline{F}_1(q^2)=1$,
$\overline{F}_2(q^2)=\overline{\Pi
}(q^2)=0$, and with the form factor
$\mathcal{A}_{\pi^0\gamma^{*}\gamma^{*}}(l^2,(P-l)^2)$
reduced to its expression for a pointlike pion, i.e.
a constant, $\mathcal{A}_{\pi^0\gamma^{*}\gamma^{*}}^{LO} = -N_C/12\pi
^2F_\pi$, fixed by the chiral anomaly.
The leading order expressions of the
form factors $P$, $A_\pm$ and $T$
are then given, for $N_C=3$ and according to
Eq.~(\ref{1GR}), by
\begin{eqnarray}
P^{LO}(x,y) &=&0, \nonumber\\
A_{\pm }^{LO}(x,y) &=&-\frac{\mathrm{i}e^3}{4\pi ^2F_\pi M_{\pi^0}^2}
\cdot\frac 1x, \nonumber\\
T^{LO}(x,y) &=&-\frac{2\mathrm{i}me^3}{4\pi ^2F_\pi M_{\pi^0}^2}
\cdot\frac 1x.
\end{eqnarray}
Note that in the limit $m\rightarrow 0$, only the form factors
$A_{\pm }^{LO}$ survive. The square of the leading
invariant amplitude summed over polarizations is, according
to~(\ref{MMbar}),
\begin{equation}
\overline{|\mathcal{M}_{\pi^0\rightarrow e^+ e^- \gamma}^{LO}|^2}
=\frac 1{32}\frac{e^6}{\pi
^4F_\pi ^2}\,\frac{(1-x)^2}{x^2}\,[M_{\pi^0}^2 x(1+y^2)+4m^2]
\end{equation}
and the corresponding partial decay rates read
\begin{align}
\frac{\mathrm{d}\Gamma ^{LO}}{\mathrm{d}x\mathrm{d}y}
&=\frac{\alpha^3}{(4\pi )^4}\frac{M_{\pi^0}}{F_\pi ^2}\,
\frac{(1-x)^3}{x^2}\,[M_{\pi^0}^2 x(1+y^2)+4m^2],\notag \\
\frac{\mathrm{d}\Gamma ^{LO}}{\mathrm{d}x} &=
\frac{\alpha^3}{(4\pi)^4} \frac{8}{3} \frac{M_{\pi^0}}{F_\pi^2}\,
\frac{(1-x)^3}{x^2}\,\sigma_e(x M_{\pi^0}^2) \,(x M_{\pi^0}^2  + 2
m^2). \label{GammaLeading}
\end{align}

The next-to-leading corrections to the differential
decay rates will be described as
\begin{equation}
\begin{split}
\frac{\mathrm{d}\Gamma }{\mathrm{d}x\mathrm{d}y} &=\delta(x,y)\,
\frac{
\mathrm{d}\Gamma ^{LO}}{\mathrm{d}x\mathrm{d}y}, \label{ddef} \\
\frac{\mathrm{d}\Gamma }{\mathrm{d}x} &=\delta(x)\,
\frac{\mathrm{d} \Gamma ^{LO}}{\mathrm{d}x}.
\end{split}
\end{equation}
Knowledge of the corrections $\delta(x,y)$ and $\delta(x)$
to the Dalitz plot distributions allows to extract
information on the QCD part of the form factor
${\cal F}_{\pi^0\gamma\gamma^{*}}(q^2)$ from the experimentally
measured decay distribution. For instance, if the
form factor is approximated by a constant plus
linear term
\begin{equation}
{\cal F}_{\pi^0\gamma\gamma^{*}}(q^2)\,=\,
{\cal F}_{\pi^0\gamma\gamma^{*}}(0)\,
\big[
1\,+\,a_\pi\,\frac{q^2}{M_{\pi^0}^2}\,+\cdots
\big]\, ,
\end{equation}
the slope parameter $a_\pi$ is obtained from
\begin{equation}
\frac{\mathrm{d} \Gamma ^{exp}}{\mathrm{d}x}\,-\,
\delta_{QED}(x)\,\frac{\mathrm{d} \Gamma ^{LO}}{\mathrm{d}x}
\,=\,
\frac{\mathrm{d} \Gamma ^{LO}}{\mathrm{d}x}\,
[1 + 2x\,a_\pi],
\label{defa_pi}
\end{equation}
where the QED part, $\delta_{QED}(x)$, of the corrections
$\delta(x)$ will be specified below.

For the purpose of the following subsections, we introduce
functions $\delta^i(x,y)$ and $\delta^i(x)$,
$i = 1\gamma R, 1\gamma IR, \ldots$, measuring the
magnitude of various ${\cal O}(e^5)$ and/or ${\cal O}(e^3p^2)$
corrections $\mathrm{d}\Gamma^i$ to the leading order decay rate.
In terms of the corresponding corrections to the invariant amplitudes,
$T(x,y)=T^{LO}(x,y) + (\delta^i T)(x,y)$, etc.,
one has\footnote{As usual, the NLO corrections
to the decay rate arise from the interference between
the leading and NLO amplitudes. This explains why
there is no contribution involving $(\delta^iP)(x,y)$
in these expressions, given that $P^{LO}(x,y)$ vanishes.}
\begin{eqnarray}
\delta ^i(x,y) &=&-4\pi^2\frac{M_{\pi^0}^2 F_\pi }{e^3}
\frac{x}{M_{\pi^0}^2 x(1+y^2)+4m^2} \times \text{Im}\{
8m(\delta^iT)(x,y) \label{deltaPAT}
 \\
&& +[M_{\pi^0}^2 x(1+y)^2-4m^2] (\delta^i A_{+})(x,y)
+[M_{\pi^0}^2 x(1-y)^2-4m^2] (\delta^i A_{-})(x,y) \}  \nonumber
\end{eqnarray}
and
\begin{equation}
\delta^i(x) \,=\,\frac{3}{8}\,
\frac{1}{\sigma_e(xM_{\pi^0}^2)}
\int_{-\sigma_e(xM_{\pi^0}^2)}^{+\sigma_e(xM_{\pi^0}^2)}
\mathrm{d}y \,\frac{M_{\pi^0}^2 x(1+y^2)+4m^2}{xM_{\pi^0}^2  + 2m^2}
\,\delta^i(x,y) .
\end{equation}

\subsection{NLO one-photon reducible corrections}

The computation of the corrections belonging to the
one-photon reducible type of topology requires the
evaluation of several quantities beyond leading
order. Thus, next-to-leading corrections,
of  orders ${\cal O}(p^6e^0)$
and ${\cal O}(p^4e^2)$, to
${\cal F}_{\pi^0\gamma\gamma^{*}}(q^2)$,
as well as
corrections of orders ${\cal O}(p^2e^0)$
and ${\cal O}(p^0e^2)$ to the electromagnetic form
factors ${\overline F}_1(xM_{\pi^0}^2)$,
${\overline F}_2(xM_{\pi^0}^2)$, and to
the vacuum polarization function ${\overline\Pi}(xM_{\pi^0}^2)$,
have to be evaluated within the framework of (extended)
ChPT. These corrections involve one loop
graphs with virtual pions, photons and electrons, and
local contributions given in terms of counterterms.
The interested reader may find the details of these calculations
in Appendices \ref{App:cepppv} and \ref{App:NLO}.
The corresponding NLO corrections to the Dalitz distribution read
\begin{eqnarray}
\delta^{1\gamma R}(x,y) &=& 2\mathrm{Re}\bigg[
a_{NLO}^{ChPT}(xM_{\pi ^0}^2)-\overline{\Pi }(xM_{\pi^0}^2)+
\overline{F}_1(xM_{\pi ^0}^2)-1
\nonumber\\
&&\quad
+\frac{2xM_{\pi ^0}^2}{M_{\pi^0}^2x(1+y^2)+4m^2}
\overline{F}_2(xM_{\pi ^0}^2)
\bigg]\label{d1grxy}
\end{eqnarray}
and
\begin{equation}
\delta^{1\gamma R}(x)=2\mathrm{Re}\Bigl[ a_{NLO}^{ChPT}(xM_{\pi
^0}^2) -\overline{\Pi }(xM_{\pi^0}^2)+ \overline{F}_1(xM_{\pi
^0}^2)-1 +\frac 32\frac{xM_{\pi ^0}^2}{M_{\pi^0}
^2x+2m^2}\overline{F}_2(xM_{\pi ^0}^2) \Bigr].\label{d1gr}
\end{equation}
The expressions of the various quantities appearing in
these formulae are displayed in Eqs. (\ref{achptnlo}),
(\ref{appF2}), (\ref{Pibar}) and (\ref{F1bar}) of Appendix
\ref{App:cepppv} and \ref{App:NLO}.
Let us just mention here that at NLO
the Dirac form factor ${\overline F}_1(s)$ develops
an infrared singularity,
\[
\overline{F}_1(s)_{IR\,div} \,=\,\frac \alpha {2\pi }\ln \left(
\frac{m^2}{m_\gamma ^2}\right) \left\{ 1+(s-2m^2)\frac 1{s\sigma
_e(s)}\left[ \ln \left( \frac{1-\sigma _e(s)}{1+\sigma
_e(s)}\right) +\mathrm{i}\pi \right] \right\},
\]
where $m_\gamma$ is a small photon mass introduced as
an infrared regulator.
Thus, the infrared divergent part of the one-photon reducible
corrections reads
\begin{equation}
\delta^{1\gamma R}(x,y)_{IR\,div}=\frac{e^2}{(2\pi )^2}\ln
\left( \frac{m^2}{m_\gamma ^2}\right) \left\{ 1+\left(
1-\frac{2m^2}{xM_{\pi ^0}^2}\right) \frac 1{\sigma _e(xM_{\pi
^0}^2)}\ln \left( \frac{1-\sigma _e(xM_{\pi ^0}^2)}{1+\sigma
_e(xM_{\pi ^0}^2)}\right) \right\} .  \label{deltaxyNLOIRdiv}
\end{equation}

\subsection{One-photon irreducible contributions}

The evaluation of the contribution $\delta^{1\gamma IR}(x,y)$
involves the QCD form factor
$\mathcal{A}_{\pi^0\gamma^{*}\gamma^{*}}(q_1^2,q_2^2)$ for
arbitrary virtualities. This in turn addresses non perturbative
issues beyond the low energy range covered by ChPT. While the
asymptotic regime can be reached through the short distance
properties of QCD and the operator product expansion
\cite{Wilson69,SVZ}, there still remains the intermediate energy
region, populated by resonances at the 1 GeV scale, to be
accounted for. If one restricts oneself to approaches with a clear
theoretical link to QCD, the large-$N_C$ framework is almost the
only available possibility\footnote{Large scale numerical
simulations on a discretized space-time might become an
alternative in the future.}. In Refs. \cite{knecht01,knechtcpt},
the form factor
$\mathcal{A}_{\pi^0\gamma^{*}\gamma^{*}}(q_1^2,q_2^2)$ has been
investigated within a well defined approximation to the
large-$N_C$ limit of QCD, which consists in retaining only a
finite number of resonances in each channel. Details of this
approach, as far as the form factor
$\mathcal{A}_{\pi^0\gamma^{*}\gamma^{*}}(q_1^2,q_2^2)$ is
concerned, are to be found in Appendix~\ref{App:ppp}. Thus, upon
inserting the expression of Eq. (\ref{LMD}) in the form
\begin{eqnarray*}
\mathcal{A}_{LMD}(l^2,(P-l)^2) &=&\frac{F_\pi }{3
M_V^4}l^2(l-P)^2\left[
\frac{\kappa_V}{l^2(l-P)^2}-\frac{M_V^2+ \kappa_V}{(l^2- M_V^2)(l-P)^2}\right. \\
&&\left. -\frac{M_V^2+\kappa_V}{l^2[(l-P)^2-M_V^2]}+\frac{2 M_V^2+
\kappa_V}{(l^2- M_V^2)[(l-P)^2- M_V^2]}\right]
\end{eqnarray*}
into (\ref{1FRL}) and (\ref{M1GIR2}), the integral over the loop
momentum can be done and expressed in terms of the standard
one-loop functions $B_0$, $C_0$ and $D_0$ defined in
Appendix~\ref{App:loop}. It is, however, much easier, and
equivalent\footnote{The results obtained within the two approaches
will differ by terms of the order ${\cal O}(m^2/M_V^2)$.}, to
proceed within the framework of an effective lagrangian approach
using ChPT with explicit photons and leptons \cite{neufeld}, that
we now briefly describe. Thus, we take in~(\ref{1FRL}) and
(\ref{M1GIR2})
the leading order constant expression
\begin{equation}
\mathcal{A}_{\pi^0\gamma^{*}\gamma^{*}}^{LO} = -N_c/12\pi
^2F_\pi .
\label{WZampLO}
\end{equation}
This is a good approximation only for sufficiently low loop
momentum, $l^2\ll\Lambda _H^2$, where $\Lambda _H\sim$ 1 GeV is the
hadronic scale typical for the non Goldstone resonance states.
The intermediate and asymptotic ranges are, however, not
treated properly using this effective vertex. One of the
consequences is that the loop integral (\ref {1FRL}) with
this constant form factor is ultraviolet divergent. Within the
framework of an effective low energy theory, the
difference between the exact and the low energy effective vertex can
be taken into account by a counterterm contribution stemming from
the Lagrangian \cite{savage}, \cite{knechtcpt}
\begin{multline*}
\mathcal{L}_{Pe^{-}e^{+}} =\frac{3\mathrm{i}}{32}\left( \frac
\alpha \pi \right) ^2\overline{\psi }\gamma ^\mu \gamma ^5\psi
[\chi _1\langle Q^2(D_\mu UU^{+}-D_\mu U^{+}U\rangle +\chi
_2\langle U^{+}QD_\mu UQ-UQD_\mu U^{+}Q\rangle ].
\end{multline*}
Let us note that these counterterms are also necessary to cure the
ultraviolet divergence that arise in the loop integral of Eq. (\ref{1FRL})
with a constant form factor.
$\mathcal{L}_{Pe^{-}e^{+}}$ generates a local $\pi
^0(p)\rightarrow e^{-}(q)e^{+}(q')$ vertex of the form
\[
\mathrm{i}\Gamma _{\pi ^0e^{-}e^{+}}^{CT}(q,q')=\frac{4\chi }{
F_\pi }\left( \frac \alpha {4\pi }\right)
^2(\slashed{q}+\slashed{q}') \gamma ^5,
\]
or, in terms of the decomposition (\ref{Peeformfactors}),
\begin{eqnarray}
P_{\pi ^0e^{-}e^{+}}^{CT}(q_1^2,q_2^2 ) &=&-\mathrm{i}\frac{8m\chi }{
F_\pi }\left( \frac \alpha {4\pi }\right) ^2,  \nonumber \\
A_{\pi ^0e^{-}e^{+}}^{CT}(q_1^2,q_2^2 ) &=&-\mathrm{i}\frac{4\chi }{F_\pi }
\left( \frac \alpha {4\pi }\right) ^2  \nonumber, \\
T_{\pi ^0e^{-}e^{+}}^{CT}(q_1^2,q_2^2 ) &=&0.  \label{Peecounterterm}
\end{eqnarray}
In the above formulae, $\chi$ stands for the relevant combination of
the effective couplings,\footnote{We use the convention where both
the loop functions and bare couplings are renormalization scale
dependent -- see also Appendix~\ref{App:loop}.}
\begin{equation}
\chi =-\frac{\chi _1+\chi _2}4=\chi ^r(\mu )+3 \left[ \frac
1{d-4}-\frac 12(\ln 4\pi -\gamma +1)\right] , \label{chi1}
\end{equation}
which decomposes into a finite, but scale dependent, renormalized
part $\chi ^r(\mu )$, and a well defined divergent part \cite{savage}.
We therefore split the amplitude
$\mathcal{M}_{\pi ^0\rightarrow e^{+}e^{-} \gamma}^{1\gamma IR}$
into two parts
\[
\mathcal{M}_{\pi ^0\rightarrow e^{+}e^{-} \gamma}^{1\gamma IR}=
\mathcal{M}_{\pi ^0\rightarrow e^{+}e^{-} \gamma}^{1\gamma
IR;\,loop}+\mathcal{M}_{\pi ^0\rightarrow e^{+}e^{-} \gamma}^{1\gamma IR;\,CT},
\]
corresponding to the loop (computed with the constant form factor
$\mathcal{A}_{\pi^0\gamma^{*}\gamma^{*}}^{LO}$) and counterterm
contributions, respectively. Because $\mathcal{M}_{\pi
^0\rightarrow e^{+}e^{-} \gamma}^{1\gamma IR;\,CT}$ is gauge
invariant, it can be decomposed into form factors according to
(\ref{Formfactors}), with
\begin{eqnarray*}
P^{1\gamma IR;\,CT}(x,y) &=&-\mathrm{i}\frac{8m\chi }{F_\pi }\left( \frac
\alpha \pi \right) ^2\frac e{M_{\pi^0} ^4(1-x)^2(1-y^2)}, \\
A_{\pm }^{1\gamma IR;\,CT}(x,y) &=&0, \\
T^{1\gamma IR;\,CT}(x,y) &=&-\mathrm{i}\frac{2m\chi }{F_\pi
}\left( \frac \alpha \pi \right) ^2\frac e{M_{\pi^0}
^2(1-x)(1-y^2)}.
\end{eqnarray*}
For the corresponding decomposition
$\delta ^{1\gamma IR}(x,y) =
\delta ^{1\gamma IR;\,CT}(x,y) + \delta ^{1\gamma IR;\,loop}(x,y)$,
one finds, upon using Eq. (\ref{deltaPAT}),
\begin{equation}
\delta ^{1\gamma IR;\,CT}(x,y)=16\chi \left( \frac \alpha \pi
\right) \frac{m^2x}{\left( 1-x\right) \left( 1-y^2\right) \left[
M_{\pi^0} ^2x(1+y^2)+4m^2\right] } . \label{chi2}
\end{equation}
The interference term of the loop amplitude with the lowest order
one-photon reducible amplitude
$\mathcal{M}_{\pi^0\rightarrow e^{+}e^{-} \gamma}^{LO}$ results in
\begin{equation}
\begin{split}
& \delta ^{1\gamma IR;\,loop}(x,y)=\left( \frac \alpha \pi \right) \frac
x{[x(1+y^2)+\nu ^2]}\frac 1{(1-x)^2} \\
\times\mathrm{Re}\,&\Bigl\{\, \frac
18(x-1)^2\bigl((x-1)^2(y^4-1)-4\nu ^2y^2\bigr) M_{\pi^0} ^4D_0
\\
& +\frac{(x-1)((x-1)(y-1)(y^2+1)(xy-x-y-1)-4\nu ^2(y^2-y+1))}{4(y-1)}
M_{\pi^0}^2C_0^{-0} \\
& -\frac{(x-1)((x-1)(y+1)(y^2+1)(xy+x-y+1)-4\nu ^2(y^2+y+1))}{4(y+1)}
M_{\pi^0}^2C_0^{+0} \\
& +\frac{(x-1)((y-1)^2((x-1)^2(y^2+1)+\nu ^2(x-2))+\nu ^4)}{4(y-1)}
M_{\pi^0}^2C_0^{-} \\
& -\frac{(x-1)((y+1)^2((x-1)^2(y^2+1)+\nu ^2(x-2))+\nu ^4)}{4(y+1)}
M_{\pi^0}^2C_0^{+} \\
& -\nu ^2\frac{((x-1)(y-1)^2(x(3y-5)+2)+\nu ^2(y-1)(x(y-4)+3)-\nu ^4)}{8(y-1)^2}
\frac{M_{\pi^0}^2}{m_{-}^2}B_0^{-} \\
& +\nu ^2\frac{((x-1)(y+1)^2(x(3y+5)-2)-\nu ^2(y+1)(x(y+4)-3)+\nu ^4)}{8(y+1)^2}
\frac{M_{\pi^0}^2}{m_{+}^2}B_0^{+} \\
& -\frac{\nu ^2}{16(-1+y^2)^2}\Bigl[2(x-1)^2(y^2-1)^2(x(7+3y^2)-10) \\
& \hspace{2.6cm}+\nu ^2(x-1)(y^2-1)((4+x)y^2+11x-16) \\
& \hspace{2.6cm}-\nu ^4(1+y^2(x(7+y^2)-9)+\nu ^6(1+y^2))\Bigr]
\frac{M_{\pi^0}^4}{m_{+}^2m_{-}^2}B_0^0 \\
& +(x-1)x+\frac{5\nu ^2(x(3+y^2)-4)}{2(y^2-1)}-
\frac{\nu ^4((9x-8)y^2+7x-8)}{4(x-1)(y^2-1)^2} \\
& -\frac{\nu ^6x}{32(x-1)}\left(
\frac{M_{\pi^0}^2}{(y+1)m_{+}^2}-\frac{M_{\pi^0}
^2}{(y-1)m_{-}^2}\right) \Bigr\}.
\end{split}
\label{delta1gI}
\end{equation}
In this formula we have used the shorthand notation
\begin{alignat*}{2}
B_0^0& \equiv B_0(0,m^2,m^2),&B_0^{\pm}& \equiv B_0(m_{\pm }^2,0,m^2),\\
C_0^{\pm }& \equiv C_0(0,m_{\pm }^2,m^2,m^2,m^2,0),&
C_0^{\pm 0}& \equiv C_0(m^2,M_{\pi^0}^2,m_{\pm }^2,m^2,0,0),\\
D_0& \equiv D_0(m^2,0,m^2,M_{\pi^0}^2,m_{-}^2,m_{+}^2,0,m^2,m^2,0),
\end{alignat*}
where
\begin{equation*}
m_{\pm }^2 = m^2+\frac 12(1-x)(1\pm y)M_{\pi^0}^2 ,
\end{equation*}
and $B_0$, $C_0$ and $D_0$ are the standard scalar loop functions
(bubble, triangle and box) listed in Appendix~\ref{App:loop}.
Both $\delta ^{1\gamma IR;\,CT}(x,y)$ and  $\delta ^{1\gamma IR;\,loop}(x,y)$
contain divergences, in the form of poles at $d=4$, contained either
in the bare counterterm $\chi$, or in the loop function $B_0$.
From Eqs. (\ref{B0a}) and (\ref{B0b}), one deduces
\begin{equation}
\delta ^{1\gamma IR;\,loop}(x,y)\vert_{div}\,=
\,-48\,\frac{1}{d-4}\, \left(
\frac \alpha \pi \right) \frac{m^2x}{\left( 1-x\right) \left(
1-y^2\right) \left[ M_{\pi^0} ^2x(1+y^2)+4m^2\right]} .
\end{equation}
As follows from Eqs. (\ref{chi1}) and (\ref{chi2}),
these divergences cancel in the sum, so that
$\delta ^{1\gamma IR}(x,y)$ is both finite and
independent of the renormalization scale $\mu$.

In the literature, the explicit calculation of $\mathcal{M}_{\pi
^0\rightarrow e^{+}e^{-} \gamma}^{1\gamma IR}= \mathcal{M}_{\pi
^0\rightarrow e^{+}e^{-} \gamma}^{1\psi R}+\mathcal{M}_{\pi
^0\rightarrow e^{+}e^{-} \gamma}^{1PI}$, when considered at all, was
discussed in the approximation $m=0$ and assuming the pion to be
pointlike, cf.~\cite{tpe}, i.e. $\mathcal{A}_{\pi^0\rightarrow
e^{+}e^{-} \gamma}(l^2,(P-l)^2)= \mathcal{A}_{\pi^0\rightarrow
e^{+}e^{-} \gamma}^{LO}$, see~(\ref{WZampLO}). Let us note that in
this case the ultraviolet divergent part of $\mathcal{M}_{\pi
^0\rightarrow e^{+}e^{-} \gamma}^{1\gamma IR}$ vanishes. Indeed, the
divergent part, for $m=0$, is contained in the expression
\begin{multline}
\mathcal{M}_{\pi ^0\rightarrow e^{+}e^{-} \gamma;\,div}^{1\gamma IR} =
\mathrm{i}e^5
\mathcal{A}_{\pi^0\rightarrow e^{+}e^{-} \gamma}^{LO}\int \frac{\mathrm{d}^4l}{(2\pi
)^4}\varepsilon ^{\mu \nu \alpha \beta }l_\alpha P_\beta \Bigl(
\frac 1{l^2+\mathrm{i}\varepsilon }\Bigr) ^3  \\ \times
\overline{u}(p_-)\Bigl\{ -\Bigl( \frac{\gamma ^\rho
(\slashed{p}_- +\slashed{k})\gamma _\mu \slashed{l} \gamma _\nu
}{2(k\cdot p_-)}\Bigr) + \Bigl( \frac{\gamma _\mu
\slashed{l}\gamma_\nu (\slashed{p}_+ +\slashed{k})\gamma ^\rho
}{2(k\cdot p_+)}\Bigr) \Bigr\} v(p_+)\varepsilon _\rho ^{*}(k).
\end{multline}
Upon using the identity
\[
\varepsilon ^{\mu \nu \alpha \beta }l_\alpha P_\beta \gamma ^\mu
\slashed{l}\gamma ^\nu =2\mathrm{i}[\,\slashed{l}\,(l\cdot
P)-\slashed{P}\, l^2]\gamma_5
\]
and an effective substitution $l_\alpha l_\beta \rightarrow Cl^2
g_{\alpha \beta }$ (where $C$ depends on the cut-off prescription
used to regularize the divergent integral), one obtains
\begin{multline}
\mathcal{M}_{\pi ^0\rightarrow e^{+}e^{-} \gamma;\,div}^{1\gamma IR}
=-2e^5\mathcal{A}_{\pi^0\rightarrow e^{+}e^{-} \gamma}^{LO}
(C-1)\int \frac{\mathrm{d}^4l}{(2\pi
)^4}\Bigl( \frac 1{l^2+\mathrm{i}\varepsilon }\Bigr)^2 \\
\times \overline{u}(p_-)\Bigl\{ -\Bigr( \frac{\gamma ^\rho
(\slashed{p}_- +\slashed{k})\,\slashed{P}}{2(k\cdot p_-)}\Bigr)
+\Bigl( \frac{\slashed{P}\,(\slashed{p}_+ +\slashed{k})\gamma
^\rho }{2(k\cdot p_+)}\Bigr) \Bigr\} v(p_+)\varepsilon _\rho
^{*}(k).
\end{multline}
The two terms in the curly brackets cancel each other as a
consequence of the identities
\begin{eqnarray*}
(\slashed{p}_- +\slashed{k})\slashed{P}\, v(p_+) &=&2(k\cdot p_-)v(p_+), \\
\overline{u}(p_-)\slashed{P}(\slashed{p}_++\slashed{k})
&=&2(k\cdot p_+)\overline{u}(p_-)
\end{eqnarray*}
and thus the ultraviolet divergences are absent in the limit
$m\rightarrow 0$. This limit appears at first sight
to be a very good approximation, because the relevant
dimensionless parameter, $\nu ^2=(2m/M_{\pi^0})^2\simeq 5.7\times
10^{-5}$, is tiny. This indeed turns out to be the case
as far as the corrections to the total decay rate are
concerned. However, when considering the differential
decay rate, this simple argument can sometimes be misleading,
as we discuss in the following subsection.

\subsection{The Low approximation}\label{TLA}

Let us first briefly comment on the possible approximation of the above
result by the application of the Low theorem.
Since we are dealing with a radiative three
body decay, we can borrow from general results~\cite{lows}
and obtain
\begin{equation}
\mathcal{M}_{\pi^0\rightarrow e^+ e^-\gamma } \, = \,
\mathcal{M}_{\pi^0\rightarrow e^+ e^-\gamma }^{Low}
\, + \, {\cal O}(k,(q_i\cdot k))
,
\label{low1}
\end{equation}
with
\begin{eqnarray}
\mathcal{M}_{\pi^0\rightarrow e^+ e^-\gamma }^{Low} &=&
e\varepsilon _\mu (k)^{*}\overline{u}[\frac{ 2p_{-}^{\mu
}-\frac{a_e}m(p_{-}^{\mu }\, \slashed{k}-\gamma ^\mu (p_-\cdot k))-
\mathrm{i}\sigma ^{\mu \nu }k_\nu (1+a_e)}{2(p_-\cdot k)} \\
&& \!\!\!\!\! +\frac{-2p_{+}^{\mu }+\frac{a_e}m(p_{+}^{\mu }\,
\slashed{k}-\gamma ^\mu (p_+ \cdot k))-\mathrm{i}\sigma ^{\mu \nu
}k_\nu (1+a_e)}{2(p_+\cdot k)} ]\gamma_5 v P_{\pi
^0e^{-}e^{+}}(m^2,m^2), \label{lowAmplitude}
\nonumber
\end{eqnarray}
where the important point is the absence of contributions
that are independent of $k_\mu$ in the difference
$\mathcal{M}_{\pi^0\rightarrow e^+ e^-\gamma } -
\mathcal{M}_{\pi^0\rightarrow e^+ e^-\gamma }^{Low}$
(see Appendix~\ref{App:sps}).

Note that the
on-shell amplitude $P_{\pi ^0e^{+}e^{-}}(m^2,m^2)\,$, evaluated to
the order under consideration, reads
\[
P_{\pi ^0e^{+}e^{-}}(m^2,m^2)=P_{\pi
^0e^{+}e^{-}}^{loop}(m^2,m^2)+P_{\pi ^0e^{+}e^{-}}^{CT}(m^2,m^2),
\]
where
\[
P_{\pi ^0e^{+}e^{-}}^{loop}(m^2,m^2)=-\mathrm{i}\left( \frac
\alpha {2\pi }\right) ^2\frac m{F_\pi }[5+3B_0(0,m^2,m^2)-M_{\pi^0
}^2C_0(m^2,M_{\pi^0} ^2,m^2,m^2,0,0)].
\]
This means, using (\ref{LowPAT}) and (\ref{deltaPAT}), that the
Low amplitude $\mathcal{M}_{\pi^0\rightarrow e^+ e^-\gamma
}^{Low}$ generates the following correction
\begin{multline*}
\delta ^{Low}(x,y) = 2\left( \frac \alpha
\pi \right) \frac{\nu ^2x}{[x(1+y^2)+\nu ^2]}\frac 1{(1-x)(1-y^2)} \\
\times \mathrm{Re}[5 + 2 \chi + 3 B_0(0,m^2,m^2)-
M_{\pi^0}^2C_0(m^2,M_{\pi^0}^2,m^2,m^2,0,0)],
\end{multline*}
which corresponds exactly to the single pole part of the complete
one-photon reducible amplitude
$\mathcal{M}_{\pi ^0\rightarrow e^{+}e^{-} \gamma}^{1\gamma IR}=
\mathcal{M}_{\pi ^0\rightarrow e^{+}e^{-} \gamma}^{1\gamma IR;\,loop}
+\mathcal{M}_{\pi ^0\rightarrow e^{+}e^{-} \gamma}^{1\gamma IR;\,CT}$
for $x\rightarrow 1$. When
integrated, the Low contribution to $\delta (x)$ becomes
\begin{multline*}
\delta^{Low}(x) = 6\left( \frac \alpha \pi
\right) \frac{\nu ^2x}{(2x+\nu ^2)}\frac 1{(1-x)}
\frac 1{\sigma _e(xM_{\pi^0}^2)}
\ln \left( \frac{1+\sigma _e(xM_{\pi ^0}^2)}{1-\sigma _e(xM_{\pi^0}^2)}\right) \\
\times \mathrm{Re}[5 + 2 \chi + 3B_0(0,m^2,m^2)-
M_{\pi^0} ^2C_0(m^2,M_{\pi^0}^2,m^2,m^2,0,0)]\mathrm{.}
\end{multline*}
Notice that $\delta^{Low}$ is suppressed by
the factor $\nu^2$ and vanishes in the limit $m\rightarrow 0$.
This was the argument for the conjecture that in this limit
$\delta^{1\gamma IR}$ does not develop a pole when $x\rightarrow
1$ \cite{tpe} and the contributions of $1\gamma IR$ topologies can
be safely omitted. In fact this conjecture is not quite true for
several reasons we shall briefly discuss now.

First, the Low correction is not numerically relevant for almost
the whole phase space. Because of the suppression factor $\nu ^2$,
the corrections $\delta ^{Low}(x,y)$ and $\delta ^{Low}(x)$ become
important only in the experimentally irrelevant region where
$1-x\sim \nu ^2$ (when $y$ is fixed), or where $1-y^2\sim \nu ^2$
(when $x\gg \nu^2$ is fixed), \emph{i.e.} for $|y|\sim
y_{\max}(x)=\sqrt{1-\nu ^2/x}$. This is in fact no surprise,
because it is precisely this corner of the phase space where
Low's theorem is applicable. Indeed, the standard textbook
derivation of the Low amplitude involves (and assumes the
existence of) the power expansion of the form factors
corresponding to the off-shell $\pi
^0e^{+}(\widetilde{q}_1)e^{-}(\widetilde{q}_2)$ vertices $\Gamma
_{\pi ^0e^{-}e^{+}}(\widetilde{q}_2,q_1)$ and $\Gamma _{\pi
^0e^{-}e^{+}}(q_2,\widetilde{q}_1)$ in powers of
$\widetilde{q}_i^2$ at the points $\widetilde{q}_i^2=m^2$, where
$\widetilde{q}_i=q_i+k$. This means, that the relevant expansion
parameter is
\[
\Delta _{\pm }=\frac{\widetilde{q}_{1,2}^2}{m^2}-1=\frac{2k\cdot
q_{1,2}}{m^2}=\frac 2{\nu ^2}(1-x)(1\pm y) .
\]
Therefore, the $O(k)$ terms in the formula
\begin{equation}
\mathcal{M}_{\pi ^0\rightarrow e^{+}e^{-} \gamma}
=\mathcal{M}_{\pi ^0\rightarrow e^{+}e^{-} \gamma}^{Low}+O(k)
\label{Low theorem}
\end{equation}
are small for $\Delta _{\pm }\ll 1$, and not just for $1-x\ll 1$, as one could
naively expect.

There is another subtlety connected with such an expansion.
According to Low's theorem, in the region of its applicability one
would gather from (\ref{Low theorem})
\[
\delta ^{1\gamma IR}(x,y)-\delta ^{Low}(x,y)=O(1)
\]
with the ${\cal O}(1)$ term independent of $k$ (recall that the
leading order amplitude $\mathcal{M}_{\pi ^0\rightarrow e^{+}e^{-}
\gamma}^{LO}$ is of the order ${\cal O}(k)$). However, the points
$\widetilde{q}_i^2=m^2$ do not belong to the domain of analyticity
of our $\pi^0e^{+}(\widetilde{q}_1)e^{-}(\widetilde{q}_2)$
amplitude, because of the branch cuts starting at
$\widetilde{q}_i^2=m^2$ due to the intermediate $e^{\pm }\gamma $
states. As a result, the asymptotics of the amplitude for
$x\rightarrow 1$ will also contain, apart of the pole terms, non
analytical pieces, like non integer powers and logarithms. This
means we can expect
\[
\mathcal{M}_{\pi ^0\rightarrow e^{+}e^{-} \gamma}=
\mathcal{M}_{\pi ^0\rightarrow e^{+}e^{-} \gamma}^{Low}+
{\cal O}(k\ln k)+{\cal O}(k)+\ldots
\]
rather than (\ref{Low theorem}) and, as a result
\[
\delta ^{1\gamma IR}(x,y)-\delta ^{Low}(x,y)=
{\cal O}(\ln (1-x))+
{\cal O}(1)+\ldots .
\]
The Low correction therefore does not saturate the singular part of
$\delta ^{1\gamma IR}(x,y)$ for $x\rightarrow 1$. This can be
verified explicitly at lowest order. Using the asymptotic form of
the loop functions (cf. Appendix~\ref{App:loop}) we find from
(\ref{delta1gI}), for $\Delta _{\pm }\ll 1$,
\begin{multline*}
\delta ^{1\gamma IR}(x,y)-\delta ^{Low}(x,y) =\left( \frac \alpha
\pi \right) \frac 2{\nu ^2}\frac 1{1+y^2+\nu ^2} \Bigl[ ((1-y)^2-\nu
^2)\ln \Bigl( \frac 12(1-x)(1-y)\frac{M_{\pi^0} ^2}{m^2} \Bigr)\\
+((1+y)^2-\nu ^2)\ln \Bigl( \frac 12(1-x)(1+y)\frac{M_{\pi^0}
^2}{m^2} \Bigr) \Bigr] + {\cal O}(1)+\ldots .
\end{multline*}
To conclude, the Low amplitude does not provide us with a numerically
relevant estimate of $\delta ^{1\gamma IR}(x,y)$ in the kinematical region
of interest.

On the other hand, for $\Delta _{\pm }\gg 1$, which is satisfied
practically in the whole relevant domain of $x\,$and $y$ (with the
exception of the region where $x \sim 1-\nu^2$ or $|y|\sim y_{\max
}(x)=\sqrt{1-\nu ^2/x}$ with $x\gg \nu ^2$), we can approximate
the correction $\delta ^{1\gamma IR}(x,y)$ with its $m\rightarrow
0$ ($x<1$ and $|y|<$ $y_{\max }(x)$ fixed) limit with very good
accuracy. Note that in the case $m=0$ the loop
integration is infrared finite for $k\neq 0$,
and the ultraviolet divergences as well as the counterterm
contributions vanish. Using the corresponding asymptotic formulae
for the loop functions (see Appendix~\ref{App:loop}), the limit
can be easily calculated with the result (cf. \cite{tpe2} where
this approximative formula was published for the first time)
\begin{multline}
\delta ^{1\gamma IR}(x,y)|_{m\rightarrow 0} =\Bigl( \frac \alpha
\pi \Bigr) \Bigl[ -\frac x{(1-x)(1+y^2)}+\frac{\pi ^2}6 -\ln
\Bigl( \frac 12(1-x)(1-y)\Bigr) \ln \Bigl( \frac
12(1-x)(1+y)\Bigr)\\
-\mathrm{Li}_2\Bigl( 1-\frac 12(1-x)(1-y)\Bigr)
-\mathrm{Li}_2\Bigl( 1-\frac 12(1-x)(1+y)\Bigr) \Bigr] .
\label{dxym0}
\end{multline}
Notice the presence of the $\sim (1-x)^{-1}$ term, which stems from the
following part of the expression (\ref{delta1gI}) for $\delta ^{1\gamma
IR}(x,y)$
\[
\delta _{pole}^{1\gamma IR}(x,y)=\left( \frac \alpha \pi \right)
\frac{x^2}{[x(1+y^2)+\nu ^2]}\frac 1{(1-x)}.
\]
Because the limits $m\rightarrow 0$ and $x\rightarrow 1$ are not
interchangeable, as pointed out in Ref. \cite{tpe2}, for $m\neq 0$
the contribution of such a term is cancelled by the expansion, in
powers of $(x-1)$, of another term, namely
\[
-\left( \frac \alpha \pi \right) \frac{x^2}{[x(1+y^2)+\nu
^2]}\frac{\nu ^6x}{32(x-1)}\left( \frac{M_{\pi^0}
^2}{(y+1)m_{+}^2}-\frac{M_{\pi^0} ^2}{(y-1)m_{-}^2} \right),
\]
so that the only pole terms which survive are
suppressed by a factor $\nu ^2$ according to Low's theorem.
In the limit $m\rightarrow 0$ we also obtain \cite{tpe2}
\begin{multline}
\delta^{1\gamma IR}(x)|_{m\rightarrow 0}=-\Bigl( \frac \alpha \pi
\Bigr) \Bigl[ \ln ^2(1-x)+\frac{2x}{(1-x)^2}\ln(1-x)
\\+\frac{x(2x^2-3x+3)}{(1-x)^3}
\Bigl( \frac{\pi^2}6-\mathrm{Li}_2(x)\Bigr) -\frac{x(5x+3)}{4(1-x)^2} \Bigr],
\label{dxm0}
\end{multline}
which provides an excellent approximation to the exact
$m\neq 0$ result in the whole relevant range of $x$.

\subsection{Soft photon bremsstrahlung}

\label{spb} The virtual photon corrections described in the
previous subsections produce infrared divergences, which were
regularized by introducing the soft photon mass $m_\gamma $. As
usual, an infrared finite result is obtained at this order upon
adding to the decay rate the real photon bremsstrahlung
correction. This corresponds to the radiative process
$\pi^0\rightarrow e^{+}e^{-}\gamma \gamma _B$, which cannot be
distinguished from the non-radiative one for energies of the
bremsstrahlung photon smaller than the detector resolution $\Delta
E$. In the soft photon approximation, the amplitude of the
radiative decay is related to the leading matrix element by
\[
\mathcal{M}_{\pi^0\rightarrow e^{+}e^{-}\gamma \gamma _B} =e\left(
\frac{p_-\cdot \varepsilon _B^{*}(k_B)}{p_-\cdot
k_B}-\frac{p_+\cdot \varepsilon _B^{*}(k_B)}{p_+\cdot k_B}\right)
\mathcal{M}_{\pi^0\rightarrow e^{+}e^{-}\gamma}^{LO},
\]
where $k_B$ and $\varepsilon _B^{*}(k_B)$ are the momentum and
polarization vector of the bremsstrahlung photon, respectively.
Squaring the amplitude and summing over polarizations, one obtains
\[
\overline{|\mathcal{M}_{\pi^0\rightarrow e^{+}e^{-}\gamma \gamma
_B}|}^2 =e^2\left( \frac{2(p_+\cdot p_-)}{(p_+\cdot k_B)(p_-\cdot
k_B)}-\frac{m^2}{(p_+\cdot k_B)^2}-\frac{m^2}{(p_-\cdot
k_B)^2}\right) \overline{|\mathcal{M}_{\pi^0\rightarrow
e^{+}e^{-}\gamma}^{LO}|}^2.
\]
The corresponding correction appearing in~(\ref{ddef}) is then
\[
\delta ^B(x,y)=e^2\int_{|\mathbf{k}_B|<\Delta
E}\frac{d^3\mathbf{k}_B}{(2\pi )^32k_B^0}\left( \frac{2(p_+\cdot
p_-)}{(p_+\cdot k_B)(p_-\cdot k_B)}-\frac{m^2}{(p_+\cdot
k_B)^2}-\frac{m^2}{(p_-\cdot k_B)^2}\right) ,
\]
where $k_B^0=\sqrt{\mathbf{k}_B^2+ m_\gamma ^2}$. The correction
$\delta^B(x,y)$ can be expressed, in terms of the standard
integral
\[
J(q,q')=\int_{|\mathbf{k}_B|<\Delta E}\frac{d^3\mathbf{k}_B}{(2\pi
)^32k_B^0}\frac 1{(q\cdot k_B)(q'\cdot k_B)},
\]
as
\begin{eqnarray*}
\delta ^B(x,y) &=&e^2\left( 2(p_+\cdot
p_-)J(p_+,p_-)-m^2J(p_+,p_+)-m^2J(p_-,p_-)\right) \\
&=&e^2\left( (xM_{\pi
^0}^2-2m^2)J(p_+,p_-)-m^2J(p_+,p_+)-m^2J(p_-,p_-)\right) .
\end{eqnarray*}

Let us note that the integral $J(q,q')$ is not Lorentz invariant
and the result is therefore frame dependent. On the other hand,
the infrared divergent part of $J(q,q')$ is given by the invariant
expression
\begin{eqnarray*}
J_{IR\ div}(q,q') &=&\frac 1{2(2\pi )^2}\ln \left( \frac{4\Delta
E^2}{m_\gamma ^2}\right) \int_0^1\frac{\mathrm{d}x}{[xq+(1-x)q']^2} \\
&=&\frac 1{2(2\pi )^2}\ln \left( \frac{4\Delta E^2}{m_\gamma^2}\right)
\frac 1{\lambda^{1/2}(s,q^2,q'^2 )}\ln \left(
\frac{s-q^2-q'^2 + \lambda^{1/2}(s,q^2,q'^2 )}{s-q^2-q'^2 -\lambda
^{1/2}(s,q^2,q'^2 )}\right) ,
\end{eqnarray*}
where $\lambda (x,y,z)=x^2+y^2+z^2-2xy-2xz-2yz$ is the triangle
function and $s=(q+q')^2$. The infrared finite part can be
transformed to the form
\begin{eqnarray*}
J_{IR\ fin}(q,q') &=&-\frac 1{2(2\pi )^2}\int_0^1\frac{\mathrm{d}x}{[xq+(1-x)q']^2}\frac{xq^0+(1-x)q'^0}{|x\mathbf{q}+(1-x)\mathbf{q}'|} \\
&&\times \ln \left( \frac{xq^0+(1-x)q'^0
+|x\mathbf{q}+(1-x)\mathbf{q}'|}{xq^0+(1-x)q'^0
-|x\mathbf{q}+(1-x)\mathbf{q}'|}\right) .
\end{eqnarray*}
In an arbitrary frame we can easily obtain
\[
J(q,q)=\frac 1{2(2\pi )^2}\frac 1{q^2}\left[ \ln \left(
\frac{4\Delta E^2}{m_\gamma ^2}\right) +\frac{q^0}{|\mathbf{q}|}\ln
\left( \frac{q^0-|\mathbf{q}|}{q^0+|\mathbf{q}|}\right) \right] .
\]
For $q\neq q'$, the calculation of the explicit form of $J(q,q')$
is much more complicated. In the center of mass of $q$ and $q'$, with
$q^2=q'^2=m^2$, the integral $J_{IR\, fin}(q,q')$ simplifies
considerably and we obtain
\begin{eqnarray*}
J(q,q') &=&\frac 1{(2\pi )^2}\frac 1{s\sigma (s)}\left\{ \ln
\left( \frac{1+\sigma (s)}{1-\sigma (s)}\right) [\ln \left(
\frac{4\Delta E^2}{m_\gamma ^2}\right) +\frac 12\ln \left(
\frac{1-\sigma (s)^2}4\right)
]\right. \\
&&\left. +\mathrm{Li}_2\left( \frac{1+\sigma (s)}2\right) -
\mathrm{Li}_2\left( \frac{1-\sigma (s)}2\right) -4\chi _2(\sigma (s))\right\}, \\
J(q,q) &=&J(q',q')=\frac 1{2(2\pi )^2}\frac 1{m^2}\left[ \ln
\left( \frac{4\Delta E^2}{m_\gamma ^2}\right) -\frac 1{\sigma
(s)}\ln \left( \frac{1+\sigma (s)}{1-\sigma (s)}\right) \right],
\end{eqnarray*}
where $\sigma (s)=(1-4m^2/s)^{1/2}$ and $\chi _2(x)=\frac
12[\mathrm{Li}_2(x)-\mathrm{Li}_2(-x)]$ is the Legendre
chi-function.

If we interpret $\Delta E$ as the photon energy resolution in
the center of mass of the Dalitz pair, we find
\[
\delta ^B(x,y)=\delta _{IR\ div}^B(x,y)+\delta _{IR\ fin}^B(x,y)
\]
with
\begin{equation}
\delta _{IR\ div}^B(x,y)=\frac{e^2}{(2\pi )^2}\ln \left(
\frac{4\Delta E^2}{m_\gamma ^2}\right) \left[ \left(
1-\frac{2m^2}{xM_{\pi ^0}^2}\right) \frac 1{\sigma _e(xM_{\pi
^0}^2)}\ln \left( \frac{1+\sigma _e(xM_{\pi ^0}^2)}{1-\sigma
_e(xM_{\pi ^0}^2)}\right) -1\right]  \label{deltaxyBIRdiv}
\end{equation}
and
\begin{eqnarray*}
\delta _{IR\ fin}^B(x,y) &=&\frac \alpha \pi \frac 1{\sigma _e(xM_{\pi
^0}^2)}\left\{ \left( 1-\frac{2m^2}{xM_{\pi ^0}^2}\right) \left[ \frac 12\ln
\left( \frac{1-\sigma _e(xM_{\pi ^0}^2)}{1+\sigma _e(xM_{\pi ^0}^2)}\right)
\ln \left( \frac{xM_{\pi ^0}^2}{m^2}\right) \right. \right. \\
&&\left. +\mathrm{Li}_2\left( \frac{1+\sigma _e(xM_{\pi
^0}^2)}2\right) -\mathrm{Li}_2\left( \frac{1-\sigma _e(xM_{\pi
^0}^2)}2\right) -4\chi
_2(\sigma _e(xM_{\pi ^0}^2))\right] \\
&&+\left. \ln \left( \frac{1+\sigma _e(xM_{\pi ^0}^2)}{1-\sigma _e(xM_{\pi
^0}^2)}\right) \right\} .
\end{eqnarray*}
Summing $\delta _{IR\ div}^B(x,y)$ and $\delta _{NLO}^{1\gamma
R}(x,y)_{IR\ div}$, as given by (\ref{deltaxyBIRdiv}) and
(\ref{deltaxyNLOIRdiv}), we explicitly achieve the expected
cancellation of the infrared divergences,
\begin{multline*}
\delta _{IR\ div}^B(x,y)+\delta _{NLO}^{1\gamma
R}(x,y)_{IR\ div}=\frac \alpha \pi \ln \Bigl( \frac{m^2}{4\Delta
E^2}\Bigr)\\ \times \Bigl[ 1+\Bigl( 1-\frac{2m^2}{xM_{\pi
^0}^2}\Bigr) \frac 1{\sigma _e(xM_{\pi ^0}^2)}\ln \Bigl(
\frac{1-\sigma _e(xM_{\pi ^0}^2)}{1+\sigma _e(xM_{\pi
^0}^2)}\Bigr) \Bigr] .
\end{multline*}

\section{Numerical results }
\setcounter{equation}{0}

In the previous sections we have classified the NLO corrections
according to the general topology of the corresponding Feynman
diagrams. The complete correction $\delta _{NLO}(x,y)$
(see~(\ref{ddef})) is then given by the sum of the individual
contributions of the one-photon reducible, brem\-sstrah\-lung and
one-photon irreducible graphs:
\[
\delta _{NLO}(x,y)=\delta _{NLO}^{1\gamma R}(x,y)+\delta
^B(x,y)+\delta ^{1\gamma IR}(x,y) .
\]
A similar decomposition holds for $\delta _{NLO}(x)$. The resulting
formulae contain several renormalization scale independent
combinations of the a priori unknown low energy couplings and of
chiral logarithms.

\subsection{Inputs}

The contributions of the low energy couplings to $\delta
_{NLO}(x,y)$ and $\delta _{NLO}(x)$ are contained in $\delta
^{1\gamma IR;\,CT}(x,y)$ and in $a_{NLO}^{ChPT}(xM_{\pi ^0}^2)$,
see for instance Eqs. (\ref{chi2}), (\ref{d1grxy}) and
(\ref{achptnlo}). We define the scale independent quantities
\begin{eqnarray*}
\delta _{NLO}^{LEC}(x,y) &=&2\left( {\cal C}_1 +
\frac{e^2}{64\pi^2}{\cal K}_F  +\frac 16 {\cal C}_2
\frac{M_{\pi ^0}^2}{M_{\pi ^{\pm }}^2}x\right)\\&& +4 {\overline\chi}
\left( \frac \alpha \pi \right) \frac{\nu ^2x}{\left( 1-x\right)
\left( 1-y^2\right) \left[ x(1+y^2)+\nu ^2\right] }, \\
\delta _{NLO}^{LEC}(x) &=&2\left( {\cal C}_1 +
\frac{e^2}{64\pi^2}{\cal K}_F  +\frac 16 {\cal C}_2
\frac{M_{\pi ^0}^2}{M_{\pi ^{\pm }}^2}x\right) \\
&&+3 {\overline\chi}\left( \frac \alpha \pi \right) \frac{\nu
^2x}{\left( 1-x\right) (2 x+\nu ^2)}\frac 1{\sigma _e(xM_{\pi
^0}^2)}\ln \left( \frac{ 1+\sigma _e(xM_{\pi ^0}^2)}{1-\sigma
_e(xM_{\pi ^0}^2)}\right),
\end{eqnarray*}
which contain the contributions of the low-energy constants,
with
$$
\overline{\chi} = \chi^{r}(\mu)+
\frac{3}{2}\ln\frac{M_{\pi^0}^2}{\mu^2}.
$$
The differences
$\delta _{NLO}^\text{known}=\delta _{NLO}-\delta_{NLO}^{LEC}$
are expressed in terms of the known physical
observables (and the detector resolution $\Delta E$) and represent
therefore a numerically unambiguous part of our calculations.

For the combinations ${\cal C}_1$, ${\cal C}_2^r$, and ${\cal
K}_F$, we have reasonable estimates based on resonance
approximations, as described in Appendix~\ref{App:cepppv}. For
$\chi$  the {\it LMD\/} approximation was studied in
\cite{knechtcpt}, with the result
\[
\chi _{LMD}^r(\mu )=\frac{11}4-4\pi ^2\frac{F_\pi ^2}{M_V^2}-\frac
32\ln \left( \frac{M_V^2}{\mu ^2}\right).
\]
For numerical calculation we take the {\it LMD\/} values, with
$M_V = 770\,$MeV:
\begin{eqnarray}
{\cal C}_1 &=&(2.2\pm 0.3)\times 10^{-2},  \nonumber \\
{\cal K}_F &=&-28\pm 8,  \nonumber \\
{\cal C}_2^r(\mu = M_V) &=&(1.5\pm 0.5)\times 10^{-1},  \nonumber \\
\chi^r(\mu = M_V) &=&2.2\pm 0.7.  \label{delta_a}
\end{eqnarray}

\subsection{Radiative corrections to the differential decay rate}

\begin{figure}[htb]
\center\scalebox{0.65}{\epsfig{figure=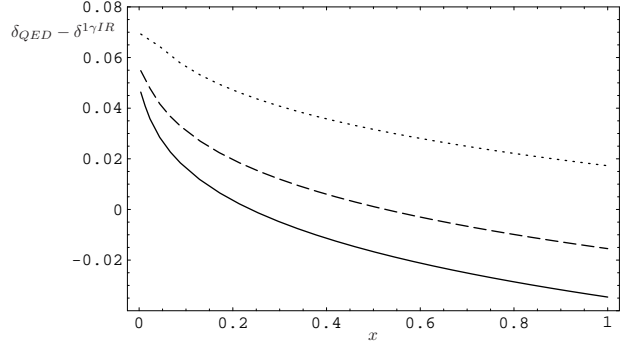}} \caption{The
traditional QED corrections (without $1\gamma IR$ contributions) for
different detector resolutions $\Delta E = 10\,$MeV (solid curve),
15\,MeV (dashed curve) and 30\,MeV (dotted curve).} \label{figure5}
\end{figure}

\begin{figure}[htb]
\center\scalebox{0.65}{\epsfig{figure=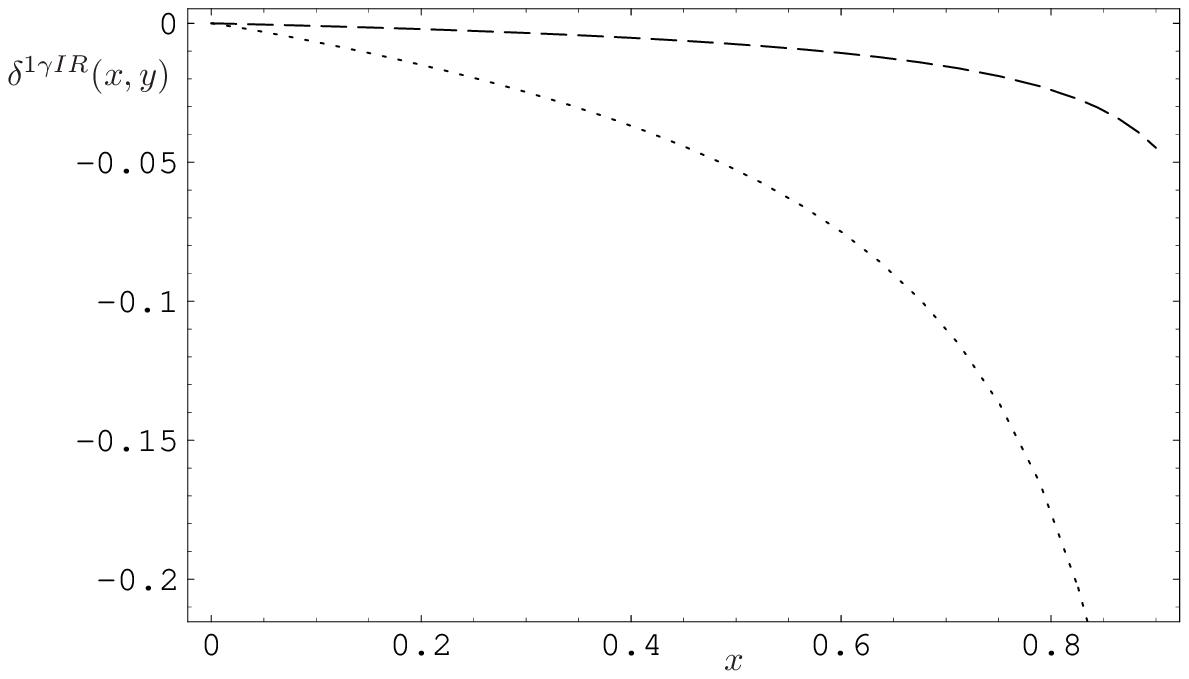}}
\scalebox{0.65}{\epsfig{figure=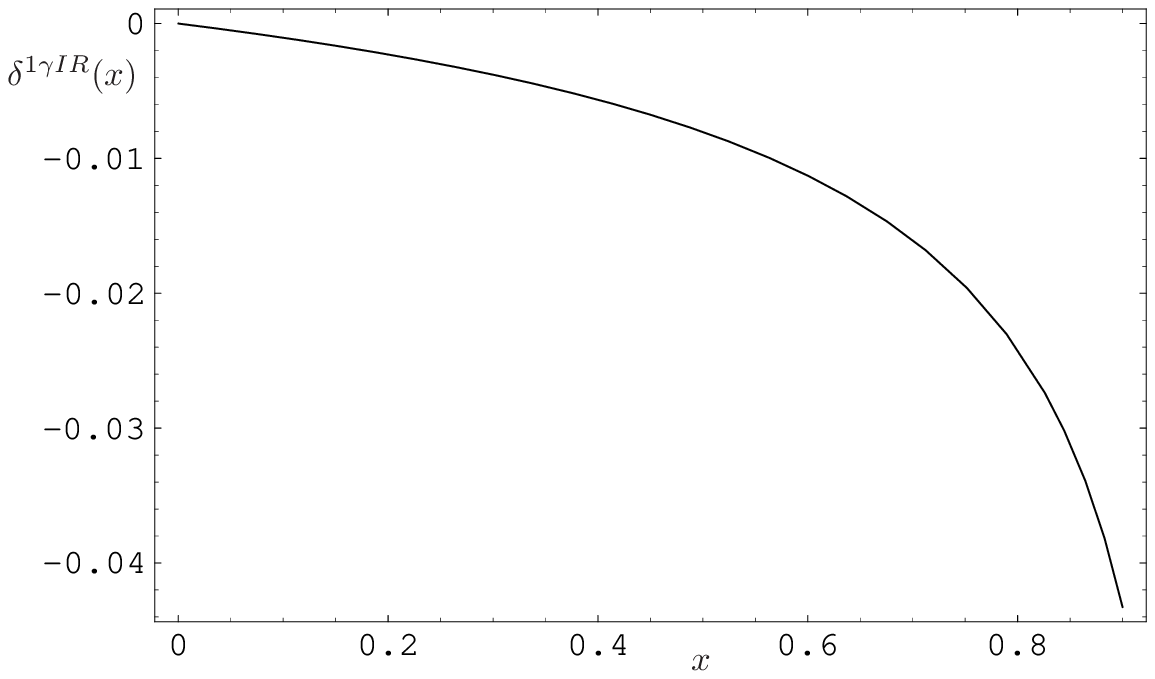}} \caption{The
one-photon irreducible corrections (triangle and box diagrams): the
dashed line represents $\delta(x,0)$, the dotted line
$\delta(x,y_{max})$, and the solid line on the right $\delta(x)$.}
\label{figure6}
\end{figure}

The traditional
point of view is to separate from the complete $NLO$ corrections
the pure electromagnetic part $\delta _{QED}$, which includes the
$1\gamma R$ graphs with the virtual fermion and photon loops only
and bremsstrahlung contribution, together with $1\gamma IR$
diagrams (the latter were usually omitted in the analysis of the
experimental data~\cite{exp1,exp2,exp3}; we shall comment on the
consequences of this omission below):
\begin{equation}
\delta _{QED}=\delta^{1\gamma R}|_{\gamma ,\psi \,loops}+\delta
^B+\delta ^{1\gamma IR},\label{QEDRB}
\end{equation}
where, cf. Eqs.~(\ref{achptnlo}) and (\ref{pp2}),
\begin{equation}
\delta^{1\gamma R}|_{\gamma ,\psi \,loops}(x)=\delta^{1\gamma R}(x)-
2\mathrm{Re}\left[a_{NLO}^{ChPT}(xM_{\pi
^0}^2)-\overline{\Pi }_{\pi ^{\pm }}(xM_{\pi ^0}^2)\right].
\end{equation}
Following this point of view we present here separate
plots\footnote{We do not show $\delta _{NLO}^{1\gamma R}\vert_{\gamma
,\psi \,loops}(x,y)+\delta ^B(x,y)$, because the $y$ dependence is
suppressed here by the factor $\nu ^2$ for $x>\nu ^2$, and is
therefore negligible in the relevant region of $x$.} for $\delta
_{NLO}^{1\gamma R}|_{\gamma ,\psi \,loops}(x)+\delta ^B(x)$, where
the experimental situation is parameterized by the detector
resolution ${\Delta }E$ (for which we take $\Delta E=10$\,MeV,
15\,MeV and 30\,MeV, see Fig.~\ref{figure5}), and for $\delta
^{1\gamma IR}(x,y)$ together with $\delta ^{1\gamma IR}(x)$
(depicted in Fig.~\ref{figure6}).


\begin{figure}[t]
\center\scalebox{0.65}{\epsfig{figure=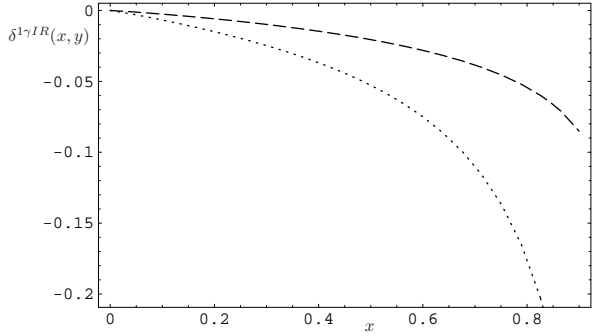}} \caption{The
difference between $\delta^{1\gamma IR}(x,y_{max})$ (dotted line)
and $\delta^{1\gamma IR}(x,y_{max})|_{m\rightarrow 0}$ (dashed
line).} \label{figure7}
\end{figure}

\begin{figure}[b]
\center\scalebox{0.65}{\epsfig{figure=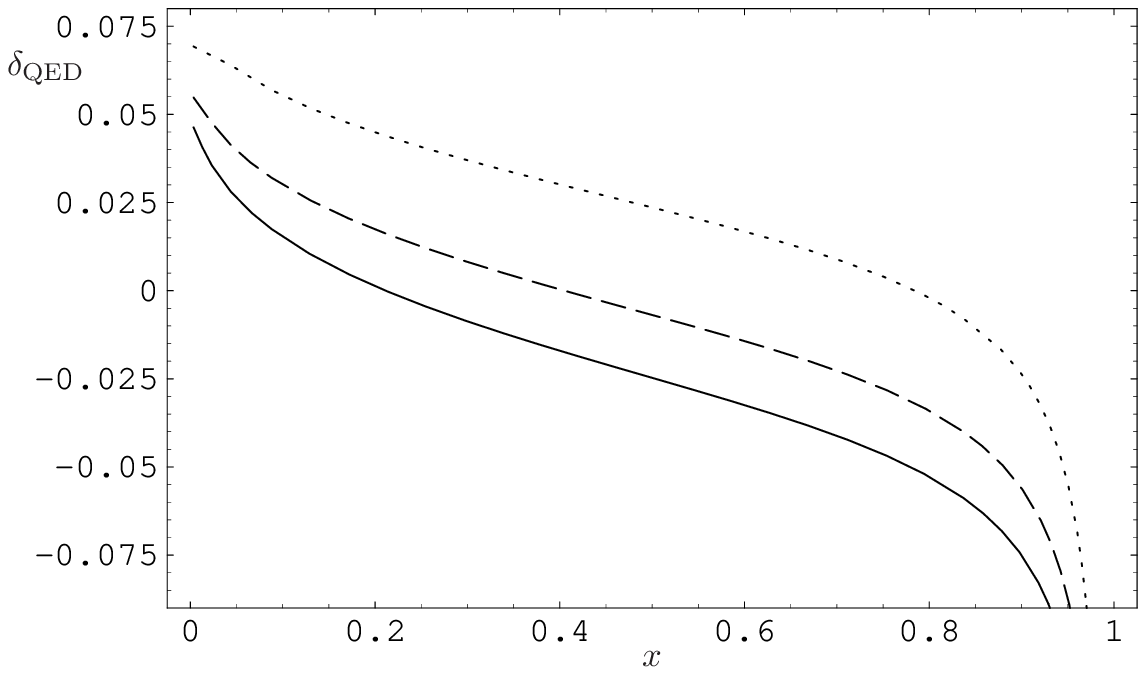}}
\scalebox{0.65}{\epsfig{figure=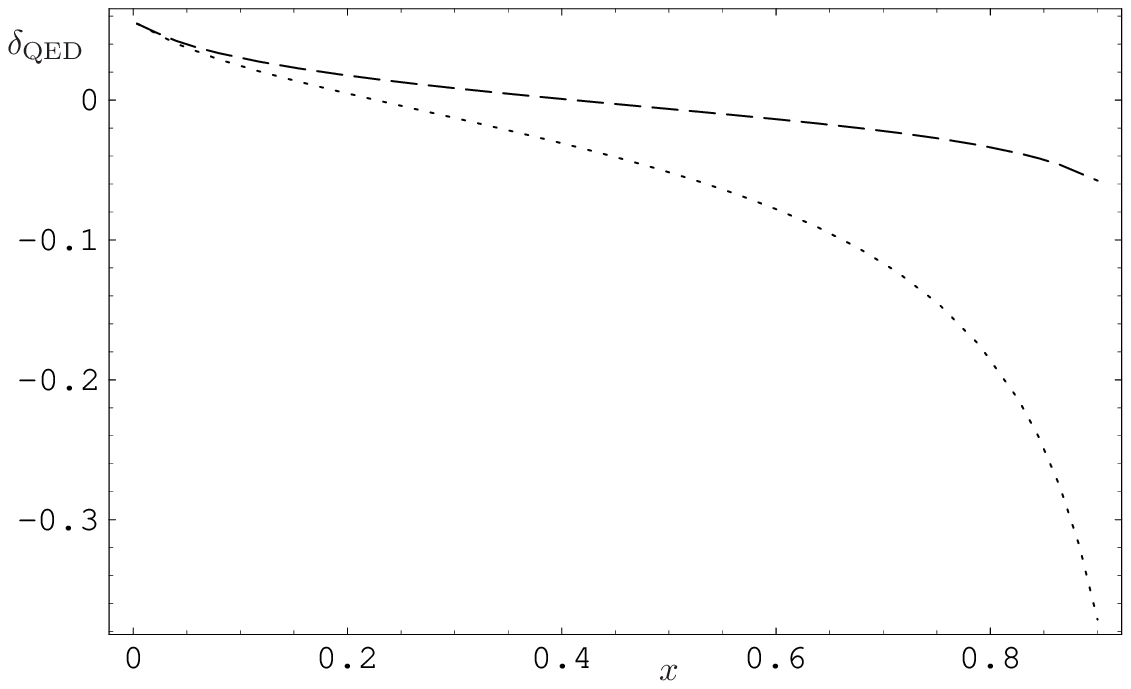}} \caption{On the left
hand side, the complete electromagnetic corrections $\delta_{QED}$
with the same detector resolutions as in Fig.~\ref{figure5}. The
right hand side shows $\delta_{QED}(x,y_{max})$ (dotted curve) and
$\delta_{QED}(x,0)$ (dashed curve).} \label{figure8}
\end{figure}

In the latter two we use the value of ${\chi }^r(M_V)$ mentioned
above. As we have discussed in the previous section, $\delta
^{1\gamma IR}(x)$ can be safely approximated with its $m\rightarrow
0$ limit (\ref{dxm0}) for almost the whole range of $x$; the same is
true for $\delta ^{1\gamma IR}(x,y)$ for $|y|<y_{\max }(x)$, the
difference between $\delta ^{1\gamma IR}(x,y)$ and (\ref{dxym0}) can
be seen for $y\sim y_{\max }(x)$ in Fig.~\ref{figure7}.
From these figures we can conclude, in agreement with \cite{tpe},
that the usually neglected $1\gamma IR$ corrections $\delta
^{1\gamma IR}(x)$ are in fact important particularly in the region
$x\gtrsim 0.6$, where they are in absolute value larger than $1\%$
(up to $\sim 4\%$ for $x\sim 0.9$), and comparable with $\delta
_{NLO}^{1\gamma R}|_{\gamma ,\psi \,loops}+\delta ^B$; the same is
true for $\delta ^{1\gamma IR}(x,y)$, which is almost independent on
$y$ (except for a very narrow region near $|y|\sim y_{\max }(x)$,
cf. previous section). The complete pure electromagnetic corrections
$\delta _{QED}(x)$ and $\delta _{QED}(x,y)$ are represented in
Fig.~\ref{figure8}.

\begin{figure}[htb]
\center\scalebox{0.65}{\epsfig{figure=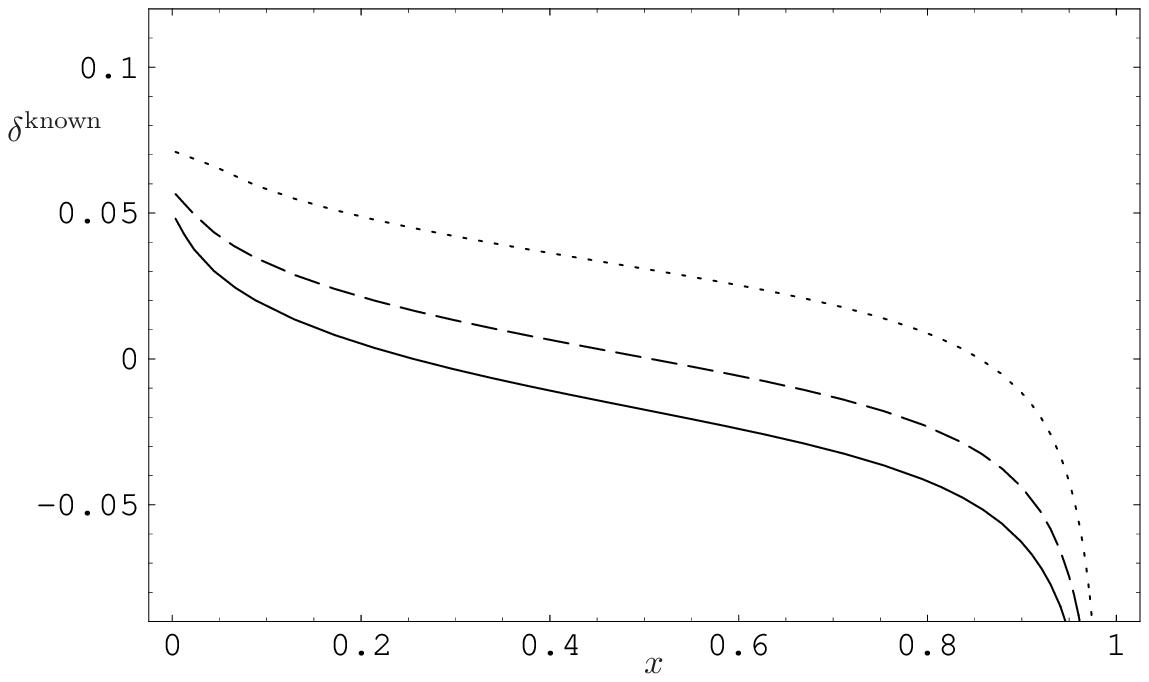}}
\scalebox{0.65}{\epsfig{figure=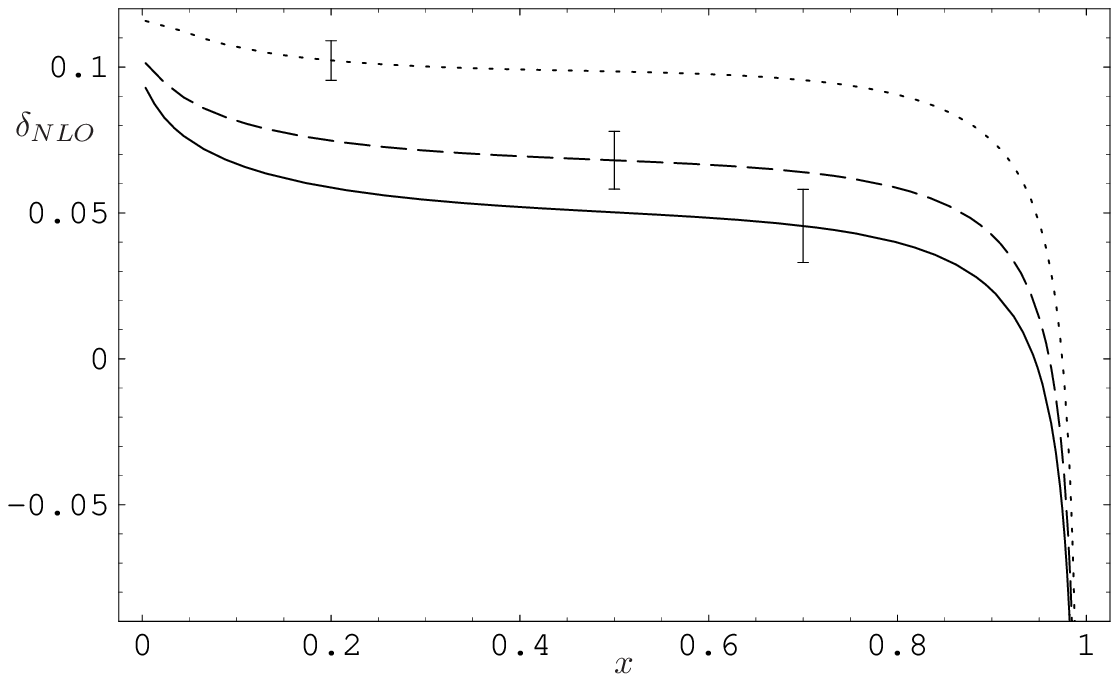}} \caption{The
NLO correction $\delta^\text{known}(x)$, without counterterm
contributions (left), and the complete correction
$\delta_{NLO}(x)$ (right). The error bars show
the uncertainties coming from the counterterm
determinations. The detector resolutions are as
in Fig.~\ref{figure5}. }
\label{figure9}
\end{figure}

Let us now change the point of view a little bit, and split the
corrections in the way we have described in the beginning of this
section, namely
\[
\delta _{NLO}=\delta^\text{known}+\delta _{NLO}^{LEC}.
\]
Here $\delta^\text{known}$ differs from $\delta _{QED}$ (with
$\overline{\chi }$ set to zero) by the contribution of
$\overline{\Pi }_{\pi ^{\pm }}(xM_{\pi ^0}^2)$ as well as by
corrections stemming from chiral pion loops,
\[
\delta^\text{known}(x,y)=\delta _{QED}(x,y)|_{\overline{\chi
}=0}+2\mathrm{Re}\left[ -\overline{\Pi }_{\pi ^{\pm }}(xM_{\pi
^0}^2)+\overline{a_{NLO}^{ChPT}}(xM_{\pi ^0}^2)\right] ,
\]
where
\begin{equation}
\overline{a_{NLO}^{ChPT}}(l^2)=\frac{M_{\pi ^{\pm }}^2}{16\pi
^2F_\pi ^2}\left[ \ln \frac{M_{\pi ^{\pm }}^2}{M_{\pi
^0}^2}-\frac{l^2}{6M_{\pi ^{\pm }}^2}\left( \frac 13+\ln
\frac{M_{\pi ^{\pm }}^2}{M_{\pi^0}^2}-16\pi ^2\sigma _{\pi
^{+}}^2(l^2){\bar J}_{\pi ^{+}}(l^2)\right) \right] .
\end{equation}
Separation of the numerically unambiguous part $\delta^\text{known}$
allows, at least in principle, to constrain the relevant
combinations of the chiral low energy constants ${\cal C}_1^r$,
${\cal C}_2^r$ and ${\cal K}_F$ from experiment. Fig.~\ref{figure9}
shows both $\delta^\text{known}$ and $\delta _{NLO}$, which allows
to appreciate the effect of the counterterms.
The difference $\delta^\text{known}(x)-\delta _{QED}(x)$ is
particularly important for $x\gtrsim 0.7$, where it represents a
correction larger than $1\%$.

Let us split further
\[
\delta _{NLO}^{LEC}=\delta _{LMD}^{LEC}+\delta _{QED}^{LEC},
\]
where $\delta _{LMD}^{LEC}=\delta _{NLO}^{LEC}|_{{{\cal K}_F}=0}$ is
the part for which we have theoretical prediction based on the {\it
LMD\/} approximation; $\delta^\text{known}+\delta _{LMD}^{LEC}$ is
shown in the Fig.~\ref{figure9}, the error band stems from the
estimate of the uncertainty of the {\it LMD\/} values
(\ref{delta_a}).

For completeness, we present our theoretical estimate of $\delta
_{QED}^{LEC} $, based on (\ref{delta_prime_a})
(cf.~(\ref{delta_a})),
\[
\delta _{QED}^{LEC}(x,y)=\delta
_{QED}^{LEC}(x)=\frac{e^2}{32\pi^2}{\cal K}_F=(-8 \pm 2)\times
10^{-3}.
\]
Note that this value is comparable with  the estimated uncertainty
of $\delta _{LMD}^{LEC}$ arising from the uncertainties in ${\cal
C}_1$ and ${\cal C}_2^r(\mu = M_V)$ as given in Eq. (\ref{delta_a}).

\subsection{Decay rate}
Let us make a brief comment on the total decay rate. At the leading
order we reproduce the old Dalitz result~\cite{dalitz51}:
$$
\frac{\Gamma^\text{LO}(\pi^0\rightarrow e^+ e^-
\gamma)}{\Gamma(\pi^0 \rightarrow 2\gamma)}= \frac{\alpha}{\pi}
\Bigl(\frac43 \ln \frac{M_{\pi^0}}{m} - \frac73 + O(\nu^2) \Bigr) =
0.01185,
$$
which should be compared with the present experimental
value~\cite{pdg}:
\begin{equation}
\frac{\Gamma^\text{exp}(\pi^0\rightarrow e^+ e^-
\gamma)}{\Gamma(\pi^0 \rightarrow 2\gamma)} = 0.01213 \pm
0.00033.\label{brpdg}
\end{equation}
As already mentioned in the Introduction, the traditional radiative
corrections to the total decay rate are tiny. The corrections
corresponding to the first and second term in the decomposition of
the QED corrections~(\ref{QEDRB}) were first numerically evaluated
in \cite{joseph60} and analytically in~\cite{lautrup71} with the
result:
\begin{align}
\frac{\Gamma^\text{NLO}_\text{QED}(\pi^0\rightarrow e^+ e^-
\gamma)}{\Gamma(\pi^0 \rightarrow 2\gamma)}&=
\Bigl(\frac{\alpha}{\pi}\Bigr)^2
\Bigl(\frac{8}{9}\ln^2{\frac{M_{\pi^0}}{m}}-\frac{19}{9}\ln{\frac{M_{\pi^0}}{m}}
+2\zeta(3) +\frac{137}{81} -\frac{2}{27}\pi^2 +
O(\nu)\Bigr)\notag\\
&= 1.04\times 10^{-4}.
\end{align}
Let us note that this formula is not based on the soft photon
approximation, but the whole energy spectrum of the bremsstrahlung
photon is included. A real photon emitted from the pion vertex is
not considered. If we take into account the remaining NLO $1\gamma
R$ (ChPT corrections) and $1\gamma IR$ electromagnetic corrections
we get an additional contribution
\begin{equation}
\frac{\Gamma^\text{NLO}_{\text{ChPT} + 1\gamma IR}(\pi^0\rightarrow
e^+ e^- \gamma)}{\Gamma(\pi^0 \rightarrow 2\gamma)}= (1.8 \pm
0.6)\times 10^{-5},
\end{equation}
where the error stems from uncertainty of ${\cal
C}_2$.\footnote{Note that the ratio $\frac{\Gamma(\pi^0\rightarrow
e^+ e^- \gamma)}{\Gamma(\pi^0 \rightarrow 2\gamma)}$ is independent
of the unknown constants ${\cal C}_1$ and ${\cal K}_F$ to the order
considered here.} Clearly these two corrections are small in
comparison with the present experimental uncertainty~(\ref{brpdg}).

Similarly we could evaluate the corrected rate in the soft-photon
approximation. The result then depends on $\Delta E$. For $\Delta E
\sim 10 MeV$, for instance,  the result reads
$\frac{\Gamma^\text{NLO}}{\Gamma_{2\gamma}}\simeq 4\times 10^{-4}$.

\subsection{Slope parameter}

We have now all the elements at hand in order to discuss
both the extraction of the slope parameter from the data,
and the prediction that can be made for it in the framework
of the low energy theory.
With the help of the formulae (\ref{d1gr}), (\ref{achptnlo}),
(\ref{Pbar}) and the definition (\ref{defa_pi}) we easily find
\begin{equation}
a_\pi=\frac{M_{\pi ^0}^2}{M_{\pi ^{\pm }}^2}\left[ \frac 16\,
{\cal C}_2 -\frac{M_{\pi ^{\pm }}^2}{96\pi ^2F_\pi ^2}\left( 1+\ln \left(
\frac{M_{\pi ^{\pm }}^2}{M_V^2}\right) \right) -\frac 1{360}\left(
\frac \alpha \pi \right) \right], \label{slope1}
\end{equation}
where the individual terms in the square bracket correspond to the
counterterm, the charged pion chiral loops, and the charged pion
vacuum polarization function contribution, respectively (the
latter we include here only for the sake of completeness,
numerically it is negligible, being of the order $10^{-6}$, and
thus can be safely neglected). Using the previous inputs, we
obtain the following theoretical prediction for the slope
parameter
\begin{equation}
a_\pi\,=\,0.029 \pm 0.005 .
\end{equation}

As we have
noted in the preceding section, previous experimental analyses,
as a rule, did not include the contribution
of the two-photon exchange (which was treated as
negligible due to the superficial arguments based on the Low
theorem). Therefore, according to the formula~(\ref{defa_pi}), the
systematic bias due to this omission can be roughly estimated as,
cf. (\ref{dxm0}),
\begin{equation}
\Delta a_\pi\bigg\vert_{1\gamma IR}=-\frac 12\frac{d\delta ^{1\gamma
IR}(x)|_{m\rightarrow 0}}{dx}\bigg\vert_{x=0}=\frac 18\left( \frac \alpha
\pi \right) (2\pi ^2-3)\doteq 0.005.
\end{equation}
This corresponds to a shift of the central values for $a_\pi$
extracted from the Dalitz decay measurements which goes into
the right direction towards the independent CELLO result.

\section{Summary and conclusions }
\setcounter{equation}{0}

The present work provides a detailed analysis of next-to-leading
order radiative corrections to the Dalitz decay amplitude. This
study involves the off-shell pion-photon transition form factor,
which requires a treatment of non perturbative strong interaction
effects. We have relied on representations of this form factor
involving zero-width vector resonances. In contrast to the simplest
vector meson dominance representation, our approach satisfies
various short distance constraints from QCD. Our analysis also
includes the one-photon irreducible contributions, which were
usually neglected. We have shown that, although these contributions
are negligible as far as the corrections to the total decay rate are
concerned, they are however sizeable in regions of the Dalitz plot
which are relevant for the determination of the slope parameter
$a_\pi$ of the pion-photon transition form factor. We have also
obtained a prediction for $a_\pi$ which is in good agreement with
the determinations obtained from the (model dependent) extrapolation
of the CELLO and CLEO data. The present difference with the central
values directly measured in the latest Dalitz decay experiments can
be ascribed to the omission of the radiative corrections induced by
the one-photon irreducible contributions. Unfortunately, the
experimental error bars on the latest values of $a_\pi$ extracted
from the Dalitz decay are still too large to make a comparison with
the CELLO and CLEO values meaningful. Nevertheless, we think that a
precise measurement of $a_\pi$ which would not rely on any kind of
extrapolation remains an interesting issue. Hopefully, future
experiments, like the one proposed by the PrimEx collaboration at
TJNAF, will improve the situation in this respect.

\indent

\indent

\noindent {\bf Acknowledgements}. We wish to thank
J.~Ho\v{r}ej\v{s}\'{\i} for comments on the manuscript, as well as
J. Schacher, L. Nemenov, and A. Bernstein for interesting
discussions and/or correspondences. We are also grateful to
B.~Moussallam for a useful discussion concerning the $SU(3)$
matching. K.~K. and J.~N. are supported in part by the program
"Research Centres", project number LC05A02A of the Ministry of
Education of the Czech Republic. The work of M. K. is supported in
part by the EC contract No. HPRN-CT-2002-00311 (EURIDICE). This work
was initiated during a stay of K. K. at Centre de Physique
Th\'eorique, partially financed by the SOCRATES/ERASMUS student
exchange program.

\indent

\appendix
\section{Form factor projectors}
\renewcommand{\theequation}{A.\arabic{equation}}
\setcounter{equation}{0}

Here we list the projectors $\Lambda _F^\mu $ which
allow to obtain the form factors
$F=P$, $A_{\pm }$ and $T$ from Eq. (\ref{projection}).
One possible choice is
\begin{eqnarray}
\Lambda _P^\mu &=&\frac 1{2\Delta ^2\delta ^2}\mathrm{i}\varepsilon
^{\mu \nu \alpha \beta }k_\nu p_{+\alpha }p_{-\beta }-\frac
m{4\Delta ^4}(k\cdot
q)(k\cdot \delta )[\gamma ^\mu (k\cdot q)-q^\mu \slashed{k}]\gamma_5
\nonumber\\
&&+\frac m{4\Delta ^4\delta ^2}[2\Delta ^2-q^2(k\cdot
\delta )^2][\gamma^\mu k\cdot \delta -\delta ^\mu \slashed{k}]\gamma_5
\nonumber\\
&&-\frac 1{4\Delta ^2\delta ^4}[2\Delta ^2-((k\cdot
q)^2-(k\cdot \delta
)^2)\delta ^2]\Delta ^\mu \gamma_5,
\label{projexpP}
\\
\Lambda _{A_{\pm }}^\mu &=&\frac m{2\Delta ^2\delta
^2}\mathrm{i}\varepsilon ^{\mu \nu \alpha \beta }k_\nu p_{+\alpha}p_{-\beta }
\mp \frac 1{16\Delta ^4}[2\Delta ^2+q^2(k\cdot (q\mp
\delta ))(k\cdot
\delta )][\gamma ^\mu (k\cdot q)-q^\mu \slashed{k}]\gamma_5
\nonumber\\
&&-\frac{q^2}{32\Delta ^4\delta ^2}[4\Delta
^2-2(k\cdot (q\mp \delta ))(k\cdot q)\delta ^2+m^{2\,}(k\cdot
(q\mp \delta ))^2][\gamma ^\mu k\cdot
\delta -\delta ^\mu \slashed{k}]\gamma_5
\nonumber\\
&&-\frac m{8\Delta ^4\delta ^2}[4\Delta ^2-2(k\cdot
(q\mp \delta ))(k\cdot
q)\delta ^2+m^{2\,}(k\cdot (q\mp \delta ))^2]\Delta ^\mu \gamma_5,
\label{projexpA}
\\
\Lambda _T^\mu &=&\frac 1{4\Delta ^2}\mathrm{i}\varepsilon ^{\mu
\nu \alpha \beta }k_\nu p_{+\alpha }p_{-\beta },
\label{projexpT}
\end{eqnarray}
where we have introduced the shorthand notation $q=p_+ +p_-$, $\Delta
=(k\cdot p_+)p_- -(k\cdot p_-)p_+$, and $\delta =p_+ - p_-$. In the
limit $m\rightarrow 0$ these expressions simplify to the form
\begin{eqnarray}
\Lambda _P^\mu &=&\frac 1{2\Delta ^2\delta ^2}\mathrm{i}
\varepsilon ^{\mu\nu \alpha \beta }k_\nu p_{+\alpha }p_{-\beta }-
\frac 1{2\Delta ^2\delta
^4}[\Delta ^2-2(k\cdot p_+)(k\cdot p_-)\delta ^2]\Delta ^\mu \gamma_5, \\
\Lambda _{A_{\pm }}^\mu &=&\mp \frac 1{16\Delta ^4}[2\Delta ^2+q^2(k\cdot
(q\mp \delta ))(k\cdot \delta )][\gamma ^\mu (k\cdot q)-q^\mu \slashed{k}]\gamma_5
\nonumber\\
&&-\frac{q^2}{32\Delta ^4\delta ^2}[4\Delta ^2-2(k\cdot (q\mp \delta
))(k\cdot q)\delta ^2][\gamma ^\mu k\cdot \delta -\delta ^\mu \slashed{k}]\gamma_5, \\
\Lambda _T^\mu &=&\frac 1{4\Delta ^2}\mathrm{i}\varepsilon ^{\mu \nu \alpha
\beta }k_\nu p_{+\alpha }p_{-\beta }.
\end{eqnarray}

\indent

\indent

\section{The pion-photon-photon vertex}\label{App:ppp}
\renewcommand{\theequation}{B.\arabic{equation}}
\setcounter{equation}{0}

As we have seen in the main text, the doubly off-shell
$\pi^0$-$\gamma ^{*}$-$\gamma ^{*}$ vertex, defined as
\[
\int d^4x\,e^{il\cdot x}\langle 0\vert
T\{j^\mu (x)j^\nu (0)\}\vert \pi ^0(P)\rangle
=-\mathrm{i}\varepsilon ^{\mu \nu \alpha \beta }l_\alpha
p_\beta \mathcal{A}_{\pi^0\gamma ^{*}\gamma ^{*}} (l^2,(P-l)^2),
\]
is a necessary ingredient for the calculation of the Dalitz decay
amplitude. While in the case of the one-photon reducible
contribution it is sufficient to use the corresponding form factor
$\mathcal{A}_{\pi^0\gamma ^{*}\gamma ^{*}} (0,l^2)$ for $l^2\lesssim
M_\pi ^2$, which is the region where ChPT (with virtual photons) is
applicable, for the leading one-fermion reducible and one-particle
irreducible contributions it is necessary to know
$\mathcal{A}_{\pi^0\gamma ^{*}\gamma ^{*}}(l^2,(P-l)^2)$ as a
function of the momentum $l$ in the full range of the loop
integration. In the present Appendix, we neglect temporarily the
electromagnetic interaction, so that $\mathcal{A}_{\pi^0\gamma
^{*}\gamma ^{*}}(l^2,(P-l)^2)$
 will refer to the \emph{strong} matrix element. We briefly
summarize some basic properties of the form factor
$\mathcal{A}_{\pi^0\gamma ^{*}\gamma ^{*}}(l^2,(P-l)^2)$
that are general consequences of QCD, as well as the results
of \cite{knecht01}, which we shall use in
the sequel. The low energy expansion of
$\mathcal{A}_{\pi^0\gamma ^{*}\gamma ^{*}}$
in the presence of both strong and electromagnetic
interactions will be the subject of the next Appendix.

General properties of the form factor $\mathcal{A}_{\pi^0\gamma
^{*}\gamma ^{*}}(l^2,(P-l)^2)$ in QCD were investigated earlier
within various approaches \cite{pi2gamma, BrodskyLepage}. In the
chiral limit, the on-shell value of the form factor is entirely
fixed by the QCD chiral anomaly.
Therefore, within ChPT the low energy behaviour is expected to be
\begin{equation}
\mathcal{A}_{\pi^0\gamma ^{*}\gamma ^{*}}(l^2,(P-l)^2)=
-\frac{N_C}{12\pi ^2F_\pi }\left[
1+{\cal O}\left(\frac{P\cdot l}{\Lambda_H^2},\frac{l^2}{\Lambda_H^2},
\frac{m_q}{\Lambda_H}\right)
\right],
\label{CHPT}
\end{equation}
where the higher order corrections come from pseudo-Goldstone
boson loops, as well as from higher order contact terms.
In particular,
\begin{equation}
\mathcal{A}_{\pi^0\gamma ^{*}\gamma ^{*}}(0,0)=
-\frac{N_C}{12\pi ^2F_\pi }\left[
1+{\cal O}\left(
\frac{m_q}{\Lambda_H}\right)
\right].
\end{equation}
Another
exact result, the leading short distance asymptotics for $l\rightarrow \infty
$, ($P$ fixed), follows from the operator product expansion (see~\cite{novikov}).
For $\lambda \rightarrow \infty $ we have
\begin{equation}
\mathcal{A}_{\pi^0\gamma ^{*}\gamma ^{*}}((\lambda l)^2,(P-\lambda l)^2)
=\frac 1{(\lambda l)^2}\frac
23F_\pi \left[ 1\,+\, \frac{1}{\lambda}\,
\frac{P\cdot l}{l^2}\,+\,\cdots
\right]  .
\label{OPE}
\end{equation}
The ellipsis stands for higher order terms in the short-distance expansion,
or for ${\cal O}(\alpha_s)$ QCD corrections to the terms that are shown.
Notice that the latter are not affected by quark mass effects, so that
Eq.~(\ref{OPE}) holds beyond the chiral limit. On the other hand, the
expression (\ref{OPE}) assumes isospin and CP invariance of the strong
interactions. The explicit form of
$\mathcal{A}_{\pi^0\gamma ^{*}\gamma ^{*}}(l^2,(P-l)^2)$
 in the intermediate energy range, however, is not known from first principles.
Among the various approaches that have been considered, models
inspired by the large-$N_C$ properties of QCD have been proven
particularly useful in order to provide parameterizations of the
form factor $\mathcal{A}_{\pi^0\gamma ^{*}\gamma ^{*}}(l^2,(P-l)^2)$
compatible with the above low and high energy behaviours predicted
by QCD. Let us give here a brief overview of the results obtained in
Ref.  \cite{knecht01} within this framework. At leading order in the
$1/N_C$ expansion, $\mathcal{A}_{\pi^0\gamma ^{*}\gamma
^{*}}(l^2,(P-l)^2)$ can be expressed as infinite sum of the
tree-level exchanges of the zero-width resonances in the various
channels. Truncating this infinite sum and keeping only the
contribution of the lowest resonances, {i.e.} the lowest vector
meson octet in the present case, we obtain the Lowest Meson
Dominance approximation ({\it LMD\/}) to the large-$N_C$
expression~\cite{knechtcpt}
\begin{equation}
\mathcal{A}_{\pi^0\gamma ^{*}\gamma ^{*}}^{LMD}(l^2,(P-l)^2)
=\frac{F_\pi
}3\frac{l^2+(P-l)^2+\kappa_V}{(l^2-M_V^2)((P-l)^2-M_V^2)}.
\label{LMD}
\end{equation}
This Ansatz satisfies all the properties of
$\mathcal{A}_{\pi^0\gamma ^{*}\gamma ^{*}}$
discussed so far, provided the constant
$\kappa_V$ is chosen such as to provide compatibility with (\ref{CHPT}),
\begin{equation}
\kappa_V = \frac{3M_V^4}{F_\pi}\,
\mathcal{A}_{\pi^0\gamma ^{*}\gamma ^{*}}(0,0) =
-\,\frac{N_C}{4\pi ^2F_\pi ^2}M_V^4 (1 + {\cal C}_1 + \cdots).
\end{equation}
The second equality involves the leading quark mass corrections
described by the combination of low energy constants given in
Eqs.~(\ref{calC12}) and (\ref{calC2r}) below, while the ellipsis
stands for higher order quark mass corrections, that will not be
considered here. Let us note that if the large-$N_C$ vector meson
mass is identified with the physical mass of the $\rho$ meson,
$M_V=M_\rho$, the form factor $\mathcal{A}_{\pi^0\gamma ^{*}\gamma
^{*}}^{LMD}(l^2,(P-l)^2)$ contains ${\cal C}_1$ as the only free
parameter, and interpolates smoothly between (\ref{CHPT}) and
(\ref{OPE}). On the other hand, at low energy, the non-analytical
contributions from Goldstone boson intermediate states are not taken
into account (note that according to the large-$N_C$ counting rules,
meson loops are suppressed in the $1/N_C$ expansion). As further
discussed in \cite{knecht01}, the simple Ansatz (\ref{LMD}) is not
sufficient to describe the full asymptotic behaviour for
$Q^2\rightarrow \infty $, where $Q^2=-(q_1^2+q_2^2)$ with fixed
$\omega =(q_1^2-q_2^2)/(q_1^2+q_2^2)=\pm 1$, given by the general
formula \cite{BrodskyLepage}
\begin{equation}
\mathcal{A}_{\pi^0\gamma ^{*}\gamma ^{*}}(q_1^2,q_2^2)=
-\,\frac{4F_\pi }3\,\frac{f(\omega
)}{Q^2}+\mathcal{O}\left( \frac 1{Q^4}\right) \,,  \label{OPE
general}
\end{equation}
with a function $f(\omega )$ that is not known explicitly. In order
to reconcile the large $N_C$ ansatz with (\ref{OPE general}), at
least one additional vector resonance is unavoidable. We thus
obtain, in the notation of \cite{knecht01}, the more general Ansatz
\begin{equation}
\mathcal{A}_{\pi^0\gamma ^{*}\gamma ^{*}}^{LMD+V}(q_1^2,q_2^2)=\frac{F_\pi
}3\frac{q_1^2q_2^2(q_1^2+q_2^2)+\kappa_1(q_1^2+q_2^2)^2+
\kappa_2q_1^2q_2^2+\kappa_5(q_1^2+q_2^2)+\kappa_7}
{(q_1^2-M_{V_1}^2)(q_1^2-M_{V_2}^2)(q_2^2-M_{V_1}^2)(q_2^2-M_{V_2}^2)}\,.
\label{fF}
\end{equation}
The chiral anomaly  fixes now
\begin{equation}
\kappa_7=\frac{3M_{V_1}^4 M_{V_2}^4}{F_\pi}\,
\mathcal{A}_{\pi^0\gamma ^{*}\gamma ^{*}}(0,0) =
-\,\frac{N_C}{4\pi ^2}\frac{M_{V_1}^4M_{V_2}^4}{F_\pi ^2}
(1 + {\cal C}_1 + \cdots),
\end{equation}
while the large $Q^2$ behaviour of
$\mathcal{A}_{\pi^0\gamma ^{*}\gamma ^{*}}(Q^2,0)$
 requires $\kappa_1=0$.
From experimental data one can also determine
\begin{equation}
\kappa_5=6.93\pm 0.26\,\mathrm{GeV}^4,  \label{h5}
\end{equation}
(one takes $M_{V_1}=769\,\mathrm{MeV}$,
$M_{V_2}=1465\,\mathrm{MeV}$, $F_\pi =92.4\,\mathrm{MeV,}$ further
details can be found in \cite{knecht01}). Finally, as pointed out
in Ref. \cite{Melnikov03}, the coefficient $\kappa_2$ is also available
from Ref.~\cite{novikov},
\begin{equation}
\kappa_2  \sim -4(M_{V_1}^2 + M_{V_2}^2) = -10 \,\mathrm{GeV}^2
,
\end{equation}
a value which lies within the range considered in
Ref. \cite{knecht01}.

\indent

\section{Chiral Expansion of the pion-photon-photon
vertex}\label{App:cepppv}
\renewcommand{\theequation}{C.\arabic{equation}}
\setcounter{equation}{0}

In this Appendix, we first summarize the results of our
recalculation of the pure QCD form factor $\mathcal{A}_{\pi^0\gamma
^{*}\gamma ^{*}}(0, l^2)$ in two-flavour Chiral Perturbation Theory
\cite{gasser, urech, kaiser00, kn} up to one loop, i.e. up to the
order $\mathcal{O}(p^6)$. After that, we describe the additional
modifications that appear if electromagnetic effects are also
included.

\subsection{${\mbox{$e=0$}}$}

The relevant chiral Lagrangian can be
written in the form
\[
\mathcal{L=L}^{(2)}+\mathcal{L}^{(4)}+\mathcal{L}_{WZW}^{(4)}+\mathcal{L}^{(6)}+\ldots,
\]
where the terms with even intrinsic parity at order $O(p^2)$ and
$O(p^4)$ are
\begin{eqnarray}
\mathcal{L}^{(2)} &=&\frac{F_0^2}4\langle D^\mu U^{+}D_\mu U+\chi
^{+}U+U^{+}\chi \rangle  \label{L_2} \\
\mathcal{L}^{(4)} &=&\frac{l_1}4\langle D^\mu U^{+}D_\mu U\rangle
^2+\frac{l_2}4\langle D^\mu U^{+}D^\nu U\rangle \langle D_\mu
U^{+}D_\nu U\rangle
\nonumber \\
&&+\frac{l_3}{16}\langle \chi ^{+}U+U^{+}\chi \rangle ^2+\frac{l_4}4\langle
D_\mu UD^\mu \chi ^{+}+D^\mu U^{+}D_\mu \chi \rangle  \nonumber \\
&&+l_5\langle \widehat{R}_{\mu \nu }U\widehat{L}^{\mu \nu
}U^{+}\rangle + \mathrm{i}\frac{l_6}2(\langle \widehat{R}_{\mu \nu
}D^\mu UD^\nu U^{+}\rangle +\langle \widehat{L}_{\mu \nu }D^\mu
U^{+}D^\nu U\rangle )
\nonumber \\
&&-\frac{l_7}{16}\langle \chi ^{+}U-U^{+}\chi \rangle ^2  \nonumber \\
&&+\frac 14(h_1+h_3)\langle \chi ^{+}\chi \rangle +\frac 14(h_1-h_3)(\det
\chi ^{+}+\det \chi )  \nonumber \\
&&-\frac 12(l_5+4h_2)\langle \widehat{R}_{\mu \nu
}\widehat{R}^{\mu \nu }+ \widehat{L}_{\mu \nu }\widehat{L}^{\mu
\nu }\rangle +\frac{h_4}4\langle R_{\mu \nu }+L_{\mu \nu }\rangle
\langle R^{\mu \nu }+L^{\mu \nu }\rangle
\nonumber \\
&&+\frac{h_5}4\langle R_{\mu \nu }-L_{\mu \nu }\rangle \langle R^{\mu \nu
}-L^{\mu \nu }\rangle ).  \label{L_4}
\end{eqnarray}
The odd intrinsic parity Wess-Zumino-Witten Lagrangian, which
accounts for the two-flavour anomaly can be written in the form
\cite{kaiser00}
\begin{eqnarray}
\mathcal{L}_{WZW}^{(4)} &=&-\frac{N_C}{32\pi ^2}\varepsilon ^{\mu
\nu \rho \sigma }[\langle U^{+}\widehat{r}_\mu U\widehat{l}_\nu
-\widehat{r}_\mu \widehat{l}_\nu +\mathrm{i}\Sigma _\mu
(U^{+}\widehat{r}_\nu U+\widehat{l}_\nu )\rangle \langle v_{\rho \sigma }\rangle
\nonumber \\
&&+\frac 23\langle \Sigma _\mu \Sigma _\nu \Sigma _\rho \rangle \langle
v_\sigma \rangle ].  \label{WZW}
\end{eqnarray}
In the above formulae, the notation is as follows:
\begin{eqnarray}
U &=&\mathrm{e}^{\frac{\mathrm{i}\phi }{F_0}},\quad \phi =\left(
\begin{array}{cc}
\pi ^0 & \sqrt{2}\pi ^{+} \\
\sqrt{2}\pi ^{-} & -\pi ^0
\end{array}
\right) , \\
D_\mu U &=&\partial _\mu U-\mathrm{i}r_\mu U+\mathrm{i}Ul_\mu
,\,\,\,\,\Sigma _\mu =U^{+}\partial _\mu U, \\
R_{\mu \nu } &=&\partial _\mu r_\nu -\partial _\nu r_\mu
-\mathrm{i}[r_\mu ,r_\nu ],\,\,\,L_{\mu \nu }=\partial _\mu l_\nu
-\partial _\nu l_\mu -
\mathrm{i}[l_\mu ,l_\nu ], \\
\widehat{R}_{\mu \nu } &=&R_{\mu \nu }-\frac 12\langle R_{\mu \nu }\rangle
,\,\,\,\,\widehat{L}_{\mu \nu }=L_{\mu \nu }-\frac 12\langle L_{\mu \nu
}\rangle, \\
\widehat{r}_\mu &=&r_\mu -\frac 12\langle r_\mu \rangle
,\,\,\,\widehat{l}
_\mu =l_\mu -\frac 12\langle l_\mu \rangle , \\
v_\mu &=&\frac 12(r_\mu +l_\mu ),\,\,\,v_{\mu \nu }=\partial _\mu v_\nu
-\partial _\nu v_\mu -\mathrm{i}[v_\mu ,v_\nu ].
\end{eqnarray}
Further relevant chiral invariant Lagrangians of order
$\mathcal{O}(p^6)$ and also the other details can be found in
 \cite{fearing, bijnens} and references therein.

\begin{figure}[htb]
\center\scalebox{1.3}{\epsfig{figure=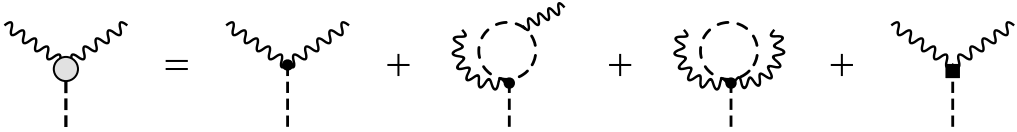}}
\caption{Next-to-leading order corrections to $\pi \gamma \gamma $
vertex.} \label{figure10}
\end{figure}

\begin{figure}[htb]
\center\scalebox{1.1}{\epsfig{figure=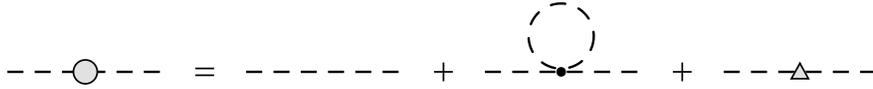}} \caption{$\pi
^0$ wave function renormalization in next-to-leading order.}
\label{figure11}
\end{figure}

The form factor $\mathcal{A}_{\pi^0\gamma ^{*}\gamma ^{*}}
(0,l^2)$ starts at the order $\mathcal{O}(p^4)$ with the tree
graph with vertex derived from the Wess-Zumino-Witten Lagrangian
(\ref{WZW}), and reproduces the anomaly result (\ref{CHPT}),
\begin{equation}
\mathcal{A}_{\pi^0\gamma ^{*}\gamma ^{*}}^{LO}(0, l^2)=
-\,\frac{N_C}{12\pi ^2F_0}, \label{AL}
\end{equation}
since $F_0$ can be identified with $F_\pi$ at this order. At the
next-to-leading order, there are two types of one-loop contributions
with one vertex from $\mathcal{L}_{WZW}^{(4)}$, namely the tadpole
and the bubble graphs (see Fig.~\ref{figure10}).
Another type of contributions correspond to the contact terms
derived from the tree graphs with one vertex from the odd intrinsic
parity part of $\mathcal{L}^{(6)}$. A last contribution comes from
the renormalization factor of the external pion leg; this one
contains the tadpole with vertex from $\mathcal{L}^{(2)}$ and
contact terms with vertices from $\mathcal{L}^{(4)}$, see
Fig.~\ref{figure11}.
Putting all these parts together we obtain the following result
\begin{equation}
\mathcal{A}_{\pi^0\gamma ^{*}\gamma ^{*}}(0,l^2)=-\frac{N_C}{12\pi ^2F_\pi
}\Bigl\{1 +{\cal C}_1 - \frac{l^2}{6M_\pi ^2}\Bigl[\left( \frac{M_\pi
}{4\pi F_\pi }\right) ^2\left( \tfrac 13-16\pi ^2\sigma _\pi
^2(l^2){\bar J}_\pi (l^2)\right) - {\cal C}_2\Bigr]\Bigr\},  \label{A_QCD}
\end{equation}
where $\sigma _P(s)\equiv \sqrt{1-4 M_P^2/s}$ and
\begin{equation}
{\bar J}_P(s)=\frac s{16\pi ^2}\int_{4M_P^2}^\infty
\frac{\mathrm{d}x}x \frac{\sigma _P(x)}{x-s}
\end{equation}
is the Chew-Mandelstam function (the scalar bubble subtracted at
$s=0$). In the above formula, we keep the neutral and charged pion
masses equal (within pure QCD their difference is an effect of
second order in the isospin breaking parameter $(m_u-m_d)$), the
isospin breaking QED corrections, which are, taking $e=O(p)$, of
the same order as the leading order terms, will be taken into account in the
next section where further details can be found. Finally,
${\cal C}_1$ and ${\cal C}_2$ represent the following
renormalization-scale independent combinations of the $O(p^6)$ low
energy constants $A_i$ and $c_i$ introduced in \cite{fearing} and
\cite{bijnens} respectively:
\begin{eqnarray}
{\cal C}_1 &=&
\frac{32}3\pi
^2\Bigl[(A_2^r-2A_3^r-4A_4^r)M_\pi ^2+
\frac{20}3(A_4^r+2A_6^r)2\bar{m}B\Bigr] \nonumber\\
&=&\frac{64}3\pi ^2\Bigl[(c_{11}^r-4c_3^r-4c_7^r)M_\pi ^2+\frac
43(5c_3^r+c_7^r+2c_8^r)2\bar{m}B\Bigr], \nonumber\\
{\cal C}_2 &=& {\cal C}_2^r(\mu) -\left( \frac{M_\pi }{4\pi
F_\pi }
\right) ^2\ln \frac{M_\pi ^2}{\mu ^2},
\label{calC12}
\end{eqnarray}
with $\bar{m}=(m_d-m_u)/2$ and
\begin{equation}
{\cal C}_2^r (\mu)\,=\,-64\pi ^2M_\pi ^2(A_2^r-4A_3^r)=-64\pi ^2M_\pi
^2c_{13}^r.
\label{calC2r}
\end{equation}
The renormalization of the external
pion line is responsible for the replacement of the constant $F_0$ with
the physical decay constant $F_\pi $ in the leading order term of
(\ref{A_QCD}).

The actual values of the constants ${\cal C}_1$ and ${\cal C}_2$
are not known from first principles. Recently the relevant
combinations of the low energy constants that occur in ${\cal
C}_2$ have been estimated in \cite{knecht01} by using the matching
of the {\it LMD\/} approximation to the large $N_C$ form factor
$\mathcal{A}_{LMD}(l^2,0)$ and the large $N_C$ approximation to
the ChPT result $\mathcal{A}^{ChPT}(l^2,0)$. Since in the
large-$N_C$ limit the contribution of meson loops is suppressed,
the chiral logarithms as well as the running of the renormalized
couplings with the renormalization scale $\mu $ are
next-to-leading order effects. Following \cite{EGPdR}, we assume
that the values of the low energy constants obtained this way
correspond to scale given by the mass scale of the non-Goldstone
resonances $\mu \sim M_V$. We thus have the following {\it LMD\/}
determination of the low energy constant ${\cal C}_2$
\cite{knecht01}:
\begin{equation}
{\cal C}_2^r(M_V)_{LMD}  = 6\Bigl[ 1 + {\cal C}_1 -\frac
1{4N_C}\Bigl( \frac{4\pi F_\pi }{M_V}\Bigr) ^2 \Bigr] \Bigl(
\frac{M_\pi }{M_V}\Bigr) ^2.
\end{equation}
The same procedure can be done with the $LMD+V$ approximation; in
this case we find \cite{knecht01}
\begin{multline}
{\cal C}_2^r(M_V)_{LMD+V} = 6\Bigl( \frac{M_\pi }{4\pi F_\pi
}\Bigr) ^2\Bigl[ (1+{\cal C}_1)\Bigl(\Bigl( \frac{4\pi F_\pi
}{M_{V_1}}\Bigr) ^2+\Bigl( \frac{4\pi F_\pi }{M_{V_2}}\Bigr)
^2\Bigr)
\\-\frac 1{4N_C}\Bigl( \frac{4\pi F_\pi }{M_{V_1}} \Bigr) ^2\Bigl(
\frac{4\pi
F_\pi}{M_{V_2}}\Bigr)^2\frac{\kappa_5}{M_{V_1}^2M_{V_2}^2}\Bigr].
\end{multline}
The issue of the quark mass corrections to
$\mathcal{A}_{\pi^0\gamma ^{*}\gamma ^{*}}(0,l^2)$,
contained in ${\cal C}_1$,
have been addressed in \cite{Moussallam95}, and more
recently in \cite{A-M2}. From \cite{A-M2}, one infers
\begin{equation}
{\cal C}_1 = \frac{m_d - m_u}{m_s - {\widehat m}}
\,(0.93\pm 0.12)\pm 0.14\cdot 10^{-2} ,
\end{equation}
with ${\widehat m}=(m_u+m_d)/2$.

Numerically, with $(m_d - m_u)/(m_s - {\widehat m})=1/43$, $M_\pi
=135\,\mathrm{MeV}$, $F_\pi =92.4\,\mathrm{MeV}$,
$M_V=M_{V_1}=770\,\mathrm{MeV}$, $M_{V_2}=1465\,\mathrm{MeV}$ and
$\kappa_5$ given by (\ref{h5}), we then have the following
determinations
\begin{eqnarray*}
{\cal C}_1 &=& (2.2\pm 0.3)\times 10^{-2}, \\
{\cal C}_2^r(M_V)_{LMD} &=&(1.5\pm 0.5)\times 10^{-1}, \\
{\cal C}_2^r(M_V)_{LMD+V} &=&(1.8\pm 0.6)\times 10^{-1}.
\end{eqnarray*}
A 30\% uncertainty, typical for a result based on a leading order
large-$N_C$ calculation, has been assigned to ${\cal C}_2$. Within
these error bars, the $LMD$ result is stable with respect to the
inclusion of a second resonance. Notice also that a variation of
the scale between, say, $M_V=M_{V_1}$ and $M_V=M_{V_2}$ gives
${\cal C}_2^r(M_{V_1}) - {\cal C}_2^r(M_{V_2}) = 0.02$, which is
well within these error bars.

\subsection{${\mbox{$e\neq 0$}}$}

In this section we shall describe the results of our calculation of the
next-to-leading $O(p^6)$ corrections to the leading order
amplitude in the expansion scheme in which the electric charge,
fermion masses, and fermion bilinears are assumed to be counted as
quantities of order $p$. Within this scheme, the $O(p^2)$
Lagrangian reads
\begin{eqnarray*}
\mathcal{L}^{(2)} &=&\frac{F_0^2}4\langle D^\mu U^{+}D_\mu U+\chi
^{+}U+U^{+}\chi \rangle +e^2ZF_0^4\langle QUQU^{+}\rangle \\
&&-\frac 14F_{\mu \nu }F^{\mu \nu }+\overline{\psi }(\mathrm{i}\gamma \cdot
(\partial -\mathrm{i}eA)-m_e)\psi ,
\end{eqnarray*}
where
\[
Q=\mathrm{diag}(2/3,-1/3)
\]
is the quark charge matrix. Then the leading order amplitude,
which corresponds to the tree graph with one vertex from
$\mathcal{L}_{WZW}^{(4)}$, is of the order $O(p^4)$. Let us also
note that electromagnetic splitting of the charged and neutral
pion masses is treated as a leading order effect.
The $O(p^4)$ Lagrangian with even intrinsic parity then reads
\[
\mathcal{L}^{(4)}=\mathcal{L}_{p^4}^{(4)}+\mathcal{L}_{e^2p^2}^{(4)}+
\mathcal{L}_{lept}^{(4)}.
\]
The $\mathcal{L}_{p^4}^{(4)}$ is the same as (\ref{L_4}) while the
explicit form of $\mathcal{L}_{e^2p^2}^{(4)}$ and
$\mathcal{L}_{lept}^{(4)}$ can be found in \cite{urech} and
\cite{neufeld}. In the following we need only
\begin{eqnarray*}
\mathcal{L}_{e^2p^2}^{(4)} &=&F_0^2\left\{ k_1\langle D^\mu U^{+}D_\mu
U\rangle \langle Q^2\rangle +k_2\langle D^\mu U^{+}D_\mu U\rangle \langle
QUQU^{+}\rangle \right. \\
&&+k_3\left( \langle D^\mu U^{+}QU\rangle \langle D_\mu U^{+}QU\rangle
+\langle D^\mu UQU^{+}\rangle \langle D_\mu UQU^{+}\rangle \right) \\
&&\left. +k_4\langle D^\mu U^{+}QU\rangle \langle D_\mu UQU^{+}\rangle
+\ldots \right\}, \\
\mathcal{L}_{lept}^{(4)} &=&e^2x_6\overline{\psi }(\mathrm{i}\gamma \cdot
(\partial -\mathrm{i}eA))\psi +e^2x_7m_e\overline{\psi }\psi +e^2x_8F_{\mu
\nu }F^{\mu \nu }+\ldots
\end{eqnarray*}

The NLO contributions within the pure QCD were presented in the
previous section. In the enlarged case there are two main
distinctions: First, because the pion mass difference is of order
$O(p^2)$ now, we have to take care of the unequal pion masses in
the loops. Second, at order $O(p^4)$, the pion decay constants
$F_{\pi ^0}$ and $F_{\pi ^{\pm }}$ are different as a consequence
of the new type of $O(p^4)$ terms coming from
$\mathcal{L}_{e^2p^2}^{(4)}$, as well as of unequal tadpole
contributions. The mass difference leads to additional terms of
the form $\ln \left( M_{\pi ^{\pm }}^2/M_{\pi ^0}^2\right) $, the
latter to the replacement of the $F_0$ with $F_{\pi ^0}$ (and not
with $F_{\pi ^{\pm }}$) in the leading term as a result of the
renormalization of the external pion line. Taking all these effects into
account leads to
\begin{multline}
\mathcal{A}^{ChPT}(l^2,0) =-\frac{N_C}{12\pi ^2F_{\pi ^0}}\\
\times \Bigl\{ 1+ {\cal C}_1 -\frac{l^2}{6M_{\pi ^{\pm
}}^2}\Bigl[\frac{M_{\pi ^{\pm }}^2}{16\pi ^2F_\pi ^2}\Bigl( \tfrac
13+\ln \frac{M_{\pi ^{\pm }}^2}{M_{\pi ^0}^2}-16\pi ^2\sigma _{\pi
^{+}}^2(l^2){\bar J}_{\pi ^{+}}(l^2)\Bigr) - {\cal C}_2
\Bigr]\Bigr\} ,
\end{multline}
with ${\cal C}_1$ and ${\cal C}_2$ now given by
\begin{eqnarray*}
{\cal C}_1 &=&\frac{32}3\pi ^2\Bigl[(A_2^r-2A_3^r-4A_4^r)M_{\pi ^0}^2+
\frac{20}3(A_4^r+2A_6^r)2\bar{m}B\Bigr] \\
&=&\frac{64}3\pi ^2\Bigl[(c_{11}^r-4c_3^r-4c_7^r)M_{\pi ^0}^2+\frac
43(5c_3^r+c_7^r+2c_8^r)2\bar{m}B\Bigr], \\
{\cal C}_2 &=&-64\pi ^2M_{\pi ^{\pm }}^2(A_2^r-4A_3^r)-
\frac{M_{\pi ^{\pm }}^2}{16\pi ^2F_\pi ^2}\ln \frac{M_{\pi ^0}^2}{\mu ^2} \\
&=&-64\pi ^2M_{\pi ^{\pm }}^2c_{13}^r-\frac{M_{\pi ^{\pm }}^2}{16\pi ^2F_\pi
^2}\ln \frac{M_{\pi ^0}^2}{\mu ^2}.
\end{eqnarray*}
Because the constant $F_{\pi ^0}$ is not known very accurately, we
use the following relation \cite{knecht98,meissner97}
\begin{equation}
F_{\pi ^0}=F_\pi \Bigl(1-\frac{M_{\pi ^{\pm }}^2}{16\pi ^2F_\pi
^2}\ln \frac{M_{\pi ^{\pm }}^2}{M_{\pi ^0}^2}-\frac{e^2}{64\pi
^2}\,{\cal K}_F
\Bigr) \label{Fp0}
\end{equation}
with
\begin{equation}
{\cal K}_F =
 (3+\tfrac 49Z)
\bar{k}_1-\tfrac{40}9Z\bar{k}_2-3\bar{k}_3-4Z\bar{k}_4
\end{equation}
to eliminate $F_{\pi ^0}$ in favour of $F_\pi$, where $F_\pi =
F_{\pi^\pm}|_{e=0}$ is measured in the charged pion decays,
see~\cite{A-M2}. We thus write
\[
\mathcal{A}^{ChPT}(l^2,0)=-\frac{N_C}{12\pi ^2F_\pi }\left(
1+a_{NLO}^{ChPT}(l^2)\right),
\]
where
\begin{eqnarray}
a_{NLO}^{ChPT}(l^2) &=& {\cal C}_1 +
\frac{e^2}{64\pi^2}\,{\cal K}_F
+\frac{M_{\pi ^{\pm }}^2}{16\pi ^2F_\pi
^2}\ln \frac{M_{\pi ^{\pm }}^2}{M_{\pi ^0}^2}
\nonumber\\
&&\quad
-\frac{l^2}{6M_{\pi ^{\pm
}}^2} \Bigl[\frac{M_{\pi ^{\pm }}^2}{16\pi ^2F_\pi ^2}\Bigl(
\tfrac 13+\ln \frac{ M_{\pi ^{\pm }}^2}{M_{\pi ^0}^2}-16\pi
^2\sigma _{\pi ^{+}}^2(l^2){\bar J }_{\pi ^{+}}(l^2)\Bigr)
-{\cal C}_2\Bigr].\label{achptnlo}
\end{eqnarray}
The $\bar{k}_i$, $i=1,\ldots ,4$ are the a priori unknown
scale-independent constants, defined in terms of the bare
low-energy constants from
$\mathcal{L}_{e^2p^2}^{(4)}$ according to the formulae
\begin{eqnarray*}
k_i &=&k_i^r(\mu )+\frac{\sigma _i}{(4\pi )^2}\left( \frac 1{d-4}-\frac
12(\ln 4\pi -\gamma +1)\right) , \\
k_i^r(\mu ) &=&\frac{\sigma _i}{2(4\pi )^2}\left( \bar{k}_i+\ln \frac{M_{\pi
^0}^2}{\mu ^2}\right) ,
\end{eqnarray*}
where
\begin{equation}
\sigma _1 =-\frac{27}{20}-\frac 15Z,
\ \sigma _2 =\sigma _4=2Z,
\ \sigma _3 =-\frac 34.
\end{equation}

In order to obtain a numerical evaluation of ${\cal K}_F$
we further express it in terms of the analogous $SU(3)$ constants
$K_i$. Several determinations of the latter are available in the
literature \cite{QEDLEC}. In the most recent work
\cite{A-M}, a new estimate of these parameters based on sum
rules involving QCD 4-point correlators (for $SU(3)$ case,
parametrized with help of the improved chiral Lagrangian with
resonances in the spirit of large-$N_C$ approximation) was made.
In order to use these results,
we first have to match the $SU(3)$ variant of the theory with the
$SU(2)$ one we have used in our calculation. This can be made as
follows.
Starting from the fact that ${\cal K}_F$ enters the formula
(\ref{Fp0}) expressing the electromagnetic difference between $F_\pi
\equiv F_{\pi ^{\pm }}|_{e=0}$ and $F_{\pi ^0}$, we can write (in
the $SU(2)$ case)
\begin{equation}
\frac{e^2}{64 \pi^2}{\cal K}_F=1-\frac{F_{\pi ^0}}{F_\pi
}-\frac{M_{\pi ^{\pm }}^2}{16\pi ^2F_\pi ^2}\ln \frac{M_{\pi ^{\pm
}}^2}{M_{\pi ^0}^2}. \label{daprime}
\end{equation}
The ratio $F_{\pi ^0}/F_\pi $ can be calculated within the $SU(3)$
version with the result\footnote{The known experimental value of
$F_{\pi^0} = 92 \pm 4$ MeV has an uncertainty too large  to provide
a useful determination of ${\cal K}_F$.} \cite{ruperts}
\begin{multline*}
\frac{F_{\pi ^0}}{F_\pi }\Bigr|_{SU(3)}=1-\frac{M_{\pi ^{\pm
}}^2}{16\pi ^2F_\pi ^2}\ln \frac{M_{\pi ^{\pm }}^2}{M_{\pi
^0}^2}-\frac{e^2Z}{32\pi ^2} \Bigl( 4\ln \Bigl( \frac{M_{\pi
^0}^2}{\mu ^2}\Bigr) +\ln \Bigl( \frac{
M_{K^{+}}^2}{\mu ^2}\Bigr) +1\Bigr)  \\
\!\!\!\!\!\!\!\!\!\!
+e^2\Bigl( \frac 43K_1^r+\frac
43K_2^r-2K_3^r+K_4^r+\frac{10}9K_5^r+\frac{10}9K_6^r\Bigr);
\end{multline*}
therefore, upon matching the two expressions, it follows
that\footnote{Let us note, that the $SU(3)$ on-shell
$\pi\gamma\gamma$ amplitude \cite{A-M2} contains (besides the
electromagnetic  difference between $F_\pi$ and $F_{\pi^0}$) an
additional ${\cal O}(e^2)$ contribution which originates in the
electromagnetic correction to the $\pi\eta$ mixing. Within the
$SU(2)$ power counting it is in fact of the order ${\cal
O}(e^2p^2)$, so that it need not to be included in the matching
procedure.}
\begin{align}
&\frac{e^2}{64 \pi^2}{\cal K}_F =1-\frac{F_{\pi ^0}}{F_\pi
}\Bigl|_{SU(3)}-\frac{M_{\pi ^{\pm }}^2}{16\pi ^2F_\pi ^2}\ln
\frac{M_{\pi ^{\pm }}^2}{M_{\pi ^0}^2}=  \label{delta_prime_a} \\
&=\frac{e^2Z}{32\pi ^2}\Bigl( 4\ln \Bigl( \frac{M_{\pi ^0}^2}{\mu
^2} \Bigr) +\ln \Bigl( \frac{M_{K^{+}}^2}{\mu ^2}\Bigr) +1\Bigr)
-e^2\Bigl( \frac 43K_1^r+\frac
43K_2^r-2K_3^r+K_4^r+\frac{10}9K_5^r+\frac{10}9K_6^r\Bigr) .\notag
\end{align}
Inserting into this expression the values
$K_1^r=-2.71\times 10^{-3}$, $K_2^r=0.69\times 10^{-3}$,
$K_3^r=2.71\times 10^{-3}$, $K_4^r=1.38\times 10^{-3}$,
$K_5^r=11.59\times 10^{-3}$ and $K_6^r=2.77\times 10^{-3}$ at
a scale $\mu =770$ MeV, obtained in \cite{A-M}
from the lowest meson dominance approximation to the large-$N_C$
limit of appropriate QCD correlators
 we find
\[
{\cal K}_F =-28 \pm 8\,.
\]
Again, we have assigned to this value an uncertainty of 30 \%,
typical for calculations based on the leading order in the
large-$N_C$ expansion. Although ${\cal K}_F$ is scale independent,
the estimates of the low energy constants $K_i^r(\mu)$ it involves
depend on the scale at which they are identified with the
resonance approximation. Varying again this scale between the
values $M_V=M_{V_1}$ and $M_V=M_{V_2}$ induces a variation in
$K_F^r$ which corresponds to these same error bars.

\indent

\section{NLO corrections to ${\overline F}_1$, ${\overline F}_2$,
${\overline \Pi}$}\label{App:NLO}
\renewcommand{\theequation}{D.\arabic{equation}}
\setcounter{equation}{0}

\subsection{Corrections to the vacuum polarization function}

The vacuum polarization function $\Pi (l^2)$ starts at $O(p^2)$
with three types of contributions,
\[
\Pi (l^2)=\Pi _{\pi ^{\pm }}(l^2)+\Pi _{e^{\pm }}(l^2)+\Pi _{CT}(l^2),
\]
where the first two correspond to the pion bubble and tadpole, and
to the fermion bubble, respectively, and the third one is a
contact term from $\mathcal{L}^{(4)}$, which is necessary to
renormalize the UV divergences. In dimensional regularization one
has
\begin{align}
\Pi _{\pi ^{\pm }}(s)&=\overline{\Pi }_{\pi ^{\pm }}(s)-\frac
\alpha {12\pi }\Bigl( \ln \frac{M_{\pi ^{\pm }}^2}{\mu ^2}+1\Bigr)
-\frac \alpha {6\pi }\Bigl[ \frac 1{d-4}-\frac 12(\ln 4\pi -\gamma
+1)\Bigr] \notag\\
\Pi _{e^{\pm }}(s)&=\overline{\Pi }_{e^{\pm }}(s)-\frac \alpha
{3\pi }\Bigl( \ln \frac{m^2}{\mu ^2}+1\Bigr) -\frac{2\alpha }{3\pi
}\Bigl[ \frac 1{d-4}-\frac 12(\ln 4\pi -\gamma +1)\Bigr],
\label{pp2}
\end{align}
where $\overline{\Pi}$ are the corresponding quantities in the
on-shell renormalization scheme in which the finite part of the
counterterms is unambiguously fixed by the condition
$\overline{\Pi }(0)=0$. We have then the standard result
\begin{equation}
\overline{\Pi }_{\pi ^{\pm }}(s)=\frac \alpha \pi \frac
s{12}\int_{4M_{\pi ^{\pm }}^2}^\infty
\frac{\mathrm{d}x}x\frac{\sigma _{\pi ^{\pm }}^3(x)}{x-s} =\frac
\alpha {18\pi }\Bigl[1+24\pi ^2\sigma _{\pi ^{\pm
}}^2(s){\bar J}_{\pi ^{\pm }}(s)\Bigl]\label{Pbar}
\end{equation}
and
\begin{equation}
\overline{\Pi }_{e^{\pm }}(s)=\frac \alpha \pi \frac
s3\int_{4m^2}^\infty \frac{\mathrm{d}x}x\frac{\sigma _{e^{\pm
}}(x)}{x-s}\Bigl( 1+\frac{2m^2} x \Bigr) =\frac \alpha {9\pi
}\Bigl[1+48\pi ^2\Bigl( 1+\frac{2m^2}s\Bigr) {\bar J}_{e^{\pm
}}(s)\Bigl].
\end{equation}
In our notation the $O(p^4)$ counterterms contribute as
\[
\Pi _{CT}(s)=16\pi \alpha \Bigl( 2h_2-\frac 19h_4-x_8\Bigr),
\]
where
\begin{eqnarray*}
h_2 &=&h_2^r(\mu )+\frac 1{12}\ \frac 1{(4\pi )^2}\Bigl( \frac
1{d-4}-\frac 12(\ln 4\pi -\gamma +1)\Bigr), \\
h_4 &=&h_4^r(\mu ), \\
x_8 &=&x_8^r(\mu )-\frac 23\frac 1{(4\pi )^2}\Bigl( \frac
1{d-4}-\frac 12(\ln 4\pi -\gamma +1)\Bigr).
\end{eqnarray*}
While $h_2$ renormalizes the divergent part of $\Pi _{\pi ^{\pm
}}(s)$, $x_8$ does the same with $\Pi _{e^{\pm }}(s)$. Note
that in this scheme $\Pi (0)\neq 0$ and, as a consequence,
renormalization of the external photon line has to be included
by means of the factor
\begin{align*}
Z_\gamma ^{1/2} &=1-\frac 12\Pi (0) \\
&=1-8\pi \alpha \left( 2h_2^r(\mu )-\frac 19h_4^r(\mu )-x_8^r(\mu
)\right) +\frac \alpha {6\pi }\left( \ln \frac{m^2}{\mu
^2}+1\right) +\frac \alpha {24\pi }\left( \ln \frac{M_{\pi ^{\pm
}}^2}{\mu ^2}+1\right) .
\end{align*}

\subsection{Corrections to the fermion self energy}

In the same way, we can recalculate the fermion self-energy
\[
\Sigma (q)=\slashed{q}\,\Sigma _V(q^2)+\Sigma _S(q^2),
\]
with the loop and counterterm contributions
\[
\Sigma _{S,V}(q^2)=\Sigma _{S,V}^{loop}(q^2)+\Sigma
_{S,V}^{CT}(q^2),
\]
where (to regularize infrared divergences we have to introduce virtual
photon mass $m_\gamma $)
\begin{eqnarray*}
\Sigma _S^{loop}(q^2) &=&-m\frac \alpha \pi \int_0^1\mathrm{d}x\ln
\frac{x(m^2 - m_\gamma  ^2)-x(1-x)q^2+ m_\gamma ^2}{m^2}-m\frac \alpha
\pi \left( \ln
\frac{m^2}{\mu ^2}+\frac 32\right) \\
&&-m\frac{2\alpha }\pi \left( \frac 1{d-4}-\frac 12(\ln 4\pi -\gamma
+1)\right) , \\
\Sigma _V^{loop}(q^2) &=&\frac \alpha {2\pi }\int_0^1\mathrm{d}x(1-x)\ln
\frac{x(m^2- m_\gamma ^2)-x(1-x)q^2+ m_\gamma ^2}{m^2}+\frac \alpha {4\pi
}\left( \ln \frac{m^2}{\mu ^2}+2\right) \\
&&+\frac \alpha {2\pi }\left( \frac 1{d-4}-\frac 12(\ln 4\pi -\gamma
+1)\right)
\end{eqnarray*}
and
\begin{alignat*}{2}
\Sigma _S^{ct}(q^2) &=-4\pi \alpha mx_7, &\qquad x_7 &=x_7^r(\mu
)-\frac 8{(4\pi )^2}\Bigl( \frac 1{d-4}-\frac 12(\ln 4\pi
-\gamma +1)\Bigr), \\
\Sigma _V^{ct}(q^2) &=-4\pi \alpha x_6, &\qquad  x_6 &=x_6^r(\mu
)+\frac 2{(4\pi )^2}\Bigl( \frac 1{d-4}-\frac 12(\ln 4\pi -\gamma
+1)\Bigr).
\end{alignat*}
{F}rom these formulae, the fermion mass renormalization follows
\begin{eqnarray*}
m &=&m_e+\Sigma _S(m^2)+m\Sigma _V(m^2) \\
&=&m_e-m\left( 4\pi \alpha (x_7^r(\mu )+x_6^r(\mu ))+\frac \alpha
\pi \left( \frac 34\ln \frac{m^2}{\mu ^2}-\frac 14\right) \right),
\end{eqnarray*}
where $m$ is the physical fermion mass. For the fermion wave
function renormalization we need
\begin{eqnarray*}
\frac{\partial \Sigma (q)}{\partial\,
\slashed{q}}|_{\slashed{q}=m} &=&2m\Sigma _S'(m^2)+\Sigma
_V(m^2)+2m^2\Sigma
_V'(m^2) \\
&=&\frac \alpha \pi \left( \frac 12\ln \frac{m^2}{m_\gamma ^2}+\frac 14\ln \frac{m^2}{\mu ^2}-4\pi ^2x_6^r(\mu )-\frac
34\right).
\end{eqnarray*}
Thus, one has
\begin{equation*}
Z_\psi ^{-1} = 1-\frac{\partial \Sigma (q)}{\partial \,
\slashed{q}}\mid_{\slashed{q}=m} =1-\frac \alpha \pi \left( \frac
12\ln \frac{m^2}{m_\gamma ^2}+\frac 14\ln \frac{m^2}{\mu ^2}-4\pi
^2x_6^r(\mu )-\frac 34\right).
\end{equation*}

\subsection{Corrections to the form factors $F_{1,2}$}

We have the following standard formula for $M_{\pi ^0}^2>s>4m^2$,
\begin{equation}
F_2(s)=\frac \alpha \pi \frac{m^2}{s\sigma _e(s)}\left[ \ln \left(
\frac{1-\sigma _e(s)}{1+\sigma _e(s)}\right) +\mathrm{i}\pi
\right]\label{appF2}
\end{equation}
and
\begin{eqnarray*}
(F_1(s)-1)\Bigr|_\text{loop} &=&-\frac 12\frac \alpha \pi \Bigl(
\frac 1{d-4}-\frac 12(\ln 4\pi -\gamma +1)\Bigr) +\frac \alpha \pi
\Bigl\{-\frac 34-\frac 14\ln
\frac{m^2}{\mu ^2}+4\pi ^2\overline{J}_e(s) \\
&&+\frac 12\Bigl( 1-m^2\frac \partial {\partial m^2}16\pi
^2\overline{J}_e(s)\Bigr) \Bigr\} \\
&&+\frac \alpha \pi \Bigl( \frac 12s-m^2\Bigr) \Bigl\{\Bigl( \frac
12\ln \frac{m^2}{m_\gamma ^2}-1\Bigr) \frac 1{m^2}\Bigl( 1-m^2\frac
\partial {\partial m^2}16\pi ^2{\bar J}_e(s)\Bigr) \\
&&+\frac 12\int_0^1\frac 1{m^2-sx(1-x)}\ln
\frac{m^2-sx(1-x)}{m^2}\Bigr\}.
\end{eqnarray*}
The counterterm contribution is
\[
F_1^{CT}(s)=4\alpha \pi x_6.
\]
We can compare now
\[
Z_1^{-1}=F_1(0)=1+\frac \alpha \pi \Bigl\{\frac 34-\frac 14\ln
\frac{m^2}{\mu ^2} -\frac 12\ln \frac{m^2}{m_\gamma ^2}+4\pi
^2x_6^r(\mu )\Bigr\}=Z_\psi ^{-1},
\]
where the last identity is a consequence of the Ward identity.

\subsection{Complete $1\gamma R$ LO+NLO form factors}

Putting the results of the previous subsections together one obtains
(note that we must include the external line renormalization
factor $Z_\gamma ^{1/2}Z_\psi$)
\begin{align}
P^{1\gamma R;LO}(x,y)+P^{1\gamma R;NLO}(x,y) &\,=\,
\frac{e^3N_C}{12\pi^2 F_\pi}\,
\frac 1{xM_{\pi^0}^2}\,\frac{\mathrm{i}}m\,F_2(xM_{\pi ^0}^2),  \notag \\
A^{1\gamma R;LO}(x,y)+A^{1\gamma R;NLO}(x,y) &\,=\,
-\frac{e^3N_C}{12\pi ^2F_{\pi ^0}}\,\frac{\mathrm{i}}{xM_{\pi
^0}^2}\Bigl[ F_1(xM_{\pi ^0}^2)-\Pi (xM_{\pi
^0}^2) \notag \\
&\qquad\qquad+ (Z_\gamma ^{1/2}-1)+(Z_\psi
-1)+a_{NLO}^{ChPT}(xM_{\pi
^0}^2)\Bigr],  \notag \\
T^{1\gamma R;LO}(x,y)+T^{1\gamma R;NLO}(x,y) & \,=\,
-\frac{e^3N_C}{12\pi ^2F_{\pi ^0}}\,\frac{\mathrm{i}}{xM_{\pi
^0}^2}\Bigl[ 2m(F_1(xM_{\pi ^0}^2)-\Pi
(xM_{\pi ^0}^2) \label{PATatNLO} \\
+ (Z_\gamma ^{1/2}&-1) +(Z_\psi -1)+\frac{xM_{\pi ^0}^2}{2m}
F_2(xM_{\pi ^0}^2)+a_{NLO}^{ChPT}(xM_{\pi ^0}^2))\Bigr]. \notag
\end{align}
Let us write $\overline{\Pi }(s)=\Pi (s)-\Pi (0)=\Pi (s)+(Z_\gamma
-1)$ and define $\overline{F}_1(s)=1+F_1(s)-F_1(0)=F_1(s)+(Z_\psi
-1)$. Explicitly, for $M_{\pi ^0}^2>s>4m^2$,
\begin{align}
\overline{\Pi }(s) &=\overline{\Pi }_{\pi^{\pm
}}(s)+\overline{\Pi}_{e^{\pm }}(s) =\frac \alpha \pi \Bigl\{
1+\frac 23\frac{2m^2-M_{\pi ^{\pm }}^2}s-\frac 16\sigma _{\pi
^{\pm }}^2(s) \bigl| \sigma_{\pi^{\pm}}(s) \bigr| \arctan \Bigl(
\frac 1{|\sigma _{\pi ^{\pm }}(s)|}\Bigr)   \notag \\
&\phantom{=\frac \alpha \pi}\;+ \frac 1{3s\sigma _e(s)}\ln \Bigl(
\frac{1-\sigma _e(s)}{1+\sigma _e(s)}\Bigr) \Bigl(
s-2m^2-\frac{8m^4}s\Bigr)
+\frac{\mathrm{i}\pi }{3s\sigma _e(s)}\Bigl( s-2m^2-\frac{8m^4}s\Bigr) \Bigr\},\label{Pibar} \\
\overline{F}_1(s) &=1+\frac \alpha \pi \Bigl\{ -1+\frac 1{s\sigma
_e(s)}\Bigl[ (2m^2-\frac 34s)\ln \Bigl(
\frac{1-\sigma_e(s)}{1+\sigma _e(s)}
\Bigr)  \notag \\
&\phantom{=1+\frac \alpha \pi}\;\;\,-(s-2m^2)\Bigl( \frac 14\ln
\Bigl( \frac{1-\sigma _e(s)^2}{4\sigma _e(s)^2} \Bigr) \ln \Bigl(
\frac{1-\sigma _e(s)}{1+\sigma _e(s)}\Bigl) +\mathrm{Li}
_2\Bigl( \frac{\sigma _e(s)-1}{2\sigma _e(s)}\Bigl)  \notag \\
&\phantom{=1+\frac \alpha \pi}\;\;\,+\frac 12\ln \Bigl(
\frac{1-\sigma _e(s)}{2\sigma _e(s)} \Bigr) \ln \Bigl(
\frac{1+\sigma _e(s)}{2\sigma _e(s)}\Bigr) -\frac{\pi
^2} 3\Bigr) \Bigr]  \notag \\
&\phantom{=1+\frac \alpha \pi}\;\;\, +\frac{\mathrm{i}\pi
}{s\sigma _e(s)}\Bigl[ (2m^2-\frac 34s)-\frac 12(s-2m^2)\ln \Bigl(
\frac{1-\sigma _e(s)^2}{4\sigma
_e(s)^2}\Bigr) \Bigr] \Bigr\}  \notag \\
&\phantom{=1+\frac \alpha \pi}\;\;\, +\frac \alpha {2\pi }\ln
\Bigl( \frac{m^2}{m_\gamma^2}\Bigr) \Bigl\{ 1+(s-2m^2)\frac
1{s\sigma _e(s)}\Bigl[ \ln \Bigl( \frac{1-\sigma _e(s)}{1+\sigma
_e(s)}\Bigr) +\mathrm{i}\pi \Bigr] \Bigr\}.  \label{F1bar}
\end{align}
Then, taking $\overline{F}_2(s)=F_2(s)$ and introducing a physical
charge $\overline{e}=eZ_\gamma ^{1/2}$ (where
$\overline{e}^2/(4\pi )=\alpha =1/137,\ldots $), we can rewrite
(\ref{PATatNLO}) in the form
\begin{align*}
P^{1\gamma R,L}(x,y)+P^{1\gamma R,NL}(x,y)
&=\frac{\overline{e}^3N_C}{12\pi ^2F_\pi }\,\frac 1{xM_{\pi
^0}^2}\,\frac{\mathrm{i}}m\,\overline{F}_2(xM_{\pi
^0}^2), \\
A^{1\gamma R,L}(x,y)+A^{1\gamma R,NL}(x,y)
&=-\frac{\overline{e}^3N_C}{12\pi ^2F_\pi
}\,\frac{\mathrm{i}}{xM_{\pi ^0}^2}\Bigl[ \overline{F}_1(xM_{\pi
^0}^2)-\overline{\Pi }(xM_{\pi^0}^2)+a_\text{\it NLO}^\text{\it ChPT}(xM_{\pi ^0}^2)\Bigr],\\
T^{1\gamma R,L}(x,y)+T^{1\gamma R,NL}(x,y)
&=-\frac{\overline{e}^3N_C}{12\pi ^2F_\pi
}\,\frac{\mathrm{i}}{xM_{\pi ^0}^2}\Bigl[
2m(\overline{F}_1(xM_{\pi^0}^2) -\overline{\Pi }(xM_{\pi^0}^2)\\
&\phantom{=}\qquad\qquad\qquad\quad +a_{NLO}^\text{\it
ChPT}(xM_{\pi^0}^2)) +\frac{xM_{\pi
^0}^2}{2m}\overline{F}_2(xM_{\pi ^0}^2)\Bigr].
\end{align*}
Identifying now the leading order amplitude with the substitution
$\overline{F}_1=1$, $\overline{F}_2=\overline{\Pi
}=a_{NLO}^{ChPT}=0$ in the above expressions, and using
(\ref{deltaPAT}) we obtain
\begin{multline*}
\delta _{NLO}^{1\gamma R}(x,y)=2\mathrm{Re}\Bigl[
\overline{F}_1(xM_{\pi ^0}^2)-\overline{\Pi }(xM_{\pi
^0}^2)+a_{NLO}^{ChPT}(xM_{\pi ^0}^2)\\+\frac{2xM_{\pi
^0}^2}{M_{\pi^0} ^2x(1+y^2)+4m^2}F_2(xM_{\pi ^0}^2)-1\Bigr]
\end{multline*}
and
\[
\delta _{NLO}^{1\gamma R}(x)=2\mathrm{Re}\Bigl[
\overline{F}_1(xM_{\pi ^0}^2)-\overline{\Pi }(xM_{\pi
^0}^2)+a_{NLO}^{ChPT}(xM_{\pi ^0}^2)+\frac 32\frac{xM_{\pi
^0}^2}{M_{\pi^0} ^2x+2m^2}F_2(xM_{\pi ^0}^2)-1\Bigr].
\]

Before concluding this section, let us give a brief survey of the IR
divergent contributions. They can be extracted from the formulae
given above and read
\begin{eqnarray*}
\overline{F}_1(s)_{IRdiv} &=&\frac \alpha {2\pi }\ln \left(
\frac{m^2}{m_\gamma ^2}\right) \left\{ 1+(s-2m^2)\frac 1{s\sigma
_e(s)}\left[ \ln \left( \frac{1-\sigma _e(s)}{1+\sigma
_e(s)}\right) +\mathrm{i}\pi \right] \right\}
, \\
\overline{F}_2(s)_{IRdiv} &=&\overline{\Pi }(s)_{IRdiv}=0.
\end{eqnarray*}
Thus, the IR divergent parts of the form factors are
\begin{eqnarray*}
\delta P_{IRdiv}(x,y) &=&0, \\
\delta A_{IRdiv}(x,y) &=&\frac 1{2m}\delta
T_{IRdiv}(x,y)=-\frac{e^3N_C}{12\pi ^2F_\pi
}\,\frac{\mathrm{i}}{xM_{\pi ^0}^2}\,\overline{F}_1(xM_{\pi
^0}^2)_{IRdiv}.
\end{eqnarray*}
Inserting these expressions into formula (\ref{deltaPAT}) yields,
after some simple algebra,
\begin{equation}
\delta _{NLO}^{1\gamma R}(x,y)_{IRdiv}=\frac{e^2}{(2\pi )^2}\ln
\left( \frac{m^2}{m_\gamma ^2}\right) \left\{ 1+\left(
1-\frac{2m^2}{xM_{\pi ^0}^2}\right) \frac 1{\sigma _e(xM_{\pi
^0}^2)}\ln \left( \frac{1-\sigma _e(xM_{\pi ^0}^2)}{1+\sigma
_e(xM_{\pi ^0}^2)}\right) \right\}.
\end{equation}

\indent

\section{Loop functions}\label{App:loop}
\renewcommand{\theequation}{E.\arabic{equation}}
\setcounter{equation}{0}

This appendix is devoted to the so-called Passarino-Veltman
\cite{PassVelt} one-loop integrals used in the main text. Generally
one defines (working in $d$ dimensions):\footnote{Notice that
according to this definition the loop functions are renormalization
scale dependent and consequently the bare LECs are also scale
dependent.}
\begin{equation}
i \pi^2 T_0(n) = (2\pi\mu)^4 \int\frac{d^d l}{(2\pi\mu)^d}
\frac{1}{[l^2 - m_1^2]\ldots[(l+p_n)^2-m_n^2]}.
\end{equation}
It is, then, common to denote these $n$-point functions in
alphabetical order, i.e. instead of $T$ one uses for $1$-point
integral the symbol $A$, for $n=2$ -- the $B$ and so on. For the
scalar functions and special combinations of arguments needed in
our work we get successively
\newcommand{\Li}{\mathrm{Li}}
\begin{equation}
B_0(0,m^2,m^2)=-2\Bigl[ \frac 1{d-4}-\frac 12(\ln 4\pi -\gamma
+1)\Bigr] -\ln \frac{m^2}{\mu ^2}-1
,\label{B0a}
\end{equation}
\begin{eqnarray}
B_0(m_{\pm }^2,0,m^2)&=& -2\bigg[ \frac
1{d-4}-\frac 12(\ln 4\pi -\gamma +1)\Bigr]
-\ln \frac{m^2}{\mu ^2}+16\pi ^2{\bar J}_{0m}(m_{\pm }^2)
\nonumber\\
& =& -2\bigg[ \frac 1{d-4}-\frac 12(\ln 4\pi -\gamma +1)\bigg] -\ln
\frac{m^2}{\mu ^2}+1-\Bigl( 1-\frac{m^2}{m_{\pm }^2}\bigg) \ln
\Bigl( 1-\frac{m_{\pm}^2}{m^2}\Bigr)\nonumber
\\
.\label{B0b}
\end{eqnarray}

\begin{equation}
C_0 (0,m_\pm^2,m^2;m^2,m^2,0) = \frac{\pi^2 - 6 \Li_2
(\frac{m_\pm^2}{m^2}+ i\epsilon)}{6(m_\pm^2-m^2)}.
\end{equation}
\begin{align}
{\rm Re}\; &C_0(m^2,M_\pi^2,m_\pm^2;m^2,0,0)\Bigl|_{m<m_\pm<M_\pi}
= \frac{1}{\sqrt{\lambda}}\Bigl\{\,2\,
\Li_2\Bigl(\frac{\sqrt{\lambda}+M_\pi^2}{M_\pi^2}\Bigr)\notag\\&+
\Li_2\Bigl(1-\frac{2\sqrt{\lambda}}{\sqrt{\lambda}-m_\pm^2+m^2+M_\pi^2}\Bigr)
-\Li_2\Bigl(1-\frac{2\sqrt{\lambda}}{\sqrt{\lambda}+m_\pm^2-M_\pi^2-m^2}\Bigr) \notag\\
&-\Li_2\Bigl( 1+\frac{2\sqrt{\lambda} m^2}{
(m_\pm^2-m^2)(\sqrt{\lambda}+m_\pm^2-m^2)-(m_\pm^2+m^2)M_\pi^2}\Bigr)\\
&+\Li_2\Bigl( 1+\frac{2\sqrt{\lambda} (m^2-m_\pm^2)}{
(m_\pm^2-m^2)(\sqrt{\lambda}+m_\pm^2-m^2)-(m_\pm^2+m^2)M_\pi^2}\Bigr)\notag\\
&-\Li_2\Bigl( 1-\frac{2\sqrt{\lambda} m_\pm^2}{
(m_\pm^2-m^2)(\sqrt{\lambda}-m_\pm^2+m^2)+(m_\pm^2+m^2)M_\pi^2}\Bigr)
-\frac{\pi^2}{6} \notag \Bigr\}\notag
\end{align}
with $\lambda = \lambda (M_\pi^2,m_\pm^2,m^2) =
(M_\pi^2-m^2-m_\pm^2)^2-4m^2 m_\pm^2$, and
$m_{\pm }^2=m^2+\delta m_{\pm
}^2$, where $\delta m_{\pm }^2=2k\cdot q_{1,2}=\frac 12(1-x)(1\pm
y)M_\pi ^2$.

The four-point function appearing in Section 4.3 is given by:
\begin{multline}
{\rm Re}\,D_0(m^2,0,m^2,M_\pi^2,m_+^2,m_-^2; 0,m^2,m^2,0) =
\frac{2y}{M_\pi^2 m^2 (y^2-1)}\\ \times\Bigl\{
\log\frac{(m_+^2-m^2)(m_-^2-m^2)}{M_\pi^2 m^2} \log y +
\Li_2(1-y)-\Li_2(1-y^{-1})\Bigr\},
\end{multline}
where $y = \frac{1}{2a}(-b + \sqrt{b^2 - 4ac})$, with
\begin{align*}
a = c =\frac{M_\pi^2}{m^2},\qquad b =
\frac{-1}{m^4}\bigl((m_+^2-m^2)(m_-^2-m^2)+2 M_\pi^2 m^2\bigr).
\end{align*}

Asymptotics of the loop functions for $k\rightarrow 0$ ($x\rightarrow 1$),
$m$ fixed, read:
\begin{align*}
&B_0(m_{\pm }^2;0,m^2)=2+B_0(0,m^2,m^2)-\frac{\delta m_{\pm
}^2}{m^2}\Bigl[ \ln \Bigl( \frac{\delta m_{\pm }^2}{m^2}\Bigr)
+\mathrm{i}\pi \Bigr] +O\Bigl( \Bigl( \frac{\delta m_{\pm
}^2}{m^2}\Bigr)^2\Bigr),\\
&C_0(0,m_{\pm }^2,m^2;m^2,m^2,0) = \frac 1{m^2}\Bigl[\ln \Bigl(
\frac{\delta m_{\pm }^2}{m^2}\Bigr) +\mathrm{i}\pi -1\Bigr] \\
&\hspace*{6cm}-\frac 14\frac{\delta m_{\pm }^2}{m^2}\Bigl[2\ln
\Bigl( \frac{\delta m_{\pm }^2}{m^2}\Bigr) +O(1)\Bigr]+O\Bigl(
\Bigl(\frac{\delta m_{\pm }^2}{m^2}\Bigr)^2 \Bigr), \\
&C_0(m^2,M_\pi ^2,m_{\pm }^2,m^2,0,0) = C_0(m^2, M_\pi^2, m^2, m^2
, 0, 0) - \frac 1{M_\pi ^2}\frac{\delta m_{\pm }^2}{m^2}\Bigl[ \ln
\Bigl( \frac{\delta m_{\pm }^2}{m^2}\Bigr) +O(1) \Bigr],\\
&D_0(m^2,0,m^2,M_\pi ^2,m_{+}^2,m_{-}^2,0,m^2,m^2,0)=\frac
1{m^2M_\pi ^2}\ln \Bigl( \frac{\delta m_{+}^2\delta m_{-}^2}{M_\pi
^2m^2}\Bigr) +O(1).
\end{align*}
Asymptotics of the loop functions  for
$m\rightarrow 0$, $\delta m_{\pm }^2$ $>0$ fixed, read:
\begin{align*}
&\mathrm{Re}\,B_0(m_{\pm }^2;0,m^2)= -2\Bigl( \frac 1{d-4}-\frac
12(\ln 4\pi-\gamma +1)\Bigr) +1-\ln \Bigl( \frac{\delta m_{\pm }^2}{\mu ^2}\Bigr) + O(m^2), \\
&\mathrm{Re}\,C_0(0,m_{\pm }^2,m^2;m^2,m^2,0) = \frac 1{\delta
m_{\pm }^2}\Bigl[ \frac 12\ln ^2\Bigl( \frac{m^2}{M_\pi ^2}\Bigr)
-\ln \Bigl( \frac{m^2}{M_\pi ^2}\Bigr) \ln \Bigl( \frac{\delta
m_{\pm }^2}{M_\pi ^2} \Bigr)\\
&\hspace*{8cm} +\frac 12\ln ^2\Bigl( \frac{\delta m_{\pm
}^2}{M_\pi
^2}\Bigr) - \frac{\pi^2}6+O(m^2)\Bigr], \\
&\mathrm{Re}\,C_0(m^2,M_\pi ^2,m_{\pm }^2,m^2,0,0)=\frac 1{M_\pi
^2-\delta m_{\pm }^2}\Bigl[ 2\mathrm{Li}_2\Bigl( 1-\frac{\delta
m_{\pm }^2}{M_\pi ^2} \Bigr) \\
& \hspace*{5cm}  +\frac 12\ln \Bigl( \frac{\delta m_{\pm
}^2}{M_\pi ^2}\Bigr) \biggl( 2\ln \Bigl( \frac{m^2}{M_\pi
^2}\Bigr) -\ln
\Bigl( \frac{\delta m_{\pm }^2}{M_\pi ^2}\Bigr) \biggr) +O(m^2)\Bigr], \\
&\mathrm{Re}\,D_0(m^2,0,m^2,M_\pi ^2,m_{+}^2,m_{-}^2,0,m^2,m^2,0)
= \frac 2{\delta m_{+}^2\delta m_{-}^2}\Bigl[ \frac 12\ln ^2\Bigl(
\frac{m^2}{M_\pi ^2}\Bigr)\\
& \hspace*{3.5cm} -\ln \Bigl( \frac{m^2}{M_\pi ^2}\Bigr) \ln
\Bigl( \frac{\delta m_{+}^2\delta m_{-}^2}{M_\pi ^4}\Bigr) +\frac
12\ln ^2\Bigl( \frac{\delta m_{+}^2\delta m_{-}^2}{M_\pi ^4}
\Bigr) -\frac{\pi ^2}3+O(m^2)\Bigr].
\end{align*}

\indent

\section{Soft photon singularities}\label{App:sps}
\renewcommand{\theequation}{F.\arabic{equation}}
\setcounter{equation}{0}

In this Appendix, we briefly address the question of soft photon
singularities, which are of relevance for the discussion in
Section 3.3. We wish in particular to elaborate in somewhat
greater detail on the statement made at the beginning of Section
3.3, concerning the absence of contributions that are independent
of $k_\mu$ in the difference $\mathcal{M}_{\pi^0\rightarrow e^+
e^-\gamma } - \mathcal{M}_{\pi^0\rightarrow e^+ e^-\gamma
}^{Low}$.
In the present context, we may arrive at this result as follows.
First, note that the Ward identity (\ref{WI1psiR}) can be solved
by the expression\footnote{Of course, the minimal solution can be
written in the form
\[
\Gamma ^{1\psi R}=e[\frac{p_-}{(p_-\cdot k)}\Gamma _{\pi
^0e^{-}e^{+}}(p_- +k,p_+)-\frac{p_+}{(p_+\cdot k)}\Gamma _{\pi
^0e^{-}e^{+}}(p_-,p_+ +k)]
\]
which takes into account only the leading order singularity for
$k\rightarrow 0$.}
\begin{multline*}
\Gamma _\mu ^{1\psi R,pole}(p_+,p_-,k)=e\frac{2p_{-\mu
}-\frac{a_e}m(p_{-\mu }\, \slashed{k}-\gamma _\mu (p_-\cdot
k))-\mathrm{i}\sigma _{\mu \nu }k^\nu (1+a_e)}{2(p_-\cdot k)}
\Gamma _{\pi ^0e^{-}e^{+}}(p_- +k,p_+) \\
-e\Gamma _{\pi ^0e^{-}e^{+}}(p_-,p_+ +k)\frac{2p_{+\mu }-\frac{a_e}
m(p_{+\mu }\, \slashed{k}-\gamma _\mu (p_+\cdot
k))+\mathrm{i}\sigma _{\mu \nu }k^\nu (1+a_e)}{2(p_+\cdot k)}
\end{multline*}
(where $a_e=\overline{F}_2(0)$ is the anomalous magnetic moment of
the fermion), which includes the leading and next-to-leading order
singularities for $k\rightarrow 0$. Indeed, for the combination
\[
\varepsilon ^\mu (k)^{*}\overline{u}\Lambda _\mu
(p_-,p_- +k)S(p_- +k)v,
\]
it is not difficult to prove that, for $k$ such that $(p_-\cdot
k)\rightarrow 0$, with $p_-^2=m^2$ and $p_-$ fixed,
\begin{multline*}
\varepsilon ^\mu (k)^{*} \overline{u}\Lambda _\mu (p_-,p_-+k)S(p_-+k) \\
=\varepsilon ^\mu (k)^{*}\overline{u}\frac{2p_{-\mu }-\frac{a_e}
m(p_{-\mu }\, \slashed{k}-\gamma _\mu (p_-\cdot
k))-\mathrm{i}\sigma _{\mu \nu }k^\nu (1+a_e)}{2(p_-\cdot k)}
+{\cal O}(1)+{\cal O}(k,(p_-\cdot k)).
\end{multline*}
Here (and in what follows), the remaining ${\cal O}(1)$ terms, which are
not written explicitly, are independent of $k$. In the same way,
for $k,\,(p_+\cdot k)\rightarrow 0$, $p_+^2=m^2$ and $p_+$ fixed,
we find
\begin{multline*}
S(-p_+ -k))\Lambda _\mu (-p_+ -k,-p_+)v\varepsilon ^\mu (k)^{*} \\
=
\frac{-2p_{+\mu }+\frac{a_e}m(p_{+\mu }\, \slashed{k}-\gamma_\mu
(p_+\cdot k))-\mathrm{i}\sigma _{\mu \nu }k^\nu
(1+a_e)}{2(p_+\cdot k)} v\varepsilon ^\mu (k)^{*}
+{\cal O}(1)+{\cal O}(k,(p_+\cdot k)).
\end{multline*}
Notice that if in the above expressions one restricts
the vertex function $\Lambda_\mu(q_1,q_2)$ to its longitudinal
part given by Eq. (\ref{longitudinal}), one arrives at the same
expression, but with $a_e$ replaced by $-\Sigma_V(m^2)$,
which is both gauge dependent and infrared divergent.
Including the contribution from the transverse part
$\Lambda_\mu^T(q',q)$ cures both problems, and yields
the anomalous magnetic moment $a_e$. This can be checked
explicitly at the one loop level with the expressions available
in Refs. \cite{Ball&Chiu1980} and \cite{Kizilersu1995}.

Let us recall that the one-particle irreducible (semi-)off-shell
$\pi ^0$-$e^{-}$-$e^{+}$
vertices $\Gamma _{\pi ^0e^{-}e^{+}}(p_- +k,p_+)$ and $\Gamma_{\pi
^0e^{-}e^{+}}$ $(p_-, p_++k)$ are free of poles for $(p_\pm\cdot
k)\rightarrow 0$. The same is true for the one-particle
irreducible (semi-)off-shell
$e^{+}$-$e^{-}$-$\gamma$ vertices $\Lambda _\mu (p_-,p_- +k)$ and
$\Lambda _\mu (-p_+ -k,-p_+)$. We have therefore, for $(p_\pm \cdot
k),\,k\rightarrow 0$ and $p_\pm$ fixed, according to
(\ref{PATamplitude}),
\begin{eqnarray*}
\mathcal{M}_{\pi^0\rightarrow e^+ e^-\gamma }^{1\psi R} &=&e\varepsilon ^\mu
(k)^{*}\overline{u}[\frac{2p_{-\mu }-\frac{a_e}m(p_{-\mu }\,
\slashed{k}-\gamma _\mu (p_-\cdot k))-
\mathrm{i}\sigma _{\mu \nu }k^\nu (1+a_e)}{2(p_-\cdot k)} \\
&&+\frac{-2p_{+\mu }-\frac{a_e}m(p_{+\mu }\,
\slashed{k}-\gamma_\mu
(p_+\cdot k))-\mathrm{i}\sigma _{\mu \nu }k^\nu (1+a_e)}{2(p_+\cdot k)}]
\gamma^5vP_{\pi^0e^{-}e^{+}}(m^2,m^2) \\
&&+{\cal O}(1)+{\cal O}(k,(p_+\cdot k)),
\end{eqnarray*}
where, as above, the implicit ${\cal O}(1)$ terms are independent
of $k$. From this formula we can read off the associated Low
amplitude given in Eq. (\ref{lowAmplitude}) which, according to
Low's theorem~\cite{lows}, corresponds to the leading singular
terms in the expansion of the complete amplitude in $k,$
$(p_\pm\cdot k)\rightarrow 0$ , (with $p_\pm$ fixed) in the sense
that the $k$-independent ${\cal O}(1)$ terms coming from
$\Gamma_\mu ^{1\psi R}$ are in fact cancelled in the complete
amplitude $\mathcal{M}_{\pi^0\rightarrow e^+ e^-\gamma }\,$ by the
corresponding ${\cal O}(1)$ terms from the $\Gamma _\mu ^{1PI}$
(let us recall that the one-photon reducible amplitude is of order
${\cal O}(k)$). On the other hand, we have\footnote{Here we use
the identities
\[
\frac{2p_{-\mu }-\frac{a_e}m(p_{-\mu }\, \slashed{k}-\gamma _\mu
(p_- \cdot k))-\mathrm{i}\sigma _{\mu \nu }k^\nu
(1+a_e)}{2(p_-\cdot k)}=\left( \gamma _\mu +\frac i{2m}a_e\sigma
_{\mu \nu }k^\nu \right) \frac 1{(\slashed{p}_- +\slashed{k})-m}
\]
and
\[
S^{-1}(p_-+k)=(1-\Sigma _V(m^2))((\slashed{p}_-
+\slashed{k})-m)+{\cal O}(p_-\cdot k).
\]
}
\[
\mathcal{M}_{\pi^0\rightarrow e^+ e^-\gamma }^{1\psi R,pole}
=\varepsilon ^\mu (k)^{*}\overline{u}\Gamma
_\mu ^{1\psi R,pole}v=
\mathcal{M}_{\pi^0\rightarrow e^+ e^-\gamma }^{Low}+
{\cal O}(1)+{\cal O}((p_+\cdot k),k).
\]
Therefore, the following subtracted quantity
\[
\Gamma _\mu ^{1\psi R,reg}=\Gamma _\mu ^{1\psi R}-\Gamma _\mu ^{1\psi
R,pole}={\cal O}(1) + {\cal O}((q_1\cdot k),k)
\]
is both transverse and with the ${\cal O}(1)$ terms independent of
$k$. It can thus be expressed in terms of  form factors $P$, $A_\pm$,
$T$, see~(\ref{Formfactors}),
\begin{align*}
\Gamma ^{1\psi R,reg}(p_+,p_-,k) &=P^{1\psi R,reg}(x,y)[(k\cdot
p_+)p_-^\mu
-(k\cdot p_-)p_+^\mu ]\gamma_5 \\
&+A_{+}^{1\psi R,reg}(x,y)[\slashed{k}\,p_+^\mu -(k\cdot
p_+)\gamma ^\mu ]\gamma_5 -A_{-}^{1\psi
R,reg}(x,y)[\slashed{k}\,p_-^\mu -(k\cdot p_-)\gamma ^\mu
]\gamma_5 \\
&-\mathrm{i}T^{1\psi R,reg}(x,y)\sigma ^{\mu \nu }k_\nu \gamma_5.
\end{align*}
Because these form factors are in fact ${\cal O}(1)$, i.e. $\Gamma
_\mu ^{1\psi R,reg}={\cal O}((p_+\cdot k),k)$ and also $\Gamma _\mu
^{1\gamma R}={\cal O}(k)$, we may conclude that the contribution of
$\Gamma _\mu ^{1\psi R,reg}$ is tiny (it is suppressed by a factor
$\alpha$ w.r.t. $\Gamma _\mu ^{1\gamma R}$) in the full kinematical
region $\nu ^2\leq x\leq 1$. On the other hand, we should expect
that the remaining gauge invariant combination, namely
\begin{equation}
\mathcal{M}_{\pi^0\rightarrow e^+ e^-\gamma }^{1PI}+
\mathcal{M}_{\pi^0\rightarrow e^+ e^-\gamma }^{1\psi
R,pole}=\mathcal{M}_{\pi^0\rightarrow e^+ e^-\gamma }^{Low}+
{\cal O}(k,(p_\pm\cdot k)),
\label{Mdominant}
\end{equation}
might be important for $x\ $sufficiently close to one (\emph{i.e}.
$k\rightarrow 0$) in spite of the suppression by a factor $\alpha$.
In the formula (\ref{Mdominant}), the one-particle irreducible
part of the amplitude
\[
\mathcal{M}_{\pi^0\gamma ^{*}\gamma ^{*}}^{1PI}  =
\overline{u}\Gamma _\mu
^{1PI}(p_+,p_-,k)v\varepsilon ^\mu (k)^{*}
\]
corresponds to the photon emission from internal lines, being therefore of
the order ${\cal O}(1)$ for $k\rightarrow 0$. Notice also, that the Low amplitude
is transverse, therefore we can decompose it in terms of $P^{Low}$, $A_{\pm
\text{ }}^{Low}$ and $T^{Low}$ form factors, where
\begin{eqnarray}
P^{Low} &=&e\,\frac{P_{\pi ^0e^{-}e^{+}}(m^2,m^2)}{(p_-\cdot
k)(p_+\cdot k)}=16e\,\frac{P_{\pi ^0e^{-}e^{+}}(m^2,m^2)}{M_{\pi^0}
^4(1-x)^2(1-y^2)},
\nonumber \\
A_{\pm \text{ }}^{Low} &=&e\,\frac{a_e}m\frac{P_{\pi
^0e^{-}e^{+}}(m^2,m^2)}{2(k\cdot
p_\pm)}=2e\,\frac{a_e}m\frac{P_{\pi ^0e^{-}e^{+}}(m^2,m^2)}{M_{\pi^0}
^2(1-x)(1\pm y)},  \nonumber \\
T^{Low} &=&e(1+a_e)P_{\pi ^0e^{-}e^{+}}(m^2,m^2)\left( \frac 1{2(p_-\cdot
k)}+\frac 1{2(p_+\cdot k)}\right)  \nonumber \\
&=&4e(1+a_e)\frac{P_{\pi ^0e^{-}e^{+}}(m^2,m^2)}{M_{\pi^0}
^2(1-x)(1-y^2)}. \label{LowPAT}
\end{eqnarray}


\end{document}